\documentclass[]{elsarticle}

\usepackage{lineno}
\modulolinenumbers[5]

\usepackage{bm}
\usepackage{amsmath,amsfonts,amssymb}
\usepackage{xcolor}
\usepackage{graphicx}
\usepackage{subcaption}
\usepackage{float}
\usepackage{siunitx}
\usepackage[hidelinks]{hyperref} 
\usepackage[capitalise]{cleveref}
\usepackage{grfext}

\usepackage{booktabs}
\usepackage{fancyhdr}

\AtBeginDocument{%
  \PrependGraphicsExtensions*{
    .mps,.MPS,.pdf,.PDF,.eps,.EPS,.ps,.PS,.jpg,.jpeg,.JPG,.JPEG,
    .png,.PNG,
    .funny,.foobar
  }%
  \PrintGraphicsExtensions 
}

\DeclareMathAlphabet{\mathcal}{OMS}{cmsy}{m}{n}

\journal{TBD}

\newcommand{\ie}{i.e.\ }
\newcommand{\eg}{e.g.\ }

\newcommand{\pd}[2]{\frac{\partial #1}{\partial #2}}
\renewcommand{\d}[2]{\frac{\mathrm{d} #1}{\mathrm{d} #2}}
\renewcommand{\vec}[1]{\bm{#1}}

\newcommand{\norm}[1]{\left\lVert#1\right\rVert}

\newcommand{\cost}{\mathcal{C}}
\newcommand{\constraint}{\mathcal{G}}
\newcommand{\desVar}{\hat{\gamma}}

\newcommand{\desVarMin}{\desVar_\text{min}}
\newcommand{\desVarMax}{\desVar_\text{max}}

\newcommand{\gammaf}{{\gamma}^*}
\newcommand{\gammap}{\gamma}
\newcommand{\neighbor}{\mathbb{N}}
\newcommand{\lagran}{\mathcal{L}}

\fancypagestyle{firstpage}{%
	\fancyfoot[R]{\small\textit{\today}}
	\fancyfoot[L]{\small\it Preprint submitted to TBD\\ \textcopyright~2023 National Research Council of Canada}
}

\begin{document}
\begin{frontmatter}
\title{Thermofluid topology optimization for cooling channel design}

\author{Farshad Navah\corref{navah}}
\cortext[navah]{Currently with BMO Financial Group}
\author{Marc-\'Etienne Lamarche-Gagnon\corref{mycorrespondingauthor}}
\cortext[mycorrespondingauthor]{Corresponding author}
\ead{marc-etienne.lamarche-gagnon@nrc-cnrc.gc.ca}
\author{Florin Ilinca}
\address{Simulation and Numerical Modeling\\
	Automotive and Surface Transportation Research Center\\
	National Research Council Canada\\
	Boucherville, QC J4B 6Y4, Canada\\}




\begin{abstract}
A framework for topology optimization of cooling channels is proposed, which paves the way towards automated design of additively-manufactured cooling channels, required in applications such as the efficient heat management of die casting molds. Combining a selection of pertinent techniques and methods, the proposed density-based approach is strengthened by systematic verification and validation steps, including the body-fitted meshing of an optimized design. Furthermore, this work features applications to simplified, yet industrially-relevant cases, as well as a detailed discussion of the effects of the hyper-parameters of the optimization problem. These enable the reader to acquire a better understanding of the control and regularization mechanisms, which are necessary for a robust development towards complex scenarios. 
\end{abstract}

\end{frontmatter}
\thispagestyle{firstpage}

\section{Introduction}
Cooling channels are a key component in many technological systems involving heat extraction such as generic heat exchangers, advanced electronic devices, injection molding and die casting. In particular, the presence of cooling channels in molds has a crucial impact on cycle time, shape deviations and residual stresses of the fabricated part \cite{fengDesignFabricationConformal2021}. Lately, with the advent of advanced fabrication technologies such as additive manufacturing, the use of channels conformal to the geometry of the part (i.e.~\textit{conformal cooling channels}) has gained interest over the use of typical straight-drilled channels, because of the increased heat removal efficiency \cite{kanburDesignOptimizationConformal2020}. This translates into a faster cooling time and can lead to a reduced warpage of the part \cite{eiamsa-ardConformalBubblerCooling2015}, thanks to a more uniform temperature distribution. However, the design of conformal channels can be challenging and time consuming, especially for complex parts. An improper design may, on the one hand, not lead to the expected enhanced cooling efficiency; on the other hand, an overly intensive heat extraction may generate important temperature gradients and thermomechanical stresses, which can lead to crack formation. Fortunately, automated, algorithmic approaches such as gradient-based topology optimization provide an avenue for devising balanced designs.

Topology optimization aims at finding how to distribute a state (e.g.\ solid or fluid material), into a design space, in order to minimize a certain objective function under some design constraints. 
Originally aimed at increasing the stiffness of a part while minimizing its weight \cite{bendsoeOptimalShapeDesign1989}, topology optimization has now found applications in a variety of other technological fields \cite{sigmundTopologyOptimizationApproaches2013}, including fluid mechanics with and without heat transfer \cite{alexandersenReviewTopologyOptimisation2020}. In conjugate heat transfer (CHT) problems, one typically seeks to find the optimal fluid-solid contact surface (\eg fluid channel) configuration which minimizes a temperature-dependent function, for instance the average or maximal temperature of a part. Hence, topology optimization has the potential to become a powerful tool in conformal cooling channel design, especially when considering a fabrication by additive manufacturing. Still, only a limited amount of work has been dedicated to this task.

Among  different available topology optimization strategies, the \textit{density approach} is by far the most utilized for fluid problems \cite{alexandersenReviewTopologyOptimisation2020}. The general concept behind this method is to consider that one state is dispersed into the other(s) by a local fraction of the total volume, hence allowing for the formation of transitional regions with intermediate properties. In fluid applications, two states, \ie solid and fluid, are generally encountered which are represented by a single \textit{porous} medium with variable properties. These properties  (such as its permeability and thermal conductivity) at a given point are evaluated through interpolation function, based on the local proportion of each state (\eg the \textit{solid fraction}) \cite{navahDevelopmentTopologyOptimization2021}. Another common approach leverages the use of a level-set function to identify the precise location of the interface between the different states, hence eliminating altogether the notion of transitional areas \cite{lamarche-gagnonComparativeStudySharp2021}.

From a recent review paper on topology optimization for fluid-based problems \cite{alexandersenReviewTopologyOptimisation2020}, it can be noted that the vast majority of studies considering heat transfer are tested in two-dimensional problems (see \cite{dedeMultiphysicsTopologyOptimization2009,matsumoriTopologyOptimizationFluid2013}, among others), and/or simplify the CHT problem to avoid the costly resolution of the Navier--Stokes equations. In particular, we note the use of Newton's law of cooling to approximate the convective fluxes \cite{coffinLevelSetTopology2016}; the modeling of the flow field using a network of one-dimensional branches, where the channels' diameters are used as design variables \cite{liuNetbasedTopologyOptimization2020}; similarly, the use of the channels' diameters and positions as design variables, while modeling both flow and heat transfer through a boundary element approach \cite{liTopologyOptimizationDesign2018}; and the use of Darcy's law (proportionality of velocity to pressure gradient) to simplify the equations and to mimic the velocity profile of a turbulent channel flow \cite{zhaoPoorManApproach2018, kambampatiLevelSetTopology2020}. Although the use of such modelings can lead to cooling designs which are intuitive and more efficient than straight-drilled channels, the outcomes are seldom compared to a design obtained without the simplifications, namely when using a full-blown CHT topology optimization approach, in three dimensions. In particular, boundary layer separation and flow recirculation may be encountered in configurations having sharp turns (especially for moderate-to-high Reynolds numbers), which translates into the presence of hot, under-cooled regions. The detection of such flow phenomena can only be ensured if the solution of the momentum equation is considered within the optimization problem. 

Over the past few years, a number of studies have utilized three-dimensional CHT approaches for cooling channel design. However, to the authors' knowledge, none were concerned with the design of \textit{conformal} cooling channels. Applications were mainly towards: the maximization of heat extraction in a uniformly-heated domain \cite{yajiTopologyOptimizationMethod2015, yuThreedimensionalTopologyOptimization2020} (\ie an extension of the canonical two-dimensional problem) or in a domain bounded by heated plates \cite{dilgenDensityBasedTopology2018}; the design of fin-type heat sinks \cite{sun3DTopologyOptimization2020, gilmoreManifoldMicrochannelHeat2021}; and the design of two-fluid heat exchangers \cite{hoghojTopologyOptimizationTwo2020, fepponBodyfittedTopologyOptimization2021, kobayashiTopologyDesignTwofluid2021}.


In this work, we propose a three-dimensional CHT topology optimization approach, targeted at the design of conformal cooling channels in applications with heated surfaces, such as die casting molds. The approach is based on a density modeling with fluid and solid states, and implies the {porosity-based} solution of the Navier--Stokes and energy equations, in laminar flow. The sensitivities with regard to the objective function are evaluated using the discrete adjoint approach. The methodology is developed in DFEM \cite{audetLibDFEMTheorieConcepts2003}, the proprietary multiphysics solver of the National Research Council Canada (NRC). The use of two different objective functions, \ie the domain-averaged and surface-averaged temperature, is investigated. Optimizations are performed for Reynolds numbers of 100 and 1,000 in a geometry representing a simplified die casting mold. The design variable, \ie the solid fraction, is parametrized using the smoothed-Heaviside filter as described \eg in \cite{zhouMinimumLengthScale2015}. The effects of varying optimization parameters such as the filter radius, the maximum volume of fluid and the pressure losses in the system are explored. Furthermore, the temperature solution of a selected design is validated against a body-fitted solver. Finally, the present work features several contributions in terms of solver verification, which is often an overlooked aspect in the topology optimization literature. Hence, regarding the verification, we note:
\begin{itemize}
	\item the introduction of a manufactured solution for the porosity-based CHT solver verification;
	\item the definition of an industrially-relevant cooling channel configuration, which extends the numerical validation of the porosity-based CHT solver and provides a basis for the calibration of the Darcy coefficient;	
	\item the verification of the adjoint sensitivities using finite differencing.
\end{itemize}

The article is structured as follows: Section \ref{sec:methodo} presents the methodology in terms of the adopted CHT modeling,  the finite element discretization and stabilization, the formulation of the optimization problem and the leveraged regularization and control mechanisms. In Section \ref{sec:solver_valid}, the porosity-based CHT solver is validated numerically against a body-fitted CHT solver (\ie based on non-porous states), for a cooling channel problem, and with particular attention to the effects of the Darcy friction coefficient and mesh resolution. Finally, the optimization cases and results are presented in Section \ref{sec:results}, with a detailed discussion of the effects of hyper-parameter variations. The article ends with a conclusion on the most relevant outcomes.

\section{Methodology}
\label{sec:methodo}
\subsection{Conjugate heat transfer modeling}
\label{sec:cht_mod}
Our modeling of the conjugate heat transfer is based on a unified set of steady-state conservation equations of mass, momentum and energy, solved  for in a medium with variable porosity and conductivity, capable of representing pure solid, pure fluid and mixed regions. The conservation of mass is governed by the equation of continuity, viz.\footnote{We adopt the index notation along with Einstein's convention of summation on repeated indices, unless explicitly summed on via $\sum$.}
\begin{equation}
\partial_j u_j = 0,
\label{eq:mass}
\end{equation}
where $u_j$ is the $j^{th} \in [1..N_\mathrm{dim}]$ ($N_\mathrm{dim}=3$), component of the velocity vector.
The conservation of momentum is described by the Navier--Stokes equations which, for a porous medium, read
\begin{equation}
 \rho \, u_j \,\partial_j u_i +\partial_i p - \partial_j \tau_{ij} +C_\alpha\,\alpha' u_i -s_i =  0,
\label{eq:mom_base}
\end{equation}
where $\rho$ is the density of the fluid, $p$ denotes the pressure, $C_\alpha\,\alpha' u_i$ is the {Darcy friction force} accounting for the resistance exerted by the solid to the free movement of the fluid in the porous medium, $C_\alpha$ designates a user-defined weight, $\alpha'$ is the impermeability of the porous medium, $s_i$ is a source term, and $\tau_{ij}$ is a component of the deviatoric stress tensor for an incompressible fluid, defined as
\begin{equation*}
\tau_{ij} := \mu\, (\partial_i u_j + \partial_j u_i),
\end{equation*}
where $\mu$ stands for the dynamic viscosity. The last governing equation, describing the conservation of energy, is expressed by
\begin{equation}
\rho \,c_p \, u_j \,\partial_j T - \partial_j (\kappa \, \partial_jT) -s_q = 0,
\label{eq:energ}
\end{equation}
where $c_p$ is the heat capacity at constant pressure, $T$ the temperature, $\kappa$ the heat conductivity and $s_q$ the source term accounting for volumetric heat generation.

\subsubsection{Interpolation schemes}
\label{sec:mixing}
We utilize interpolation functions to associate the variations of the permeability and of the conductivity to those of the porosity. The latter is inversely related to the concept of solid fraction, $\gamma(\bm{x})$, a scalar field which accounts for the local proportion of the volume occupied by the solid  at position $\bm{x}$ (hence $0\leq \gamma \leq 1)$. The interpolated quantities are computed via power laws (\ie the solid isotropic material with penalization (SIMP) interpolation\footnote{SIMP was preferred over RAMP for the permeability to ensure consistency between the conductivity and permeability interpolation functions.}), in terms of solid fraction, as follows:
\begin{itemize}
\item \textbf{Permeability ($\alpha$)}: the mixed permeability is defined as $\alpha:=\alpha_f + (\alpha_s - \alpha_f)\, \gamma^{P_\alpha}$ where $\alpha_f=1$ and $\alpha_s=0$.  The impermeability is simply set to  $\alpha':=1-\alpha$;
\item \textbf{Conductivity ($\kappa$)}: the mixed conductivity is similarly defined as $\kappa:=\kappa_f + (\kappa_s - \kappa_f)\, \gamma^{P_\kappa}$.
\end{itemize}
In these functions, $_f$ and $_s$ subscripts respectively refer to the pure fluid and the pure solid quantities. Furthermore, $P_\alpha$ and $P_\kappa$ exponents calibrate the sharpness of the variation in porous regions (more abrupt variations for higher values). One can see that the interpolation creates a dependency on the solid fraction, for the problem unknowns : e.g. $p(\alpha(\gamma))$,  $T(\kappa(\gamma))$, etc.

In Section \ref{sec:solver_valid}, the porosity-based CHT solver is validated against a CHT solver which is labeled as \textit{body-fitted}, which refers to the fact that it assumes pure solid and pure fluid regions only. This translates to the cancellation of the Darcy friction term and the velocity field in the fluid and solid regions respectively. The body-fitted approach thus relies on an interface-conformal discretization of the space, which should be accounted for while creating the finite element mesh.


\subsection{Discretization method}
In this work, we adopt a finite element discretization on linear isogeometric tetrahedra, produced by a tesselation of $\bm{\Omega}$, the open and bounded spatial domain of the problem, which is thus approximated as $\bm{\Omega}\approx \cup_K\, \bm{\Omega}_K$, where $\bm{\Omega}_K$ is the domain of element $K$. 
The $k^{th}$ governing partial differential equation, among Eqs. \eqref{eq:mass}--\eqref{eq:energ}, is multiplied by a proper test function, $\varphi^k$ with $k \in[0..4]$, and integrated over ${\bm \Omega}$. The application of integration by parts, the divergence theorem and the addition of the Petrov-Galerkin stabilization terms then yield the following  variational residual equations:
\begin{align}
\begin{split}
R^k&:=\int_{\bm \Omega} \varphi^k\,\partial_j u_j \, d\Omega 
\\
&+\sum_K \int_{\bm \Omega_K}\,\partial_i \varphi^k \tau^U ( \rho \, u_j \,\partial_j u_i  +\partial_i p - \partial_j \tau_{ij} + C_\alpha\,\alpha' u_i -s_i) \, d\Omega = 0,
\label{eq:weak_cont_frac}
\end{split}
\end{align}
for $k=0$, and
\begin{align}
\begin{split}
R^k&:=\int_{\bm \Omega}\left( \varphi^k\, \rho \, u_j \,\partial_j u_i  - \partial_i \varphi^k  \, p+  \partial_j \varphi^k \, \tau_{ij} + \varphi^k\,C_\alpha\,\alpha' u_i -\varphi^k s_i \right) \,d\Omega  \\
&+\sum_K \int_{\bm \Omega_K}\rho \, u_l\,\partial_l \varphi^k \, \tau^U\,( \rho \, u_j \,\partial_j u_i  +\partial_i p - \partial_j \tau_{ij} + C_\alpha\,\alpha' u_i -s_i) \, d\Omega  
\\&-
 \int_{\bm \Gamma_\mathrm{NS}} \varphi^k\,(\tau_{ij}n_j-p\,n_i ) \,d\Gamma =0,  
\end{split}
\label{eq:weak_mom_frac}
\end{align}
for $k=i\in [1..3]$ and where $n_i$ is the $i^{th}$ component of the outward normal unit vector on $\bm{\Gamma}_\mathrm{NS}$, the boundary of the domain where Neumann (natural) conditions are applied for the Navier--Stokes equations. 
The stabilization parameter, $\tau^U$, is defined to account for the Darcy penalization term \cite{tezduyarIncompressibleFlowComputations1992, alexandersenTopologyOptimisationNatural2014}, such that
\begin{equation}
\tau^U:= \left[ \left( \frac{2\rho \sqrt{u_i\,u_i}}{h_K} \right)^2 + \left( \frac{4 \,\mu}{m_K h_K^2} \right)^2 +(C_\alpha\,\alpha')^2 \right]^{-\frac{1}{2}},
\label{eq:tau_s_2}
\end{equation}
with $m_K$ a coefficient set to $1/3$ for linear elements and $h_K$ denoting the size of element $K$, computed as $h_K:=\frac{6\,\|{\bm\Omega}_K\|}{\|{\bm\Gamma}_K\|}$, i.e. six times the volume-to-surface ratio of the tetrahedron $K$.
Finally, for $k=4$, we obtain
\begin{align}
\begin{split}
R^k&:=\int_{\bm \Omega} \left(\varphi^k\, \rho \,c_p  u_j \,\partial_j T + \partial_j \varphi^k \, \kappa \, \partial_jT -\varphi^k s_q \right) \,d\Omega 
\\&+\sum_K \int_{\bm \Omega_K} \rho \, c_p \,  u_l \, \partial_l \varphi^k  \, \tau^T(\rho \,c_p \, u_j \,\partial_j T  - \partial_j (\kappa \, \partial_jT) -s_q ) \,  d\Omega \\
&- \int_{\bm \Gamma_\mathrm{E}} \varphi^k\,\kappa \, \partial_jT\,n_j \,d\Gamma = 0,
\end{split}
\label{eq:weak_energy_frac}
\end{align}
where $\bm{\Gamma}_\mathrm{E}$ is the boundary of the domain where Neumann conditions are applied and $\tau^T$ is the stabilization parameter of the equation of energy, defined as follows:
\begin{equation}
\tau^T:= \left[ \left( \frac{2\rho \,c_p\sqrt{u_i\,u_i}}{h_K} \right)^2 + \left( \frac{4 \,\kappa}{m_K h_K^2} \right)^2 \right]^{-\frac{1}{2}}.
\label{eq:tau_T_frac}
\end{equation}

\subsubsection{Global degrees of freedom and design parameters}
The problem unknowns, i.e. $u_i$, $p$ and $T$, are discretized via linear Lagrange interpolation functions, which have a compact support on the elements sharing a given tetrahedron vertex, to which the polynomial is associated. The vertices (nodes) have a global and a local numbering such that, for example, the same Lagrange polynomial can be equivalently noted $\phi^g(\bm{x})\equiv \phi^l_K(\bm{x})$, where $g\in[1..N_\mathrm{DOFs}]$ is the index for the global degree of freedom (per governing equation), versus a local numbering $l^{th}\in [1..4]$ on element $K$. Thus for instance, the temperature discretization can be equivalently noted $T(\bm{x}):=T^g \,\phi^g(\bm{x})$ or  $T(\bm{x}):=T^l_K \,\phi_K^l(\bm{x})$ (on element $K$). This discretization is also applied to the test function of the $k^{th}$ residual equation, $\varphi^k$, which similarly to $\phi$, is a linear Lagrange polynomial with global indexing, $\varphi^{kg}$. Thus, a discrete system of $N_\mathrm{DOFs}\times N_\mathrm{DOFs}$ algebraic equations is constituted and solved for each governing residual equation to yield for example $\bm{T}$, i.e. the vector of $T^g$, the unknown nodal values.

With regards to the optimization problem, we opt for a nodal design representation of the solid fraction, such that $\gamma(\vec{x}):=\gamma^g\,\phi^g(\bm{x})$, resulting in $\bm \gamma$, the vector of $\gamma^g$ (nodal solid fraction values). The solid fraction values further depend on design parameters, i.e. $\bm{\gamma}\equiv \bm{\gamma}(\bm\desVar)$, where $\bm \desVar$ is the vector of $N_\mathrm{DOFs}$ nodal design parameter values. In Section \ref{sec:smthdheavy}, we  further discuss the operators which define the dependency between nodal values of solid fraction and design parameters.

\subsection{Optimization problem}
\label{sec:opt-prob}
The topology optimization problem considered in this work can be formulated as follows:
\begin{align}
	\min_{\desVar} \quad & \cost({\bm\gamma},  \bm{p}({\bm\gamma}), \bm{u_1}({\bm\gamma}),\bm{u_2}({\bm\gamma}),\bm{u_3}({\bm\gamma}), \bm{T}({\bm\gamma})), \nonumber\\
	\textrm{s.t.} \quad & \constraint_i(\bm \desVar) \leq 0, \; i \in [1..N_\constraint], \label{eq:optimProblem} \\
	& \bm{R}^k({\bm\gamma},  \bm{p}({\bm\gamma}), \bm{u_1}({\bm\gamma}),\bm{u_2}({\bm\gamma}),\bm{u_3}({\bm\gamma}), \bm{T}({\bm\gamma})), \; k \in [0..4], \nonumber\\
	& \desVarMin \leq \desVar \leq \desVarMax, \nonumber
\end{align}
where $\cost$ is the cost function we seek to minimize, $\constraint_i$ is an inequality constraint (imposed through the optimizer) (see Section \ref{sec:FSconst})  and $N_\constraint$ is the total number of inequality constraints (here, $N_\constraint=1$). Let us recall that in Problem \eqref{eq:optimProblem}, $\bm{R}^k$ stands for the system of $N_\mathrm{DOFs}$ discrete residual equations, arising from the $k^{th}$ partial differential equation, and the solid fraction is a function of the design variables, i.e.\@ $\gamma(\desVar)$.
Finally, the desired bounds on the design parameters are set to  $\desVarMin=0$ and $\desVarMax=1$.

The cost function considered in this work has two components: a term which accounts for the average temperature in the domain or on a boundary of interest, and a term which penalizes the cost function by imposing a constraint on the average pressure losses. For continuous variables, it is expressed as
\begin{equation}
\cost(T,p): = \bar{T} + \zeta \, \bar{\Delta p},
\label{eq:costcomps}
\end{equation}
where $\zeta$ is a penalization coefficient for the pressure term, and
\begin{equation}
\bar{T}\equiv \bar{T}_\Omega := \frac{\int_{\bm \Omega} T d\Omega}{\Omega},
\end{equation}
or
\begin{equation}
\bar{T}\equiv  \bar{T}_\Gamma := \frac{\int_{\bm \Gamma} T d\Gamma}{\Gamma},
\end{equation}
with $\bm \Gamma$ designating a boundary, the average temperature of which we desire to minimize.
Finally, the following definitions are adopted: $\Omega\equiv\|{\bm \Omega}\|:=\int_{\bm \Omega} d\Omega$, and $\Gamma\equiv\|{\bm \Gamma}\|:=\int_{\bm \Gamma} d\Gamma$.

The overall optimization process can be summarized as follows: Subsequently to the solution of the residual Eqs. \eqref{eq:weak_cont_frac}, \eqref{eq:weak_mom_frac} and \eqref{eq:weak_energy_frac}, the cost function gradients are computed via the adjoint method, as explained in \ref{sec:adjoint}. Then, the gradient values are provided to the NLopt optimization package \cite{johnsonNLoptNonlinearoptimizationPackage2021}, which imposes the minimum and maximum constraints on the design variables as well as the inequality constraints, and at each optimization step, provides a new set of design variables by using the method of moving asymptotes \cite{svanbergClassGloballyConvergent2002}. 

In the next sections, we introduce the mechanisms adopted to improve the efficiency and robustness of the proposed optimization framework.

\subsubsection{Projection schemes}
\label{sec:smthdheavy}
The regularization of the topology optimization problem is generally desired, especially in porosity-based approaches, to avoid ill-posedness \cite{sigmundTopologyOptimizationApproaches2013} for example. Furthermore, designs obtained using a non-regularized problem are often mesh-dependent and can suffer from checkerboard patterns \cite{diazCheckerboardPatternsLayout1995, sigmundNumericalInstabilitiesTopology1998}. In order to address these issues, we leverage a density filter \cite{guestAchievingMinimumLength2004, sigmundMorphologybasedBlackWhite2007}, where each nodal value of the design parameter is weighted using values in its set of neighboring nodes, $\neighbor_i$, such that we obtain a vector of filtered nodal solid fraction values, $\gammaf_i$, i.e.
\begin{equation}
	\gammaf_i := \mathcal{D}_{ij} \desVar_j,
	\label{eq:density_filt}
\end{equation}
where $\mathcal{D}_{ij} := \omega_{ij}\,v_j / \sum_{k \in \neighbor_i} \omega_{ik}\,v_k$ (not summed on $_j$) is a sparse matrix having zero values for $j \notin \neighbor_i$, $\neighbor_i$ is defined via the sphere of radius $\varrho$, centered on $\vec{x}_i$, \ie $\neighbor_i := \{j \; | \norm{\vec{x}_j - \vec{x}_i} \leq \varrho\}$, and a linearly decaying weighting function, $\omega_{ij} := \varrho - \norm{\vec{x}_i - \vec{x}_j}$ for $j \in \neighbor_i$, is used. The nodal weight for the node $j$ is expressed by $v_j$ and computed as $v_j := \int_{\bm \Omega} \phi^j(\vec{x}) d\Omega$.  To reduce the extent of areas having intermediate solid fraction values and to generate sharper designs, the discrete values of $\gammap_i$ are computed by transforming $\gammaf_i$, using a smoothed Heaviside function \cite{wangProjectionMethodsConvergence2011, zhouMinimumLengthScale2015}, viz.
\begin{equation}
	\gammap_i := \frac{\tanh(\beta\eta) + \tanh(\beta(\gammaf_i - \eta))}  {\tanh(\beta\eta) + \tanh(\beta(1-\eta))}.
	\label{eq:heaviside_proj}
\end{equation}
Here, a threshold value of $\eta=0.5$ is used and the parameter $\beta$ is gradually increased during the optimization using a continuation strategy. The latter consists in performing a series of optimization loops, starting with $\beta=1$ and scaling this value by a factor of two at the beginning of the next loop, until $\beta=\beta_\text{max}$ is reached. In this work, $\beta_\text{max}=8$ is utilized.

The Lagrangian function gradient with regards to the nodal solid fraction values, $\pd{\lagran}{\gammap_k}$ obtained through Eq. \eqref{eq:DCDgamma_decpld}, can be complemented by taking into account the sensitivity of the solid fraction to the design parameters, via the chain rule, as
\begin{equation}
	\pd{\lagran}{\desVar_i} = \sum_{k \in \neighbor_i} \pd{\lagran}{\gammap_k} \pd{\gammap_k}{\gammaf_k} \pd{\gammaf_k}{\desVar_i},
\end{equation}
where
\begin{equation}
	\pd{\gammap_k}{\gammaf_k} = \beta \frac{1 - \tanh(\beta(\gammaf_k - \eta))^2} {\tanh(\beta\eta) + \tanh(\beta(1-\eta))},
\end{equation}
and for $k \in \neighbor_i$,
\begin{equation}
	\pd{\gammaf_k}{\desVar_i} = \mathcal{D}_{ki}.
	\label{eq:dGammaStardGammaHat}
\end{equation}
Lastly, one should note that the density filter does not ensure a minimum length scale in a strict sense, as features having a size smaller than $\varrho$ may emerge during the optimization. This filter should rather be considered as a tool which limits the geometric complexity of a design as $\varrho$ is increased.



\subsubsection{Constraint on total solid fraction}
\label{sec:FSconst}
The total (same as average) solid fraction is computed via the following formula:
\begin{equation}
\bar{\gamma} \,( \gamma \,({\bm x}) ):= \frac{\int_{\bm \Omega} \,\gamma ({\bm x})  \,d\Omega }{\Omega}.
\end{equation}
To control the amount of fluid available for the optimal design, we apply an inequality constraint (see \eqref{eq:optimProblem}), such that 
$\constraint_1(\gamma):=\bar{\gamma}_0 -\bar{\gamma} \,( \gamma)\leq0$, where $\bar{\gamma}_0$ is a user-defined minimum value for the total solid fraction. The constraint $\constraint_1(\gamma)$ is translated to $\constraint_1(\desVar)$ by considering the dependency relation $\gamma(\desVar)$ of Section \ref{sec:smthdheavy}.
\subsubsection{Constraint on pressure losses}
\label{sec:preslos}
Let us recall  Eq. \eqref{eq:costcomps} which exhibits the components of the cost function:
\begin{equation*}
\cost(T,p): = \bar{T} + \zeta \, \bar{\Delta p},
\end{equation*}
where $\zeta$ is a penalization coefficient for the pressure term. The latter is defined as
\begin{equation}
\bar{\Delta p} : = \frac{\int_{{\bm \Gamma}_\mathrm{in}} p\, d\Gamma}{\Gamma_\mathrm{in}} - \frac{\int_{{\bm \Gamma}_\mathrm{out}} p\, d\Gamma}{\Gamma_\mathrm{out}},
\end{equation}
where ${\bm \Gamma}_\mathrm{in}$ and ${\bm \Gamma}_\mathrm{out}$ respectively refer to boundaries where inlet and outlet conditions are applied.

\section{Numerical validation of the CHT solver}
\label{sec:solver_valid}

The implementation of the conjugate heat transfer model was verified using the method of manufactured solutions, as presented in \ref{sec:mms}. This verification provided a necessary level of confidence in the correctness of the implementation of the porosity-based  CHT solver. Nevertheless, this verification did not cover certain aspects such  as Neumann boundary conditions and realistic thermofluid operating conditions. Hence, we complement this verification by conducting a numerical validation (via solution verification) for a realistic cooling channel configuration, where the porous solver solutions are compared to those obtained by the body-fitted CHT solver on a fine mesh.

\begin{figure}[h!]
\centering
  \includegraphics[width=.6\textwidth,trim={7.0cm 0.0cm 20.0cm 0.cm},clip]{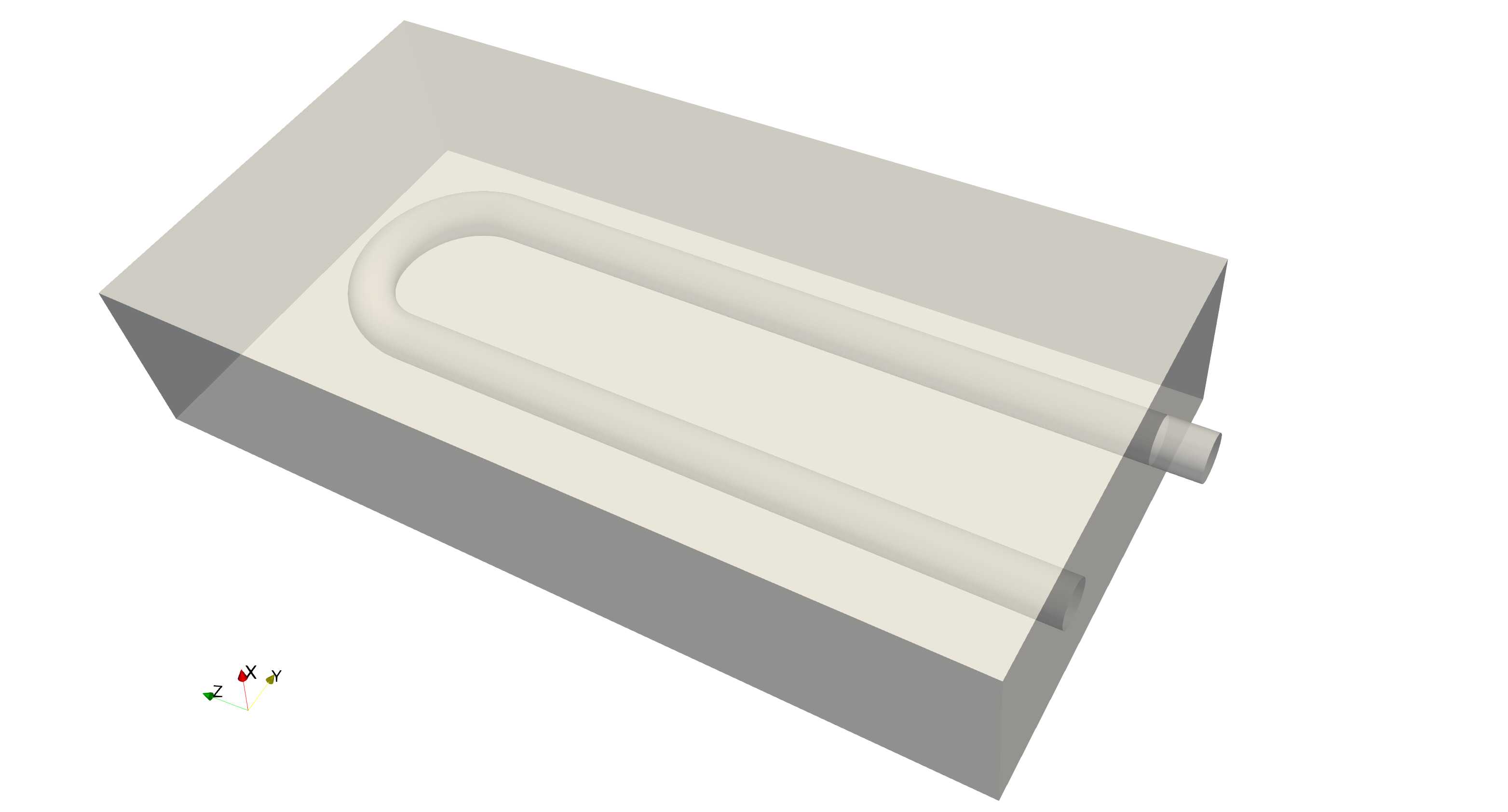}
  \caption{Geometrical configuration of the U-shaped cooling channel case.}
\label{fig:U_geomiso1}
\end{figure}

The configuration shown in Fig.~\ref{fig:U_geomiso1}, represents a simplified die casting mold. Its geometry is comprised of a solid domain of rectangular cuboid shape with dimensions\footnote{All spatial dimensions provided in this section are in meters.} $[-0.02,0.02]_x \times[-0.05,0.05]_y\times[-0.1,0.1]_z$, traversed by a U-shaped cooling fluid channel, with circular inlet and outlet, respectively centered at $(0,0.02,-0.11) $ and $(0,-0.02,-0.1) $;  and a radius of $r=0.005$. The junction between the straight portion and the circular part of the channel is located on the plane $z=0.05$.

The inlet conditions are: a parabolic profile for the velocity, viz.  $[u_x,\, u_y,\, u_z] = [0,\, 0,\,0.03887776\,( 1 - (x^2 + (y-0.02)^2)/r^2 )]$\,m/s and an ambient temperature for the fluid of $T=21\,^{\circ}\mathrm{C}$. The outlet is set free in the $z$ direction with $[u_x,\, u_y] = [0,\, 0]$\,m/s and adiabatic. No-slip conditions are applied to all other boundaries (and to the solid-fluid interface in the body-fitted solver). All outer solid boundaries are set to adiabatic, except the top surface (the plane $x=0.02$) where a heating flux of 10,000\,W/m$^2$ is imposed to account for the surface of the mold insert. The thermofluid properties correspond to those of water and steel\footnote{Since the convective effects should vanish in the pure solid regions and as we operate in steady state, $\rho$ and $c_p$ correspond to those of the fluid.} evaluated at the inlet temperature, and are set to $\rho=998$\,kg/m$^3$, $c_p=$ 4,184\,J/kg/K, $\mu=9.7 \times 10^{-4}$\,N\,s/m$^2$, $\kappa_f = 0.6$\,W/m/K and $\kappa_s = 24$\,W/m/K. Furthermore, we have $P_\alpha=100$ and $P_\kappa=100$. The corresponding Reynolds (based on the inlet diameter and average velocity) and Prandtl numbers are: $\mathrm{Re}=200$ and $\mathrm{Pr}=6.76$.

\begin{figure}[h!]
\centering
\begin{subfigure}[h]{0.48\textwidth}
  \includegraphics[width=\textwidth,trim={0.1cm 0.3cm 6.cm 0.1cm},clip]{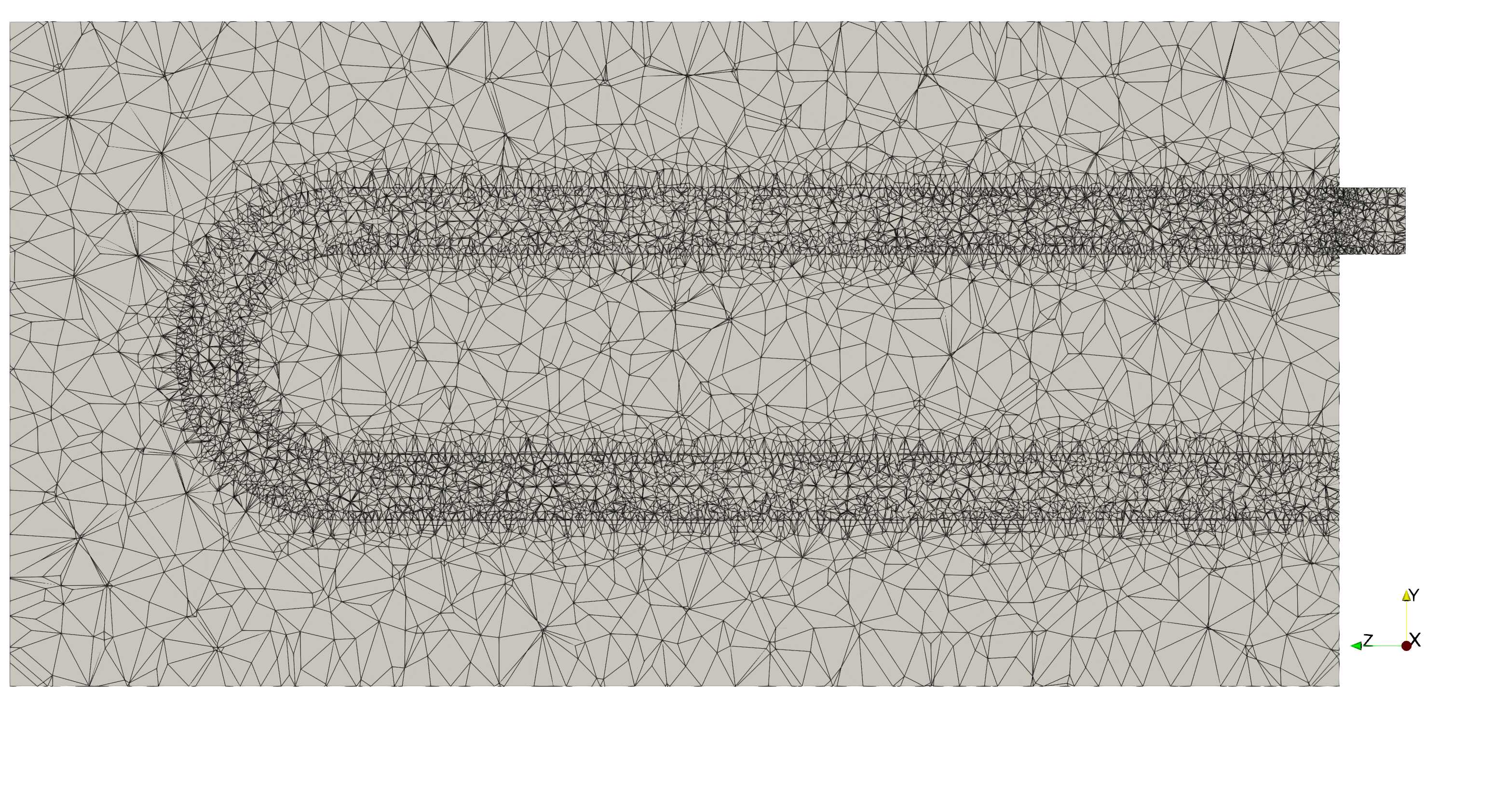}
  \caption{m2-c}
\end{subfigure}
~
\begin{subfigure}[h]{0.48\textwidth}
  \includegraphics[width=\textwidth,trim={0.1cm 0.3cm 6.cm 0.1cm},clip]{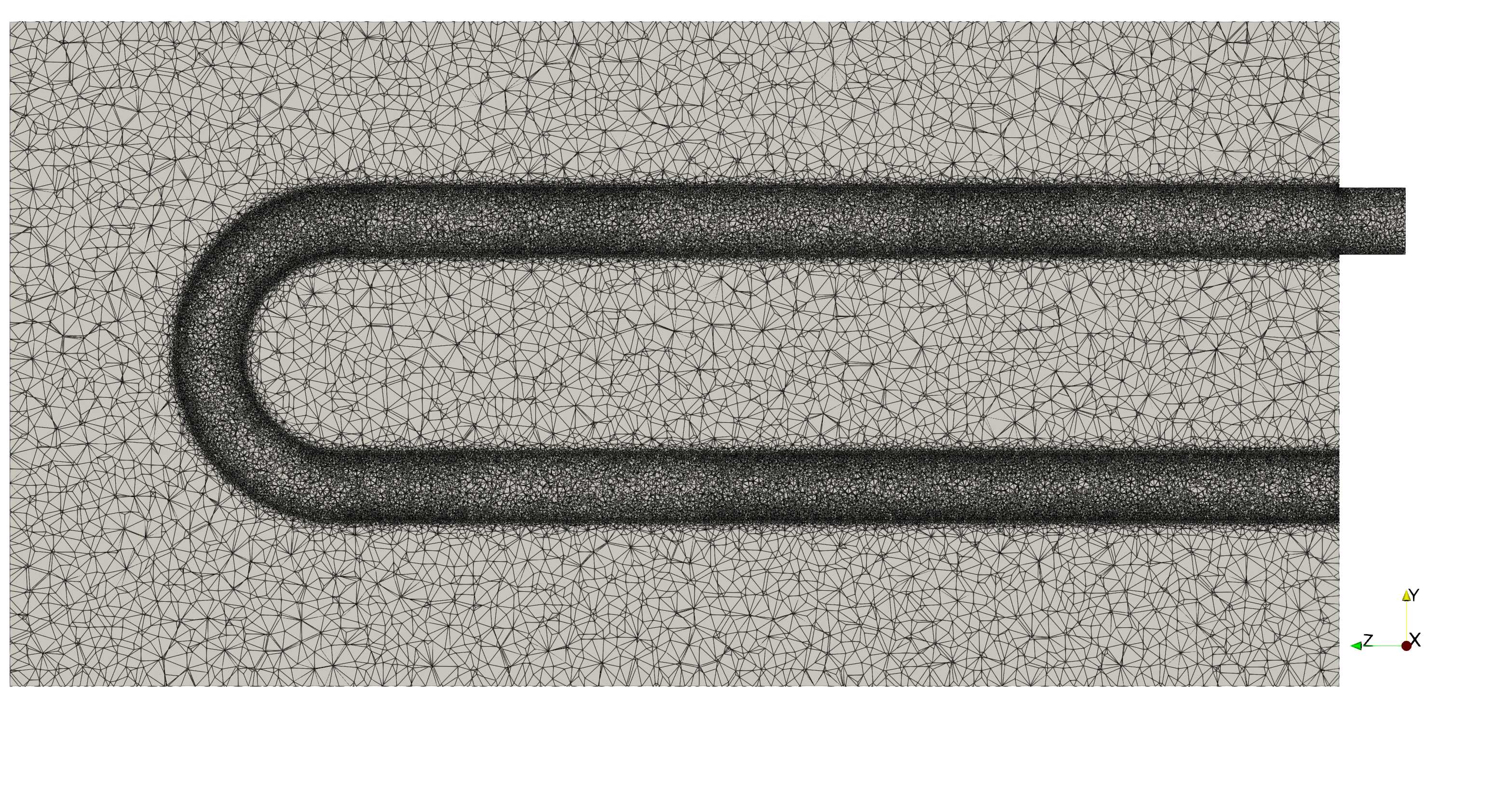}
  \caption{m4-c}
\end{subfigure}
  \caption{Conformal discretization of the U-shaped configuration on the cutting plane $x=0$ for refinement levels 2 and 4.}
\label{fig:U_meshes2cn4c}
\end{figure}

\begin{figure}[h!]
\centering
\begin{subfigure}[h]{0.48\textwidth}
  \includegraphics[width=\textwidth,trim={.1cm 0.3cm .1cm 0.1cm},clip]{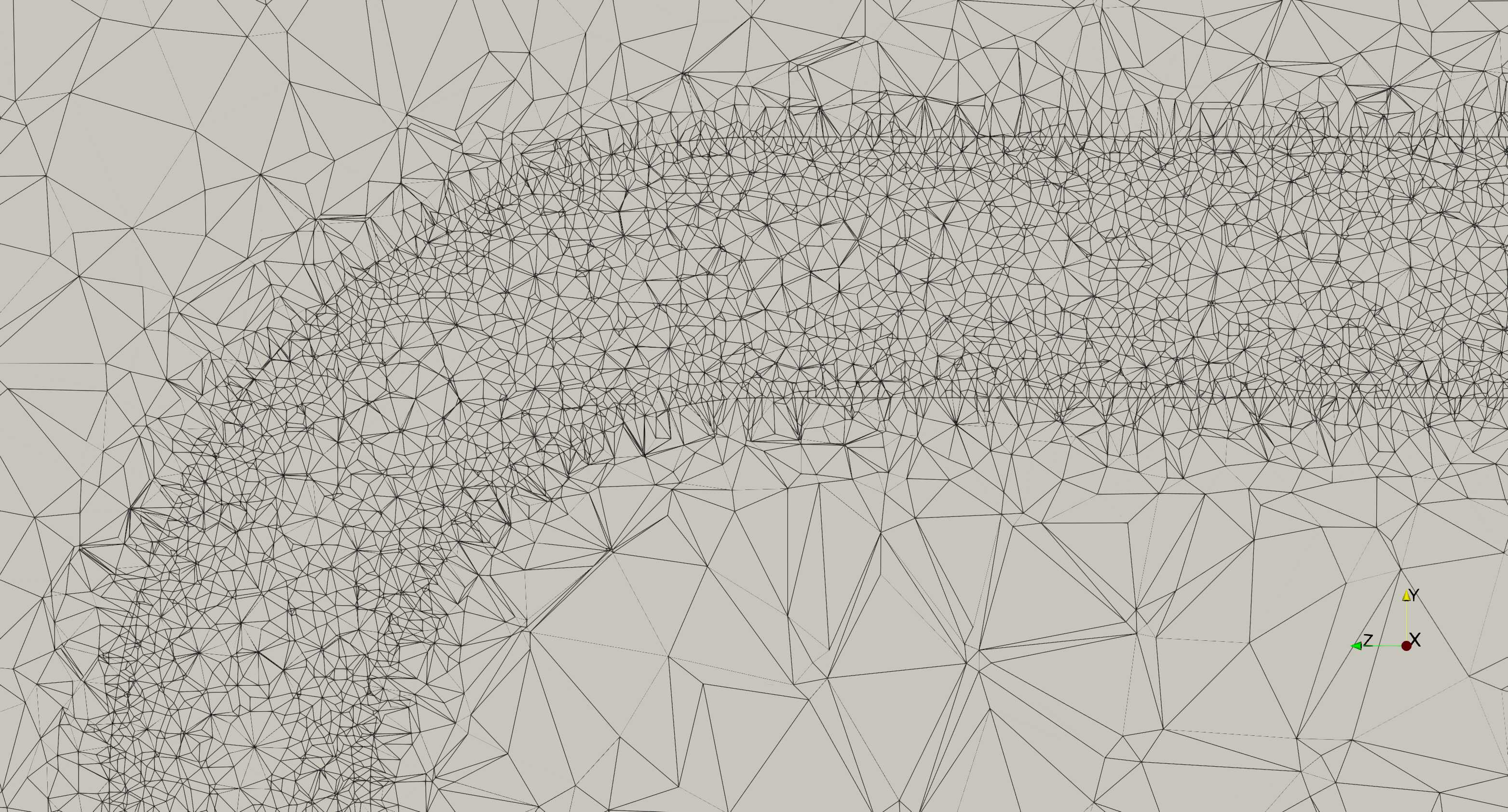}
  \caption{m3-c (conformal)}
\end{subfigure}
~
\begin{subfigure}[h]{0.48\textwidth}
  \includegraphics[width=\textwidth,trim={0.1cm 0.3cm 0.1cm .1cm},clip]{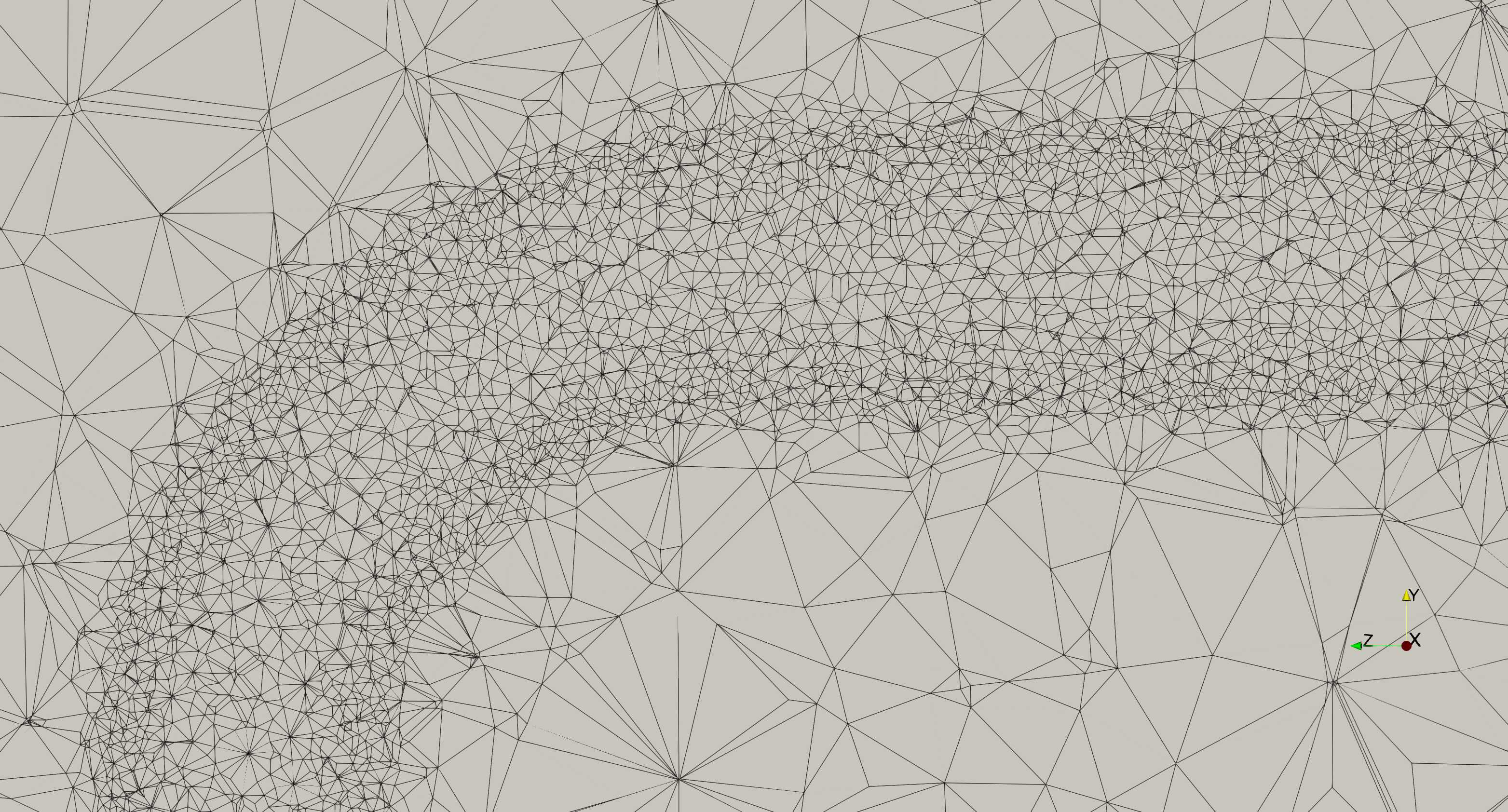}
  \caption{m3-n (non-conformal)}
\end{subfigure}
  \caption{Conformal vs non-conformal discretizations of the U-shaped configuration on the cutting plane $x=0$.}
\label{fig:U_meshes3cv3n}
\end{figure}

%

\begin{table}[!ht]  
  \centering
  \begin{tabular}{c|cccc|ccc}
    \toprule
Label           & m1-c    & m2-c   & m3-c      & m4-c      & m1-n    & m2-n     & m3-n  \\ \midrule
Number of nodes & 10,326  & 37,394 & 176,458   & 950,427   & 5,948   & 31,183   & 165,598  \\ \bottomrule
  \end{tabular}
\caption{Number of nodes for the conformal (-c) and non-conformal (-n) mesh sequences of the U-shaped configuration.}
\label{tab:Umeshes}  
\end{table}

\begin{figure}[h!]
\centering
\begin{subfigure}[h]{0.48\textwidth}
  \includegraphics[width=\textwidth,trim={0.1cm 0.3cm 0.35cm 0.cm},clip]{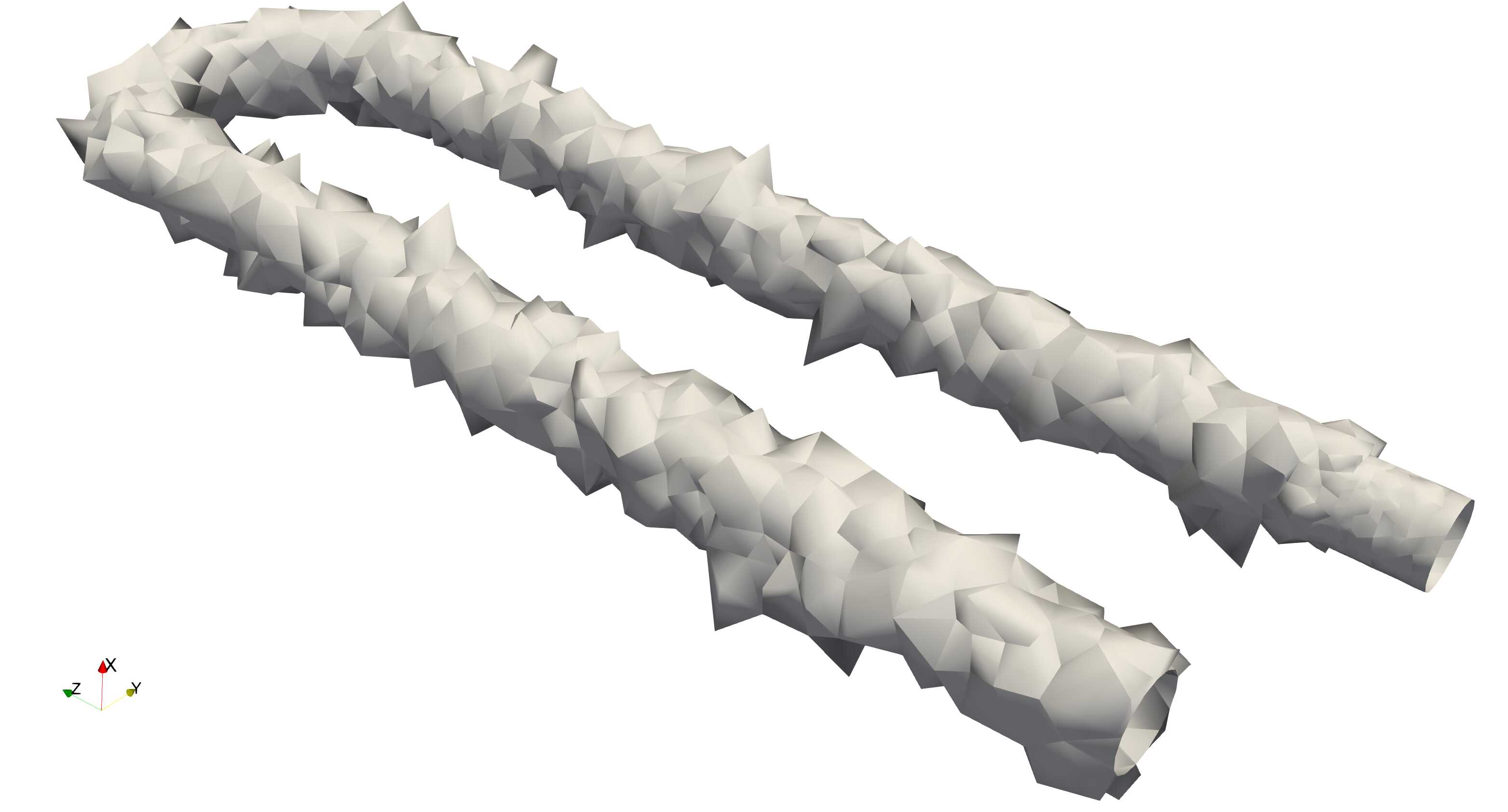}
  \caption{Isosurface of $\gamma$ = 0.9999 on m1-n}
\end{subfigure}
~
\begin{subfigure}[h]{0.48\textwidth}
  \includegraphics[width=\textwidth,trim={0.1cm 0.3cm 0.35cm 0.cm},clip]{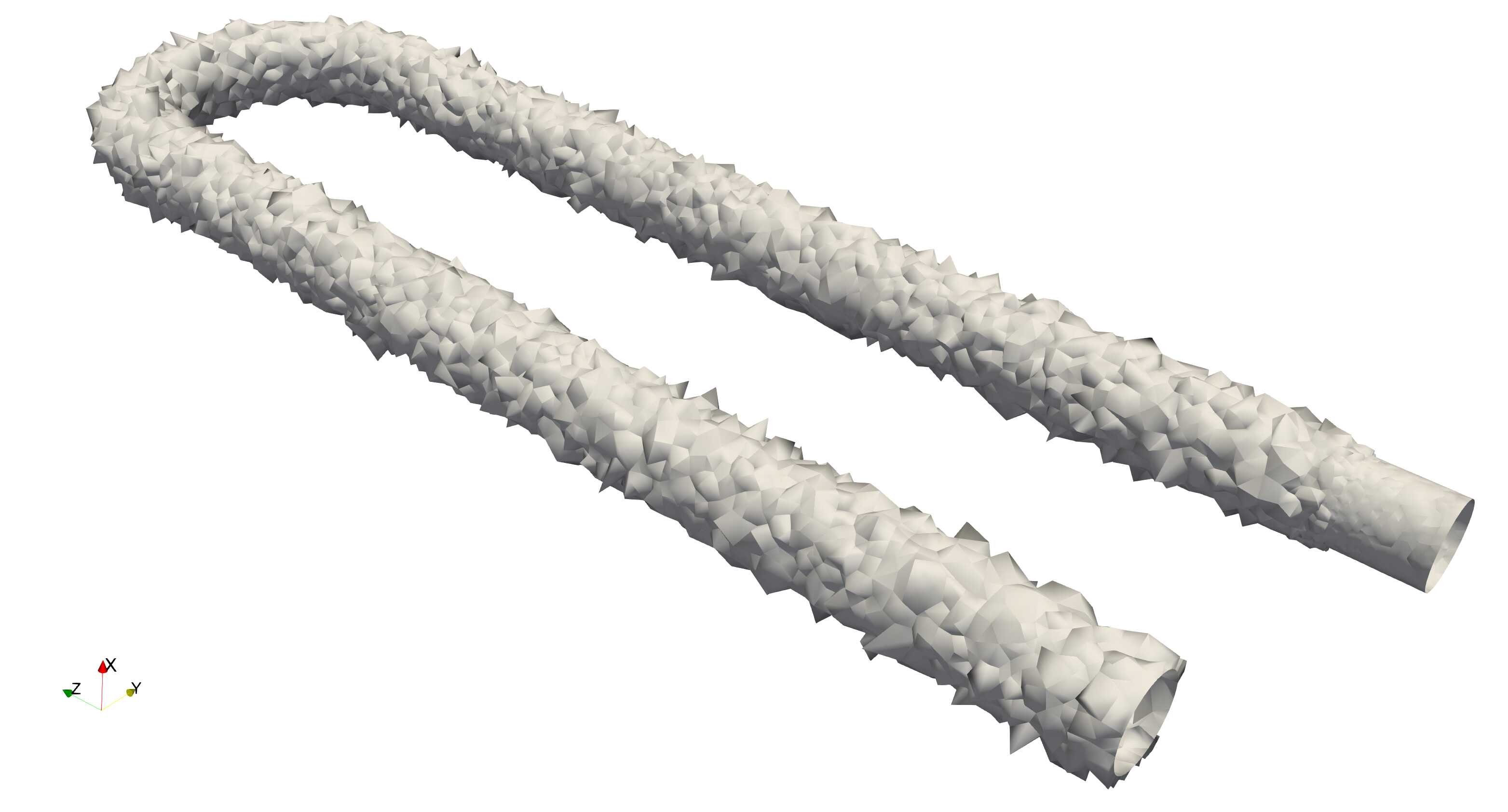}
  \caption{Isosurface of $\gamma$ = 0.9999 on m2-n}
\end{subfigure}
\\
\begin{subfigure}[h]{0.48\textwidth}
  \includegraphics[width=\textwidth,trim={0.1cm 0.3cm 0.35cm 0.cm},clip]{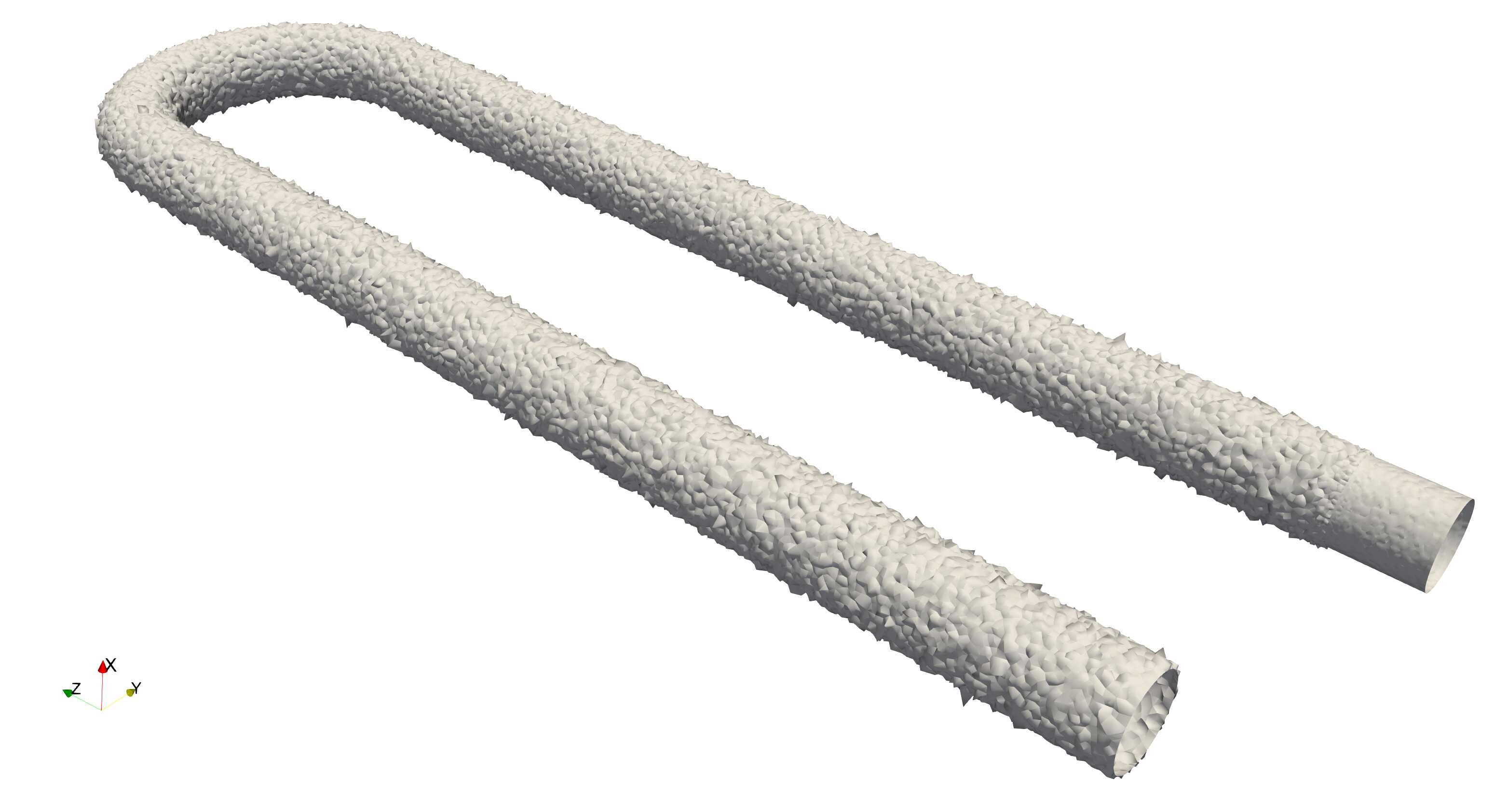}
  \caption{Isosurface of $\gamma$ = 0.9999 on m3-n}
\end{subfigure}
~
\begin{subfigure}[h]{0.48\textwidth}
  \includegraphics[width=\textwidth,trim={0.1cm 0.3cm 0.35cm 0.cm},clip]{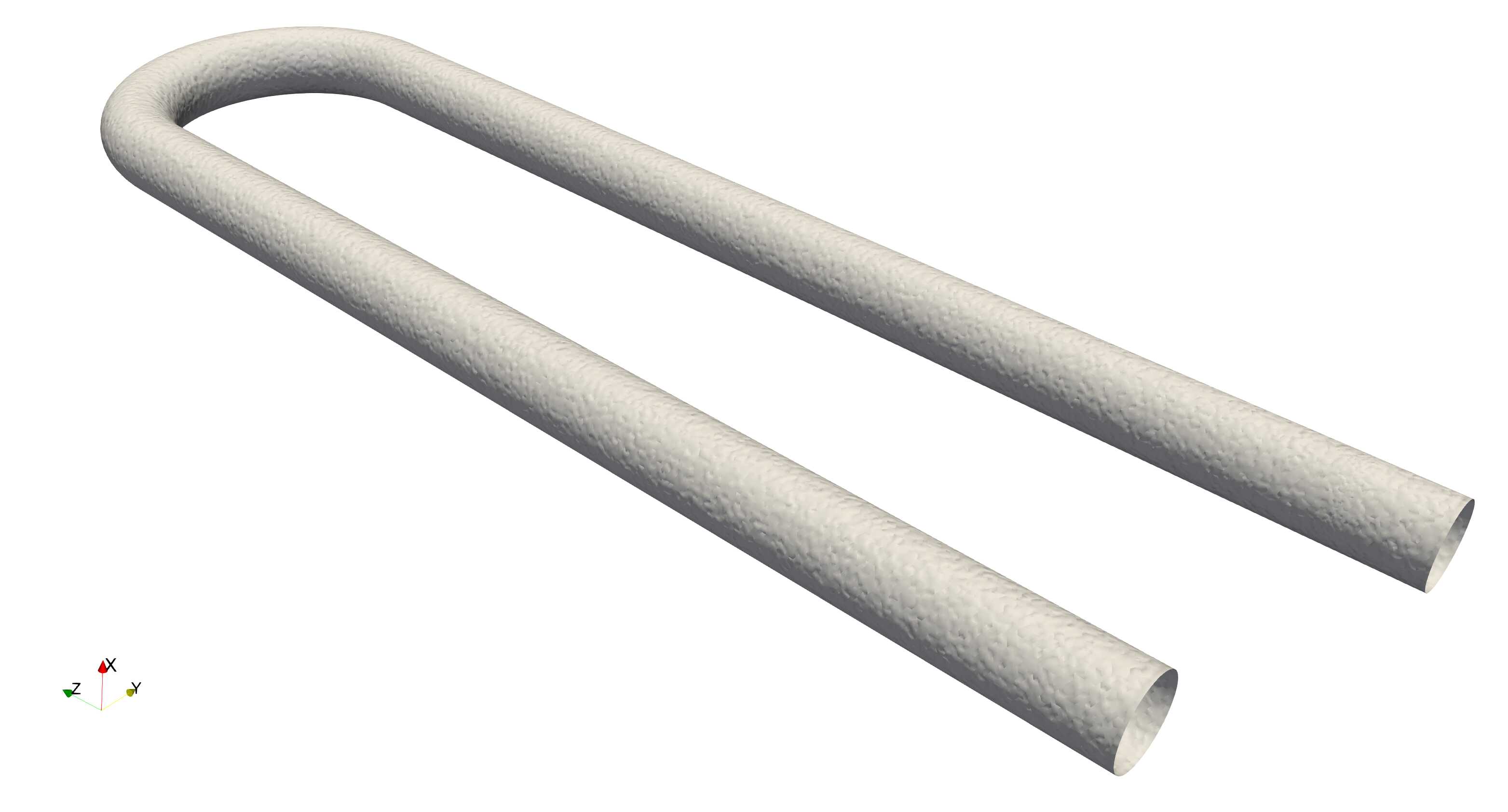}
  \caption{Isosurface of $\gamma$ = 0.9999 on m3-c}
\end{subfigure}
\\
\begin{subfigure}[h]{0.48\textwidth}
  \includegraphics[width=\textwidth,trim={0.1cm 0.3cm 0.35cm 0.cm},clip]{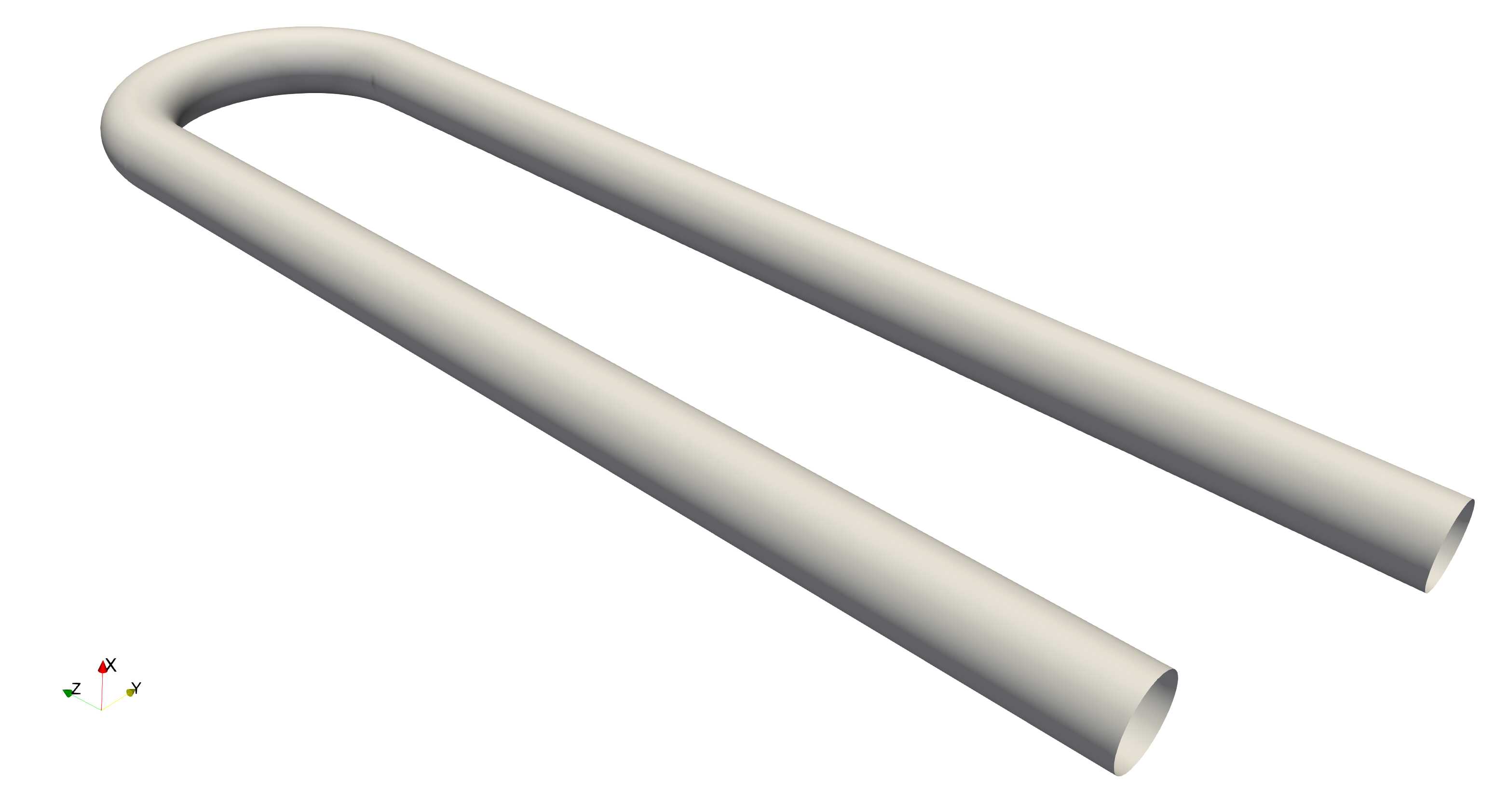}
  \caption{Isosurface of $\sqrt{u_iu_i}= 0$ on m2-c (solution via body-fitted solver)}
\end{subfigure}
  \caption{Implicit versus explicit solid-fluid interface representations.}
\label{fig:U_pipe}
\end{figure}

The domain is discretized by focusing the computational effort in the fluid region, as shown in Fig.\ \ref{fig:U_meshes2cn4c}. In order to represent the solid-fluid interface, we leverage two different approaches (see Fig.\ \ref{fig:U_meshes3cv3n}): \textit{conformal} (-c) meshes are created where the cooling channel interface with the solid is explicitly represented by the tetrahedra faces, and \textit{non-conformal} (-n) meshes,  where this interface is not explicitly accounted for in the mesh.
A sequence of self-similar and consistently refined meshes is produced in each kind using the Gmsh mesh generator \cite{geuzaineGmsh3DFinite2009}, for which the number of nodes are presented in Table \ref{tab:Umeshes}.
\begin{figure}[h!]
\centering
\begin{subfigure}[h]{0.48\textwidth}
  \includegraphics[width=\textwidth,trim={.1cm .3cm .1cm 3.cm},clip]{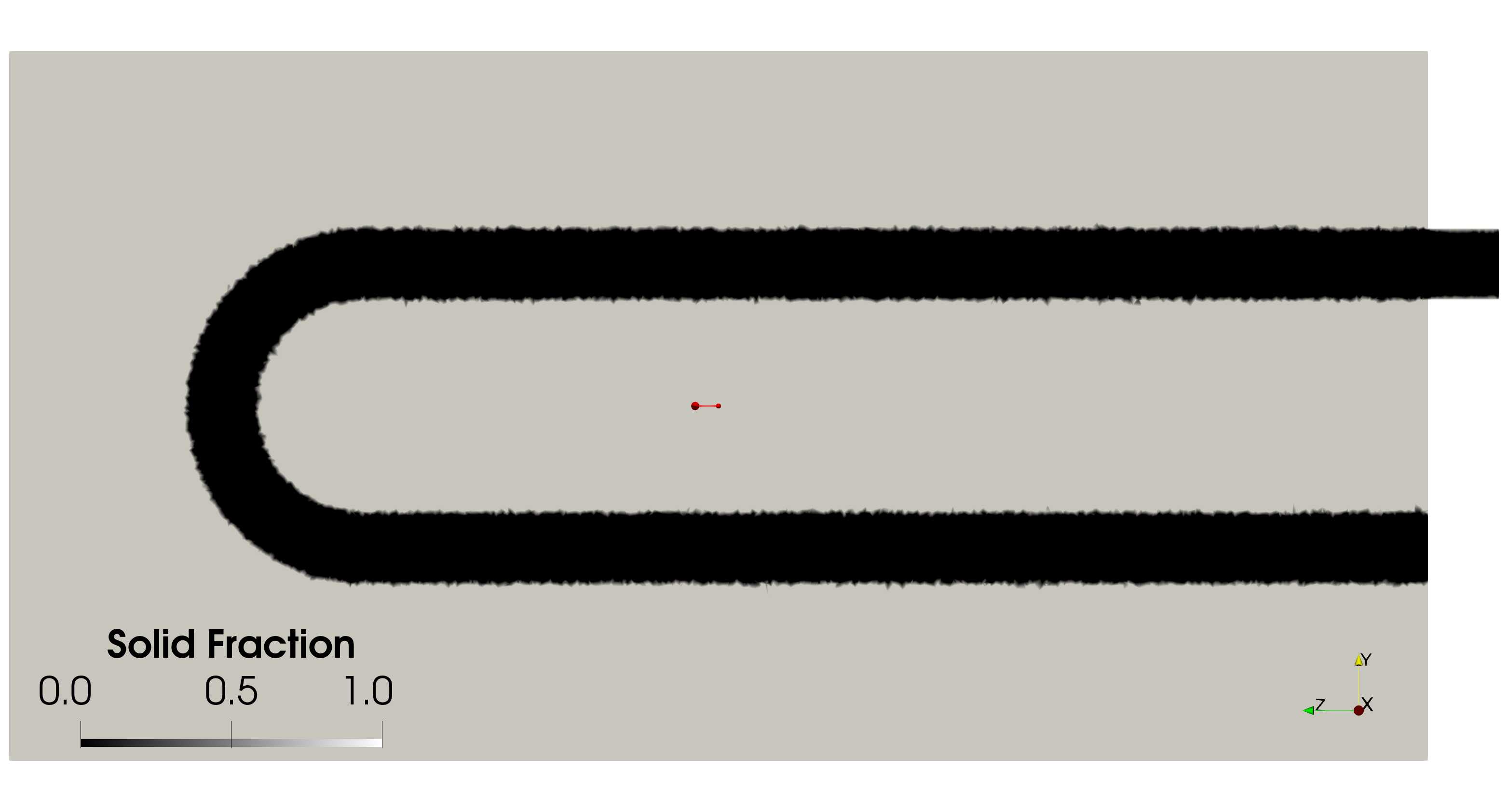}
  \caption{Full view}
\end{subfigure}
~
\begin{subfigure}[h]{0.48\textwidth}
  \includegraphics[width=\textwidth,trim={0.1cm 0.3cm 0.1cm .1cm},clip]{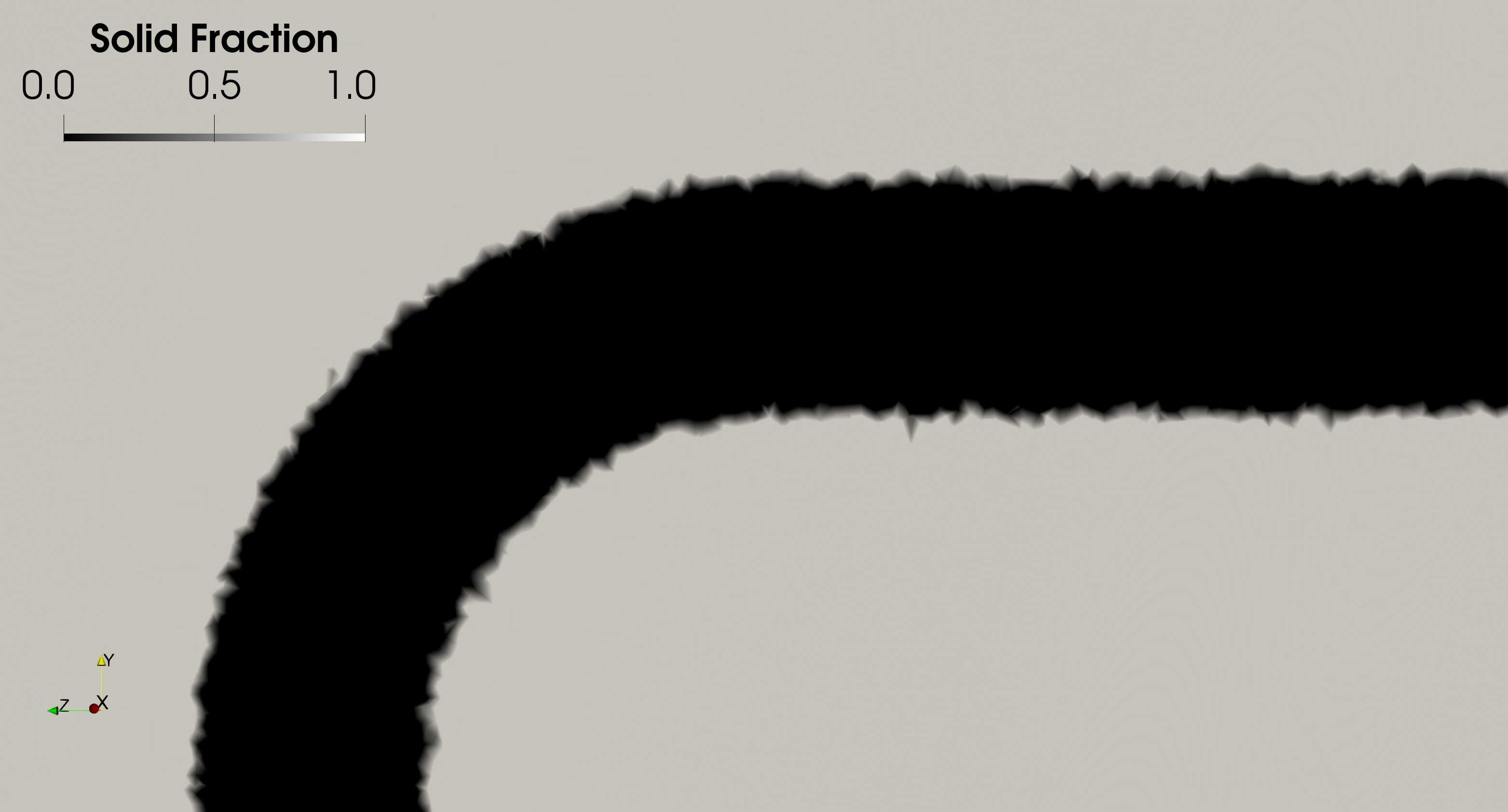}
  \caption{Zoom}
  \label{fig:U_solidfrac_zoom}
\end{subfigure}
  \caption{Linear interpolation of the solid fraction field of the U-shaped configuration on the cutting plane $x=0$ and mesh m3-n.}
\label{fig:U_solidfrac}
\end{figure}
In the case of non-conformal meshes, the interface is defined by the solid fraction field expressed by
\begin{equation}
\begin{split}
\gamma(\bm{x}):=\;
&\mathrm{step}((0.02-\mathrm{sqrt}( (z-0.05)^2 +y^2))^2 + x^2 - 0.00499^2)\,\mathrm{step}(z-0.05)  \\ 
+ \,&\mathrm{step}(\mathrm{sqrt}(x^2+(y-0.02)^2)-0.00499)\,(1-\mathrm{step}(z-0.05))\,\mathrm{step}(y) \\
+ \,&\mathrm{step}(\mathrm{sqrt}(x^2+(y+0.02)^2)-0.00499)\,(1-\mathrm{step}(z-0.05))\,\mathrm{step}(-y),
\label{eq:U_sf}
\end{split}
\end{equation}
which serves to find the nodal values of the solid fraction. The discrete values are then linearly interpolated to the Gauss quadrature points and used by the interpolation functions (see Section \ref{sec:mixing}) to compute the permeability and the thermal conductivity. The linearly interpolated field (from nodal values) is exhibited in Fig.\ \ref{fig:U_solidfrac}. The implicit interface representation comes at the cost that the effective channel interface is particularly prone to imprecision arising from the coarseness of the local mesh density. This effect is translated into an induced roughness for the implicit representation, specially on non-conformal meshes, as shown in Fig \ref{fig:U_pipe}.

\subsection{Calibration of the Darcy coefficient}
\begin{figure}[h!]
\centering
\begin{subfigure}[h]{0.49\textwidth}
  \includegraphics[width=\textwidth]{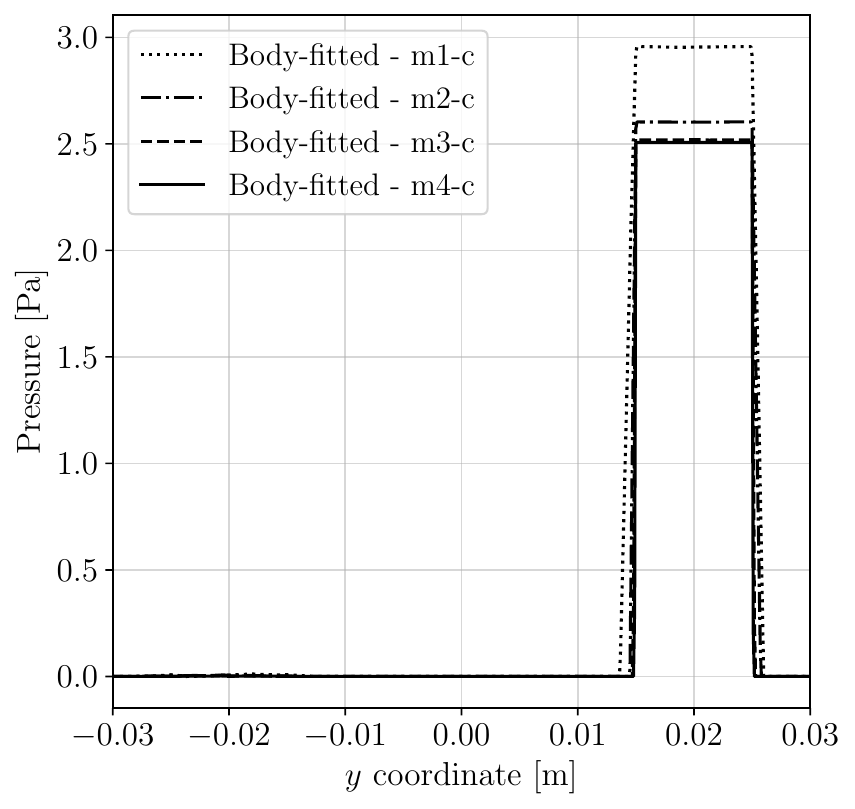}
  \caption{Pressure, $z=-0.1$}
\end{subfigure}
\begin{subfigure}[h]{0.49\textwidth}
  \includegraphics[width=\textwidth]{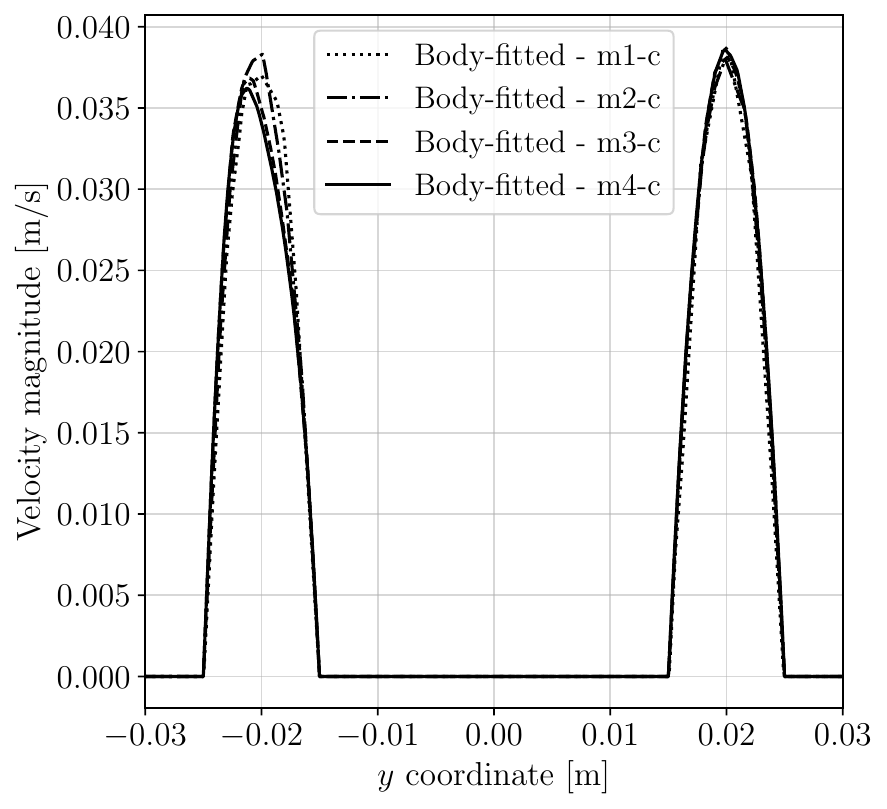}
  \caption{Velocity magnitude, $z=0$}
\end{subfigure}
  \caption{Convergence of the pressure and velocity magnitude profiles of the body-fitted CHT solver with mesh refinement along the transverse ($y$) direction, for $x=0$ and for two streamwise ($z$) locations.} 
\label{fig:U_VPprof_conv_conform}
\end{figure}

\begin{figure}[h!]
	\centering
	\begin{subfigure}[h]{0.49\textwidth}
	  \includegraphics[width=\textwidth]{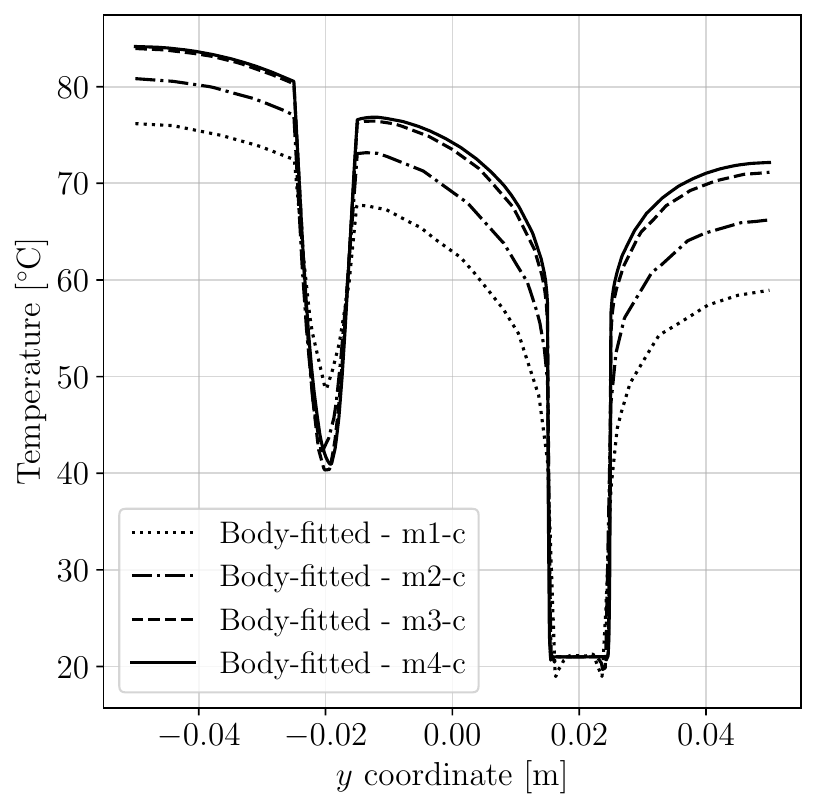}
	  \caption{$z=-0.1$}
	\end{subfigure}
	\begin{subfigure}[h]{0.49\textwidth}
	  \includegraphics[width=\textwidth]{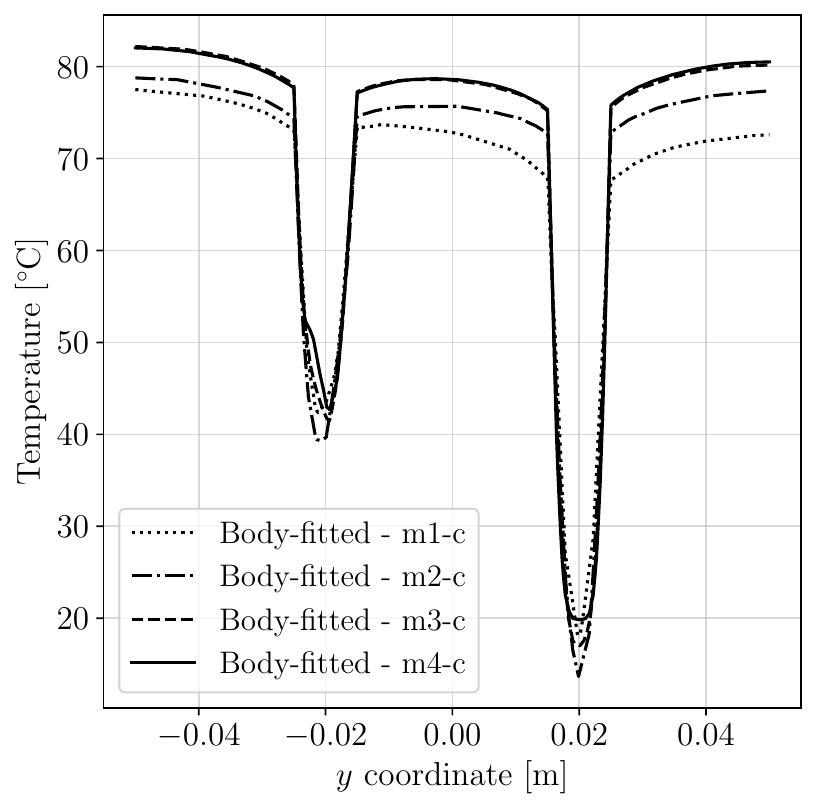}
	  \caption{$z=0$}
	\end{subfigure}
	  \caption{Convergence of the temperature profiles of the body-fitted CHT solver with mesh refinement along the transverse ($y$) direction, for $x=0$ and for two streamwise ($z$) locations.} 
	\label{fig:U_Tprof_conv_conform}
\end{figure}

A first step for the comparison of the porous vs body-fitted solutions, is to find the correct value of the Darcy coefficient, $C_\alpha$, for the problem at hand, as an insufficiently large value can result in the contamination of the solid regions by the fluid (non-zero velocities) and adversely, an exaggeratedly high value can result in spurious solutions or hinder the convergence of iterative solvers altogether. To perform the $C_\alpha$ calibration, we establish a reference solution via the body-fitted solver, on the finest considered mesh (m4-c), and ensure that the solution variations are diminishing with refinement. We recall that the Darcy model is not used by the body-fitted solver as the velocity in the solid region is set to zero. The results of this exercise are provided in Figs.\ \ref{fig:U_VPprof_conv_conform} and \ref{fig:U_Tprof_conv_conform}, where the velocity, pressure and temperature profiles are shown to converge with mesh refinement.

\begin{figure}[h!]
\centering
\begin{subfigure}[h]{0.49\textwidth}
  \includegraphics[width=\textwidth]{{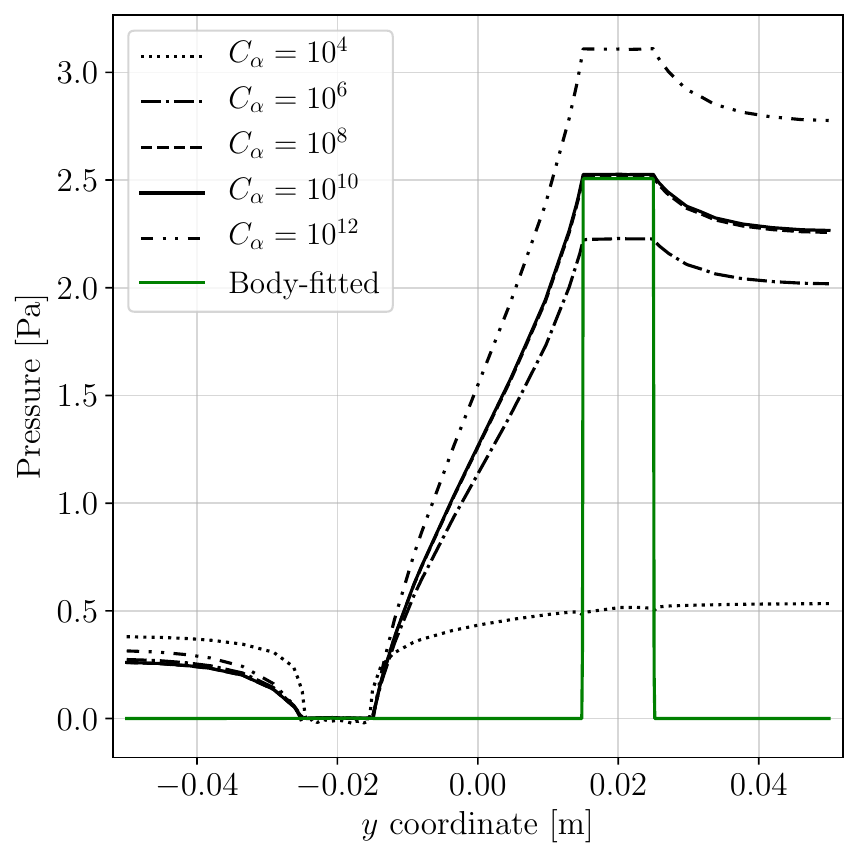}}
  \caption{Pressure, $z=-0.1$}
\end{subfigure}
\begin{subfigure}[h]{0.49\textwidth}
  \includegraphics[width=\textwidth]{{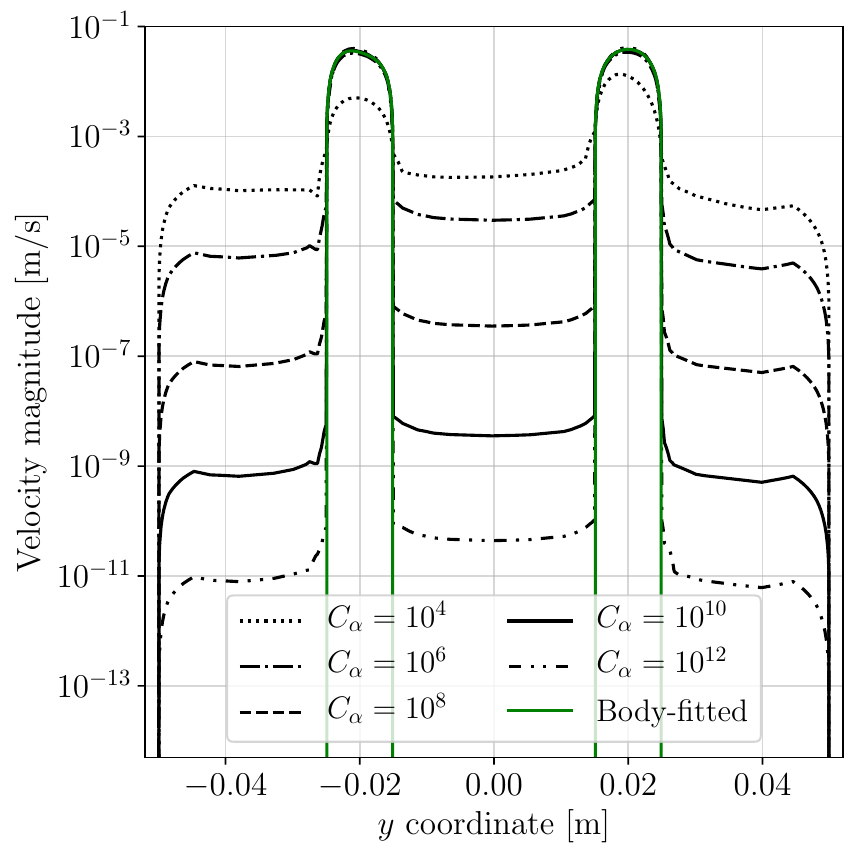}}
  \caption{Velocity magnitude, $z=0$}
\end{subfigure}
\\
\begin{subfigure}[h]{0.49\textwidth}
  \includegraphics[width=\textwidth]{{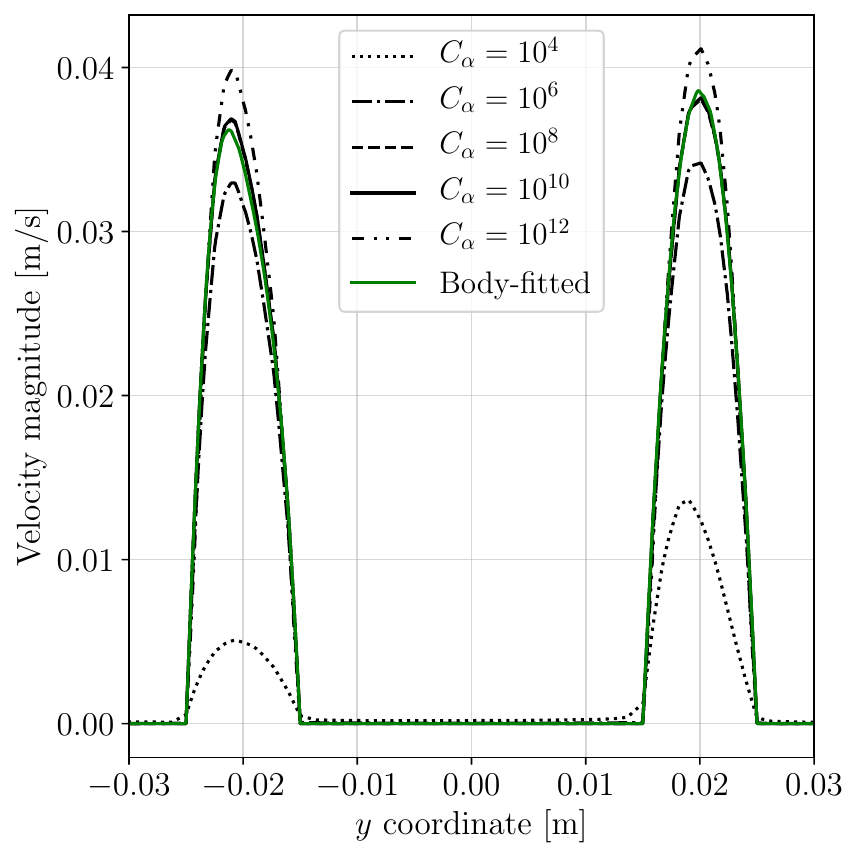}}
  \caption{Velocity magnitude, $z=0$, focus on channel flow}
\end{subfigure}
  \caption{Calibration of $C_\alpha$; Solutions of the porous CHT solver on conformal mesh m3-c vs solution of the body-fitted CHT solver on conformal mesh m4-c, along the transverse ($y$) direction, for $x=0$ and for two streamwise ($z$) locations.} 
\label{fig:U_presV_calpha_mes03c}
\end{figure}

In order to minimize the effect of the induced roughness caused by the implicit interface representation, we first consider the porous solutions on conformal mesh m3-c, such that the pressure solution is as comparable to the reference solution as possible. The results are illustrated in Fig.\ \ref{fig:U_presV_calpha_mes03c} where we can observe that by increasing the value of $C_\alpha$, the velocity magnitude in the solid regions diminishes, signifying the reduction of the fluid \textit{leakage} out of the channel. Nevertheless, a value of $C_\alpha=1\times 10^{12}$\,kg/m$^3$/s over-estimates the pressure drop as well as the velocity magnitude in the channel, which indicates a reduction in the effective channel section area. The value of $C_\alpha=1\times 10^{10}$\,kg/m$^3$/s appears to perform the best, as lower values increase the leakage and yield lower pressure drops, compared to the reference. We hence use this value on non-conformal meshes, and monitor its outcomes.

\begin{figure}[h!]
\centering
\begin{subfigure}[t]{0.49\textwidth}
  \includegraphics[width=\textwidth]{{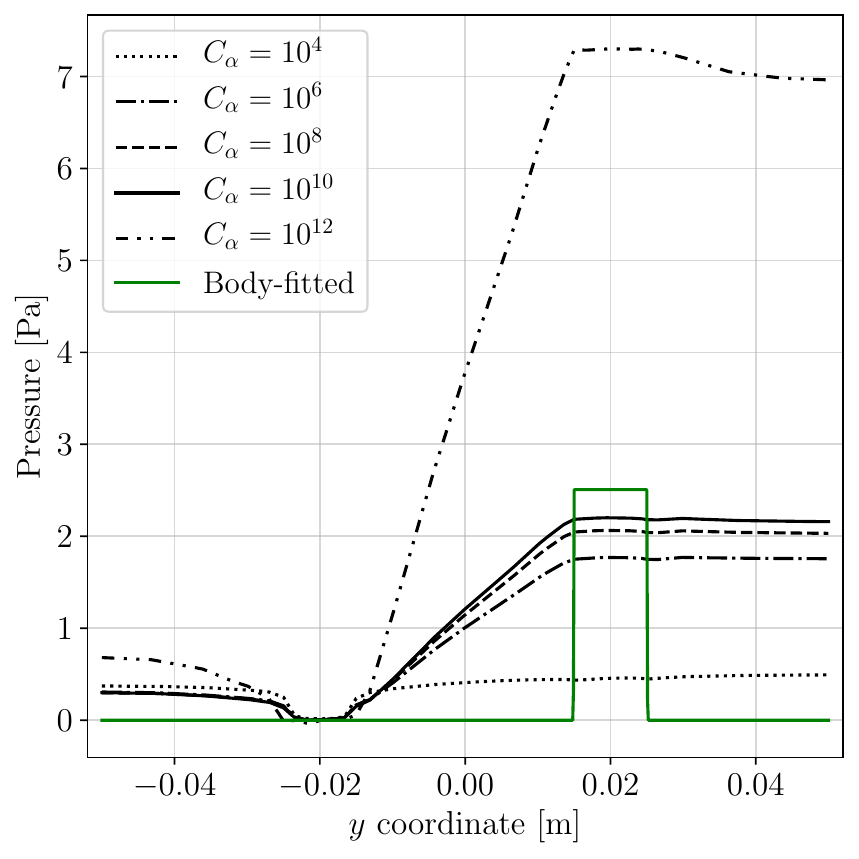}}
  \caption{Pressure, $z=-0.1$}
\end{subfigure}
\begin{subfigure}[t]{0.49\textwidth}
  \includegraphics[width=\textwidth]{{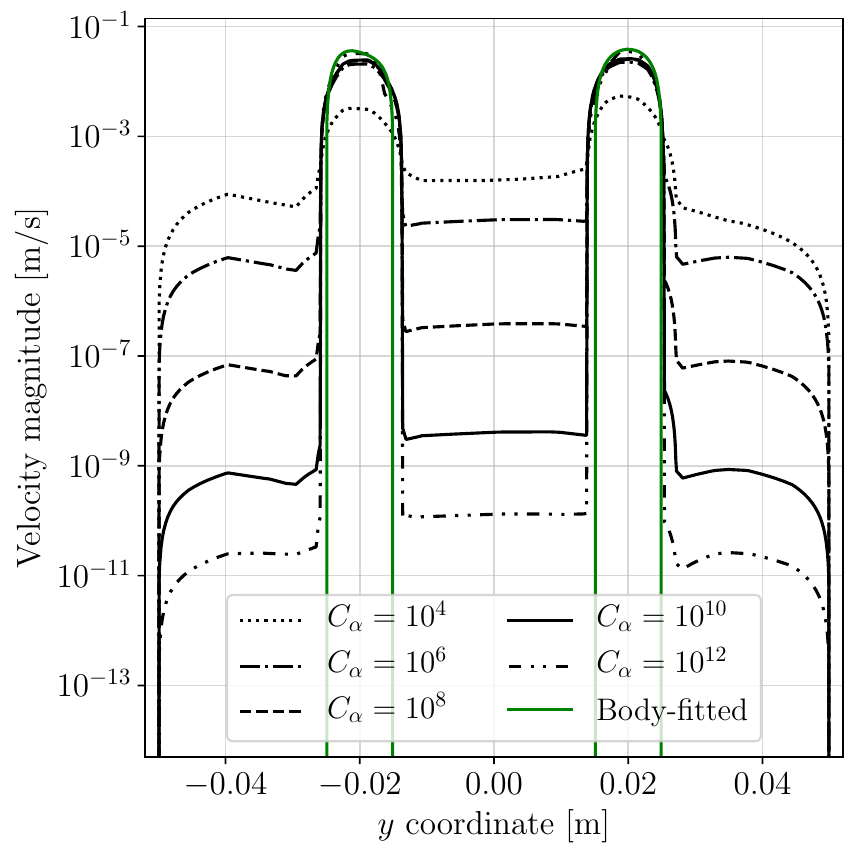}}
  \caption{Velocity magnitude, $z=0$}
\end{subfigure}
  \caption{Effect of $C_\alpha$; Solutions of the porous CHT solver on non-conformal mesh m1-n vs solution of the body-fitted CHT solver on conformal mesh m4-c, along the transverse ($y$) direction, for $x=0$ and for two streamwise ($z$) locations.} 
\label{fig:U_presV_calpha_mes01}
\end{figure}
\begin{figure}[h!]
\centering
\begin{subfigure}[h]{0.49\textwidth}
  \includegraphics[width=\textwidth]{{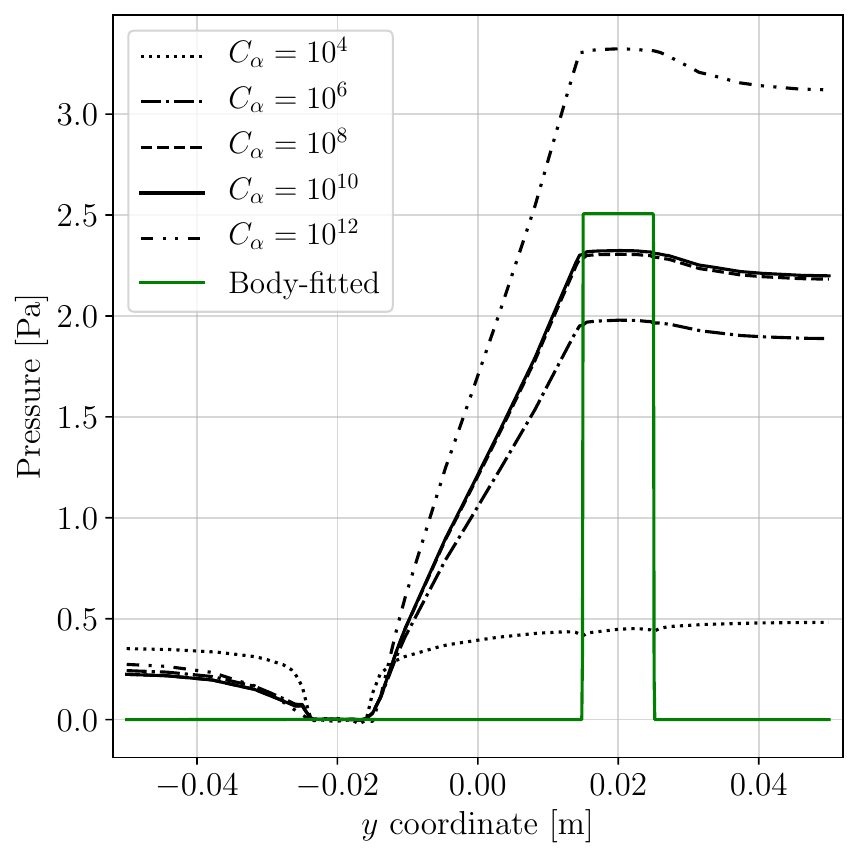}}
  \caption{Pressure, $z=-0.1$}
\end{subfigure}
\begin{subfigure}[h]{0.49\textwidth}
  \includegraphics[width=\textwidth]{{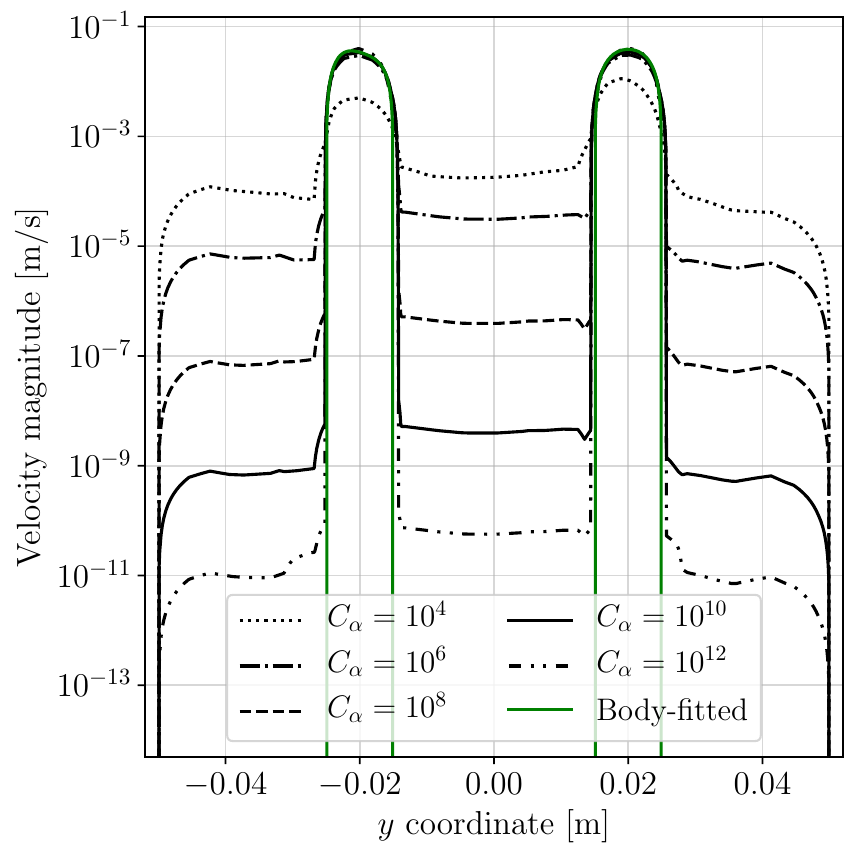}}
  \caption{Velocity magnitude, $z=0$}
\end{subfigure}
  \caption{Effect of $C_\alpha$; Solutions of the porous CHT solver on non-conformal mesh m2-n vs solution of the body-fitted CHT solver on conformal mesh m4-c, along the transverse ($y$) direction, for $x=0$ and for two streamwise ($z$) locations.} 
\label{fig:U_presV_calpha_mes02}
\end{figure}
\begin{figure}[h!]
\centering
\begin{subfigure}[h]{0.49\textwidth}
  \includegraphics[width=\textwidth]{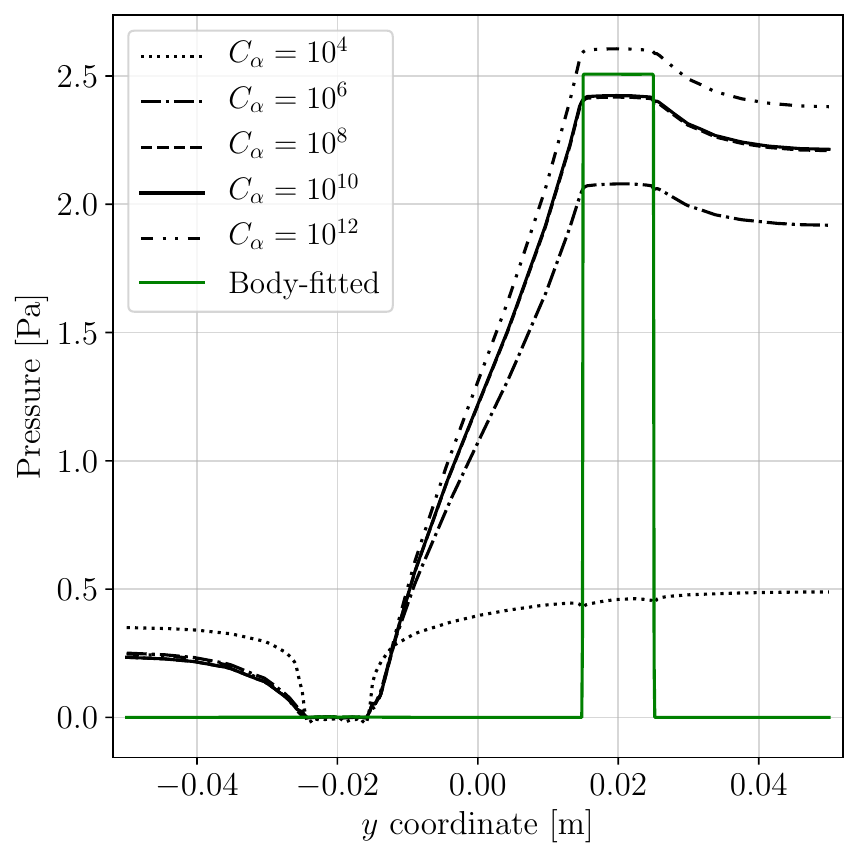}
  \caption{Pressure, $z=-0.1$}
\end{subfigure}
\begin{subfigure}[h]{0.49\textwidth}
  \includegraphics[width=\textwidth]{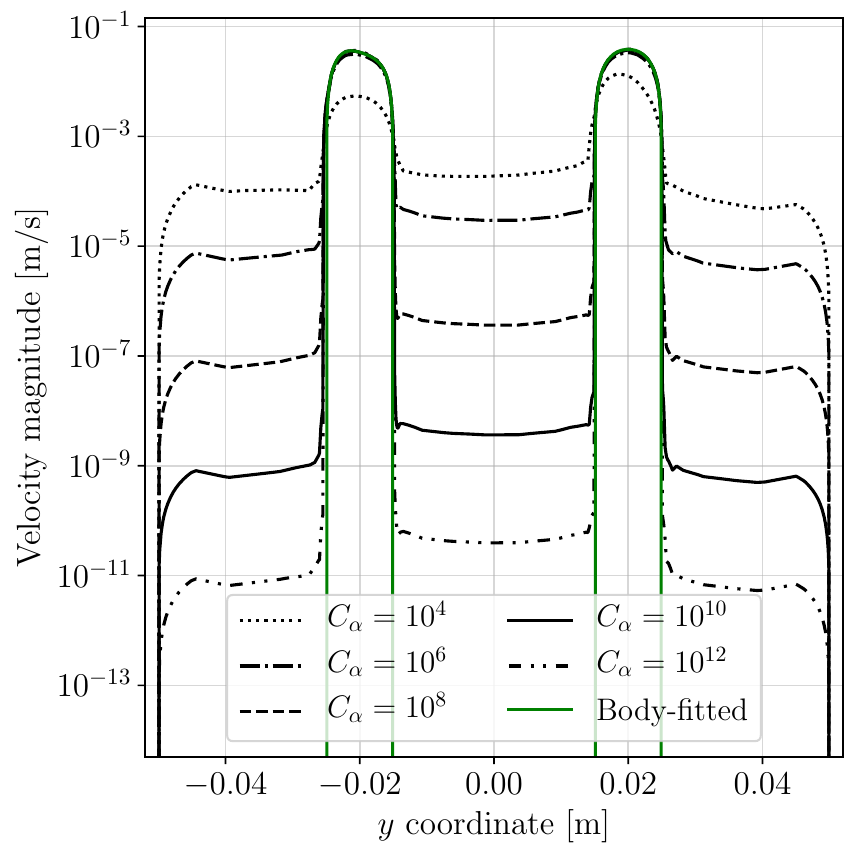}
  \caption{Velocity magnitude, $z=0$}
\end{subfigure}
  \caption{Effect of $C_\alpha$; Solutions of the porous CHT solver on non-conformal mesh m3-n vs solution of the body-fitted CHT solver on conformal mesh m4-c, along the transverse ($y$) direction, for $x=0$ and for two streamwise ($z$) locations.} 
\label{fig:U_presV_calpha_mes03}
\end{figure}
\begin{figure}[h!]
\centering
\begin{subfigure}[h]{0.485\textwidth}
  \includegraphics[width=\textwidth]{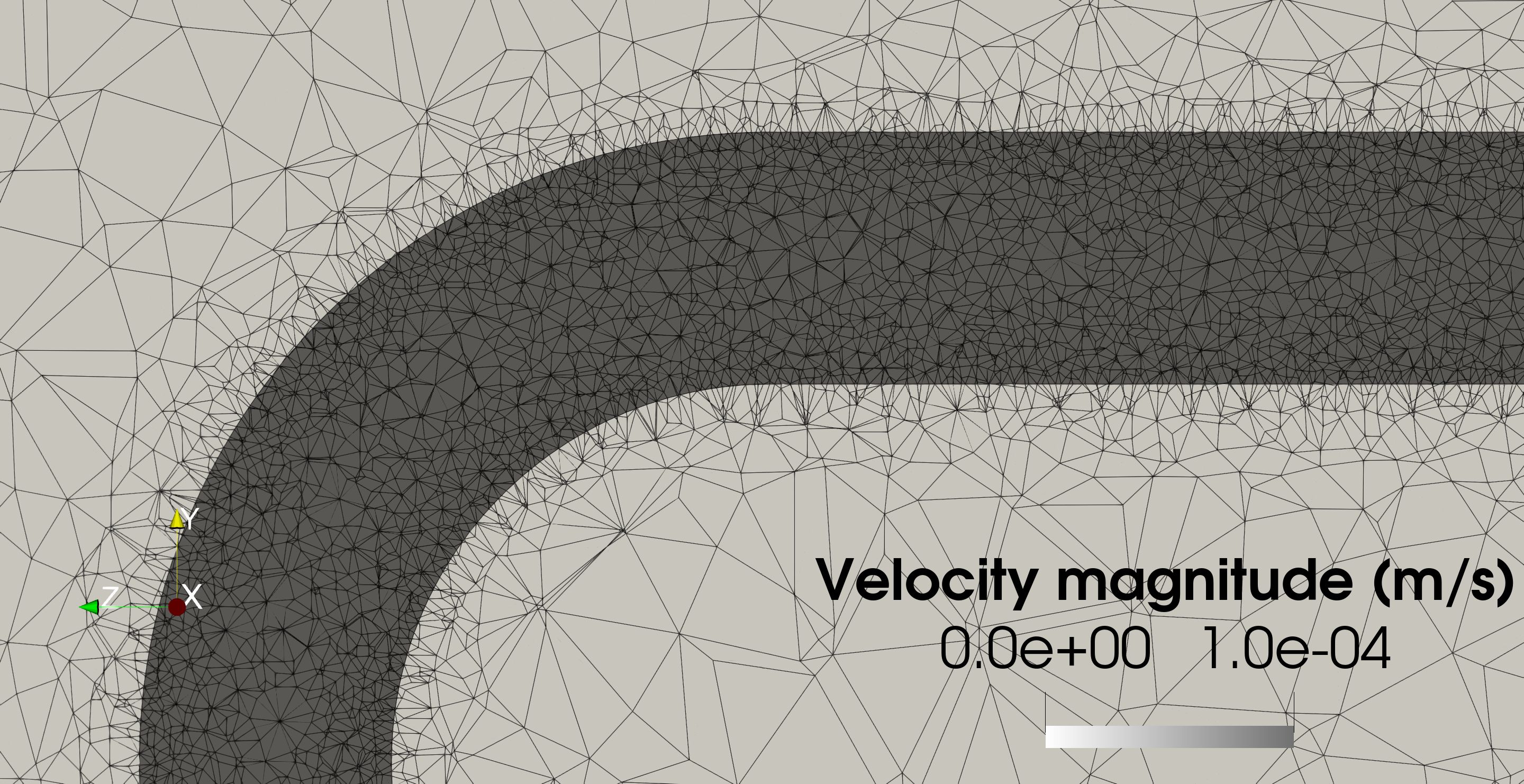}
  \caption{Mesh m3-c}
\end{subfigure}
\begin{subfigure}[h]{0.485\textwidth}
  \includegraphics[width=\textwidth]{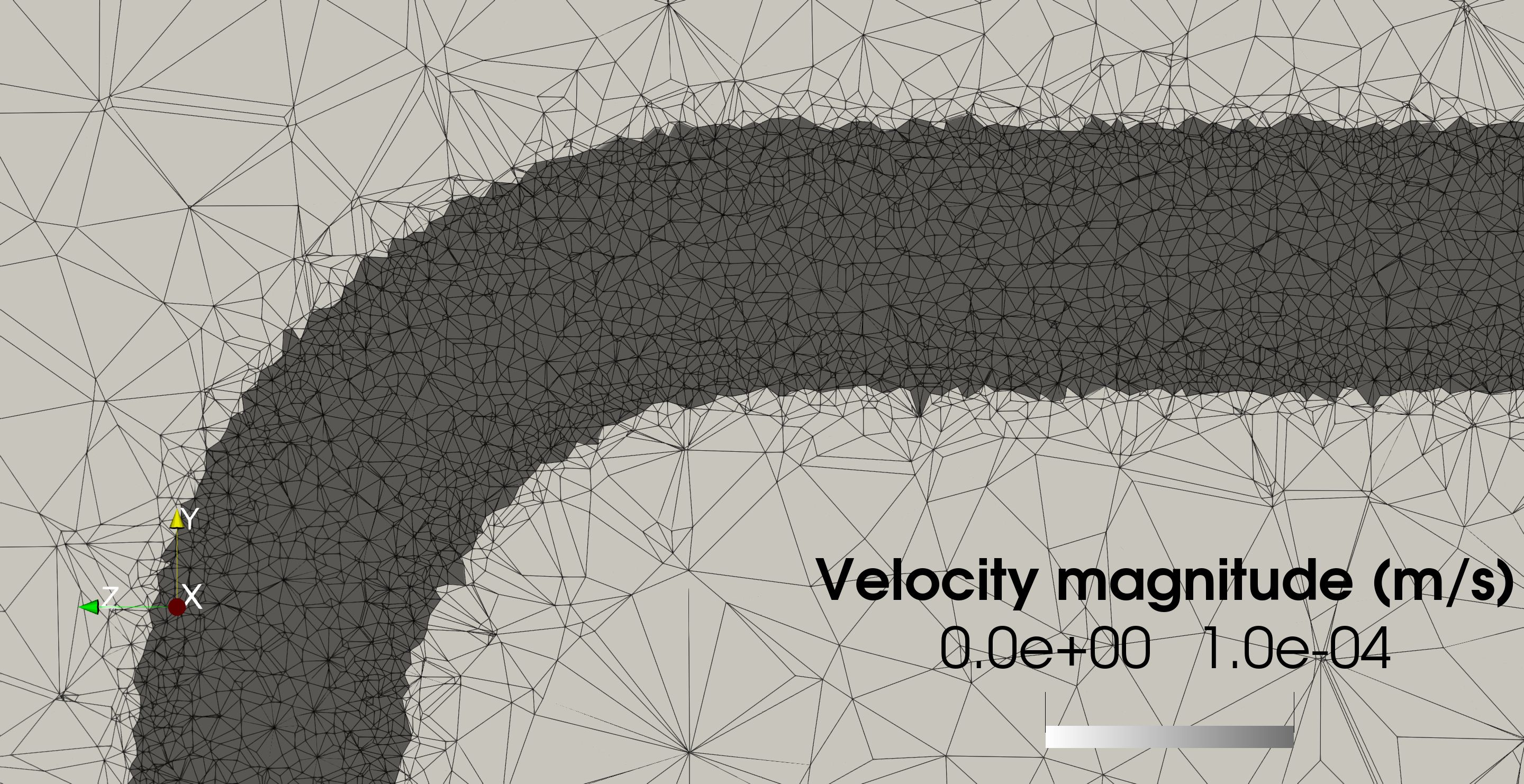}
  \caption{Mesh m3-n}
\end{subfigure}
  \caption{Effect of conformal versus non-conformal discretization on the flow speed solutions of the porous solver and the effective channel section area.}
\label{fig:confvsnonconf_Veloc}
\end{figure}

  
Through \cref{fig:U_presV_calpha_mes01,fig:U_presV_calpha_mes02,fig:U_presV_calpha_mes03}, we seek to observe the effect of  $C_\alpha$ on  pressure and velocity solutions obtained on the sequence of non-conformal meshes, compared to the reference solution, and to validate the chosen $C_\alpha$ value. These figures confirm the trends observed during the calibration phase: the value of $C_\alpha=1\times 10^{10}$\,kg/m$^3$/s is the one, among compared values, which minimizes the spurious velocity magnitude in the solid region, and furthermore is exempt of the significant pressure overshoot generated by $C_\alpha=1\times 10^{12}$\,kg/m$^3$/s. This overshoot diminishes however with mesh refinement. An interesting feature on these figures with regards to the pressure loss for $C_\alpha=1\times 10^{10}$\,kg/m$^3$/s, is that the converged value on mesh m3-n (Fig.\ \ref{fig:U_presV_calpha_mes03}) is lower than the reference value, whereas it was slightly higher on mesh m3-c (Fig.\ \ref{fig:U_presV_calpha_mes03c}). The cause of this discrepancy is the difference between the effective channel section areas on the conformal vs non-conformal meshes, when the channel is defined by nodal values of the solid fraction. This difference is visible in Fig.\ \ref{fig:confvsnonconf_Veloc}, which depicts the flow speed solution via the porous solver, clipped at $10^{-4}$\,m/s. The view is the same as in Fig.\ \ref{fig:U_solidfrac_zoom} (showing the interpolated solid fraction on mesh m3-n). From Fig.\ \ref{fig:confvsnonconf_Veloc}, one can appreciate the fact that the non-conformal position of the nodes has a twofold effect: 1. an induced roughness and 2.  a larger effective diameter of the channel. The latter causes a reduction in pressure losses, with regards to the solutions obtained on the conformal meshes.

Through the tests performed in this section, we can attest the adequacy of the value of  $C_\alpha=1\times 10^{10}$\,kg/m$^3$/s and proceed to a mesh convergence study for temperature profiles on porous solutions. 

\subsection{Mesh convergence of the porous solutions}
\begin{figure}[h!]
\centering
\begin{subfigure}[h]{0.49\textwidth}
  \includegraphics[width=\textwidth]{{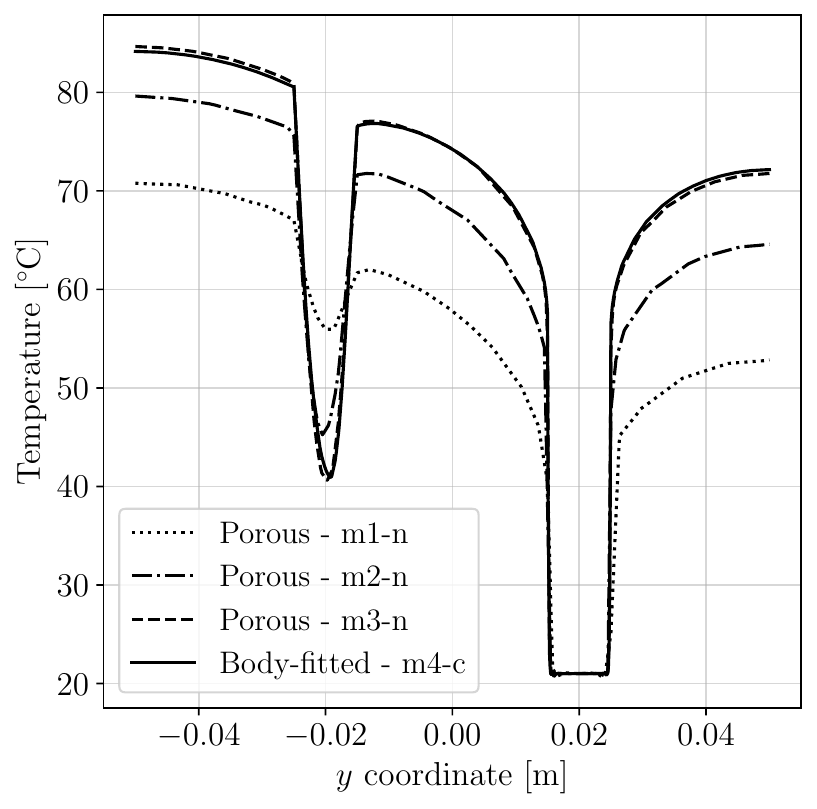}}
  \caption{$z=-0.1$}
\end{subfigure}
\begin{subfigure}[h]{0.49\textwidth}
  \includegraphics[width=\textwidth]{{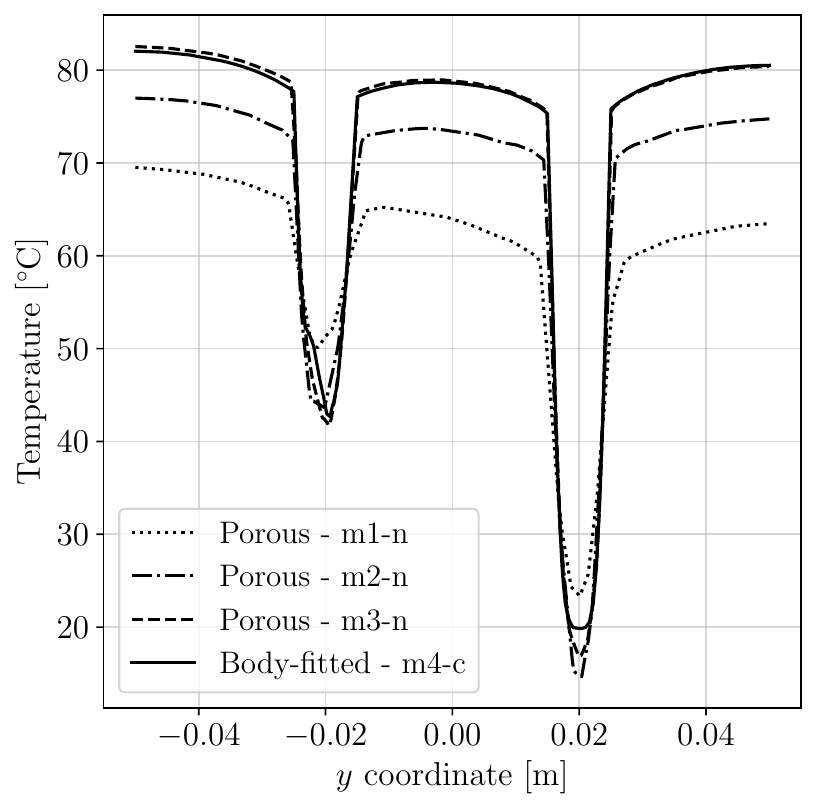}}
  \caption{$z=0$}
\end{subfigure}
  \caption{Convergence of the temperature profiles of the porous CHT solver on non-conformal meshes with mesh refinement, as compared to a body-fitted solution on m4-c, for $x=0$ and for two streamwise ($z$) positions.} 
\label{fig:U_Tprof_conv}
\end{figure}

Figure \ref{fig:U_Tprof_conv} presents the convergence, with mesh refinement, of the temperature profiles of the porous solutions on non-conformal meshes,  to the reference profiles obtained by the body-fitted solver on conformal mesh m4-c. The agreement is in general remarkable between the porous solution on m3-n and the conformal solution m4-c. This result complements the one obtained in \ref{sec:mms}, and thus enables us to conclude that the implementation of the porous CHT model is sound for  producing reliable results for complex and industrially-relevant configurations in similar operation regimes.

\section{Optimization results and discussion}
\label{sec:results}


%
%
%



\subsection{Cases definition}
\begin{figure}[h!]
	\centering
  \includegraphics[width=0.6\textwidth,trim={0.cm 0.cm 12.cm 0.cm},clip]{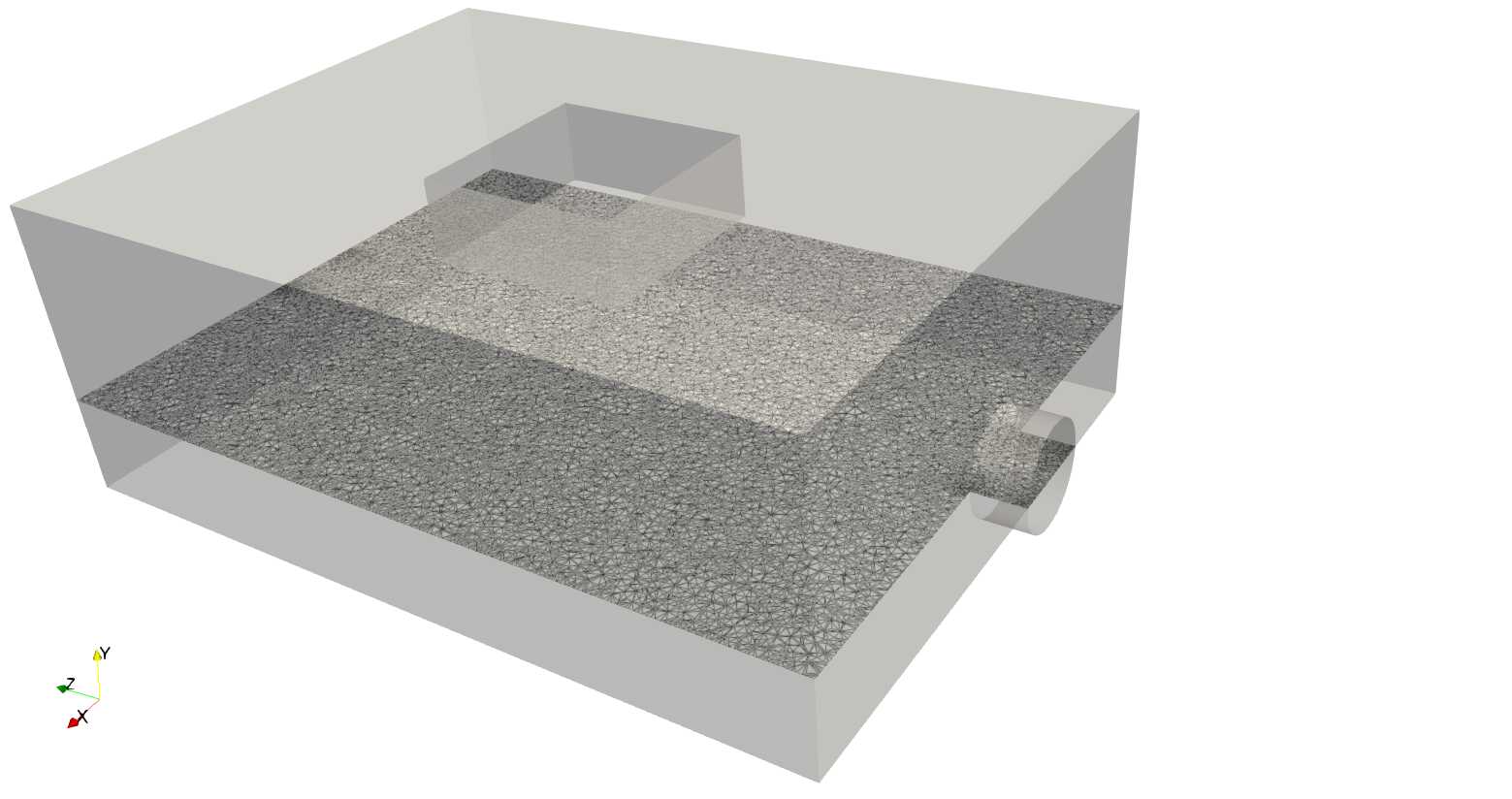}
  \caption{Domain and cut plane of the mesh for the optimization cases.}
	\label{fig:opt_domains}
\end{figure}

The considered geometry represents a simplified die casting mold with a heated cuboid cavity on the top as the die insert (see Fig.\ \ref{fig:opt_domains}). The geometry has a domain of $[-0.03,0.03]_x \times [-0.01,0.02]_y \times [0,0.08]_z$\,m$^3$ and has a small cavity on the top, with dimensions $[-0.01,0.01]_x \times [0.012,0.02]_y \times [-0.01,0.01]_z$\,m$^3$. The circular, extruded inlet is centered on $(0,0,-0.005)$ and has a radius of $r=0.005$\,m. A quasi-uniform mesh of 236,191 nodes is used to discretize the geometry.

The following conditions are imposed on the inlet: a parabolic profile for the velocity, defined by $[u_x,\, u_y,\, u_z] = [0,\, 0,\,U_\mathrm{max}\,( 1 - (x^2 + y^2)/r^2 )]$ where $U_\mathrm{max}=9.368635 \times 10^{-3}$\,m/s for a Reynolds number of $\mathrm{Re}=100$ and tenfold for $\mathrm{Re}= $ 1,000; and a temperature of $T=60\,^{\circ}\mathrm{C}$. The outlet is set free in the $z$ direction with $[u_x,\, u_y] = [0,\, 0]$\,m/s. No-slip conditions are applied to all other boundaries. All boundaries are set to adiabatic, except the inlet and the cuboid cavity on the top: a heating flux of 10\,kW/m$^2$ is imposed on the latter to account for the heat emanating from the molten metal. The thermofluid properties correspond to those of water and steel, evaluated at the inlet temperature, and are set to $\rho=993$\,kg/m$^3$, $c_p=$4,200\,J/kg/K, $\mu=4.6 \times 10^{-4}$\,N\,s/m$^2$, $\kappa_f = 0.65$\,W/m/K and $\kappa_s = 26$\,W/m/K. Furthermore, we have $P_\alpha=P_\kappa=3$ and the Darcy coefficient is set to $C_\alpha=10^{10}$\,kg/m$^3$/s. The Prandtl number is $\mathrm{Pr}=2.97$.

For each optimization step, the computation starts with the solution of the Navier--Stokes equations via the Picard (fixed point) method and finalized with the full Newton iterations, followed by the solution of the equation of energy. The residuals are minimized to machine precision for all equations.

In order to assess the topology optimization methodology developed in this work, we consider 15 cases which arise from variations of some of the most notable hyper-parameters of the optimization problem. These cases are described in Table \ref{Tab:OptCases}. Two different cost functions are considered: $\bar{T}_\Gamma$ and $\bar{T}_\Omega$ (see Section \ref{sec:opt-prob}). The former accounts for the average temperature on the heated surfaces of the top cavity whereas the latter stands for the average temperature in the whole domain. The remaining hyper-parameters of interest are: the maximum total fluid volume fraction, $1-\bar{\gamma}_0$ (see the constraint $\mathcal{G}_1$ in Section \ref{sec:FSconst}), the neighborhood radius of the geometrical filter, $\varrho$ defined in Section \ref{sec:smthdheavy}, and  the coefficient of penalization of the pressure losses in the cost function, $\zeta$ introduced in Section \ref{sec:preslos}. 
\begin{table}[!ht]  
	\centering
	\begin{tabular}{cccccc}
		\toprule
		Case  & Cost	& Re	& $1-\bar{\gamma}_0$	& $\varrho$ (m) & $\zeta$ ($^{\circ}\mathrm{C}$/Pa)	\\ 
		\midrule
		A 		& $\bar{T}_\Gamma$ & 100 	& 10\% 					& 4E-3 		& 3E-4		\\
		B 		& $\bar{T}_\Gamma$ & 100 	& 10\% 					& 4E-3 		& \textbf{3E-3}		\\
		C 		& $\bar{T}_\Gamma$ & 100 	& 10\% 					& 4E-3 		& \textbf{3E-5}		\\
		D 		& $\bar{T}_\Gamma$ & 100 	& 10\% 					& \textbf{6E-3} 		& 3E-4		\\
		E 		& $\bar{T}_\Gamma$ & 100 	& 10\% 					& \textbf{8E-3} 		& \textbf{3E-3}		\\
		F 		& $\bar{T}_\Gamma$ & 100 	& 10\% 					& \textbf{8E-3} 		& 3E-4		\\
		G 		& $\bar{T}_\Gamma$ & 100 	& \textbf{5\% }					& 4E-3 		& 3E-4		\\
		H 		& $\bar{T}_\Gamma$ & 100 	& \textbf{20\%} 					& 4E-3 		& 3E-4		\\
		I 		& $\bm{\bar{T}_\Omega}$ & 100 	& 10\% 					& \textbf{8E-3} 		& 3E-4		\\ 
		\midrule    
		J 		& $\bar{T}_\Gamma$ & \textbf{1000} 	& 10\% 					& 4E-3 		& 3E-4		\\
		K 		& $\bar{T}_\Gamma$ & \textbf{1000} 	& 10\% 					& 4E-3 		& \textbf{3E-5}		\\    
		L 		& $\bm{\bar{T}_\Omega}$ & \textbf{1000} 	& 10\% 					& 4E-3 		& \textbf{6E-5}		\\
		\bottomrule
	\end{tabular}
	\caption{The optimization cases and their defining hyper-parameters. Bold characters indicate a variation from the respective parameter value for Case A.}
	\label{Tab:OptCases}
\end{table}

In all cases, the initial condition of the optimization is $\gamma=1$, i.e. pure solid, and four loops of 100 optimization steps are performed, \ie one for every value of the smoothed-Heaviside parameter $\beta$ (cf. Section \ref{sec:smthdheavy}). The optimization is stopped if the relative cost function difference between consecutive steps is lower than $10^{-5}$. Then, we proceed to the selection of the optimal topology as follows: we look at all steps of each optimization case and cherry-pick a design which produces a fair trade-off between the cost function minimization and the continuity of the channels. In Case A for example, the design for step 342 is selected, where $\beta=8$. These aspects will be further examined in the following. For each case, the $\beta$ value of the selected step is also provided in the table of results, which is found below (Table \ref{Tab:OptResults}). 

Computations are carried on the NRC's high performance computing clusters. For Case A, the computational effort is estimated to 0.6 core-hours per optimization step when using 72 cores. By considering 400 optimization steps, this translates into a computational time of 3.3 hours. The authors are aware that the use of the density filtering is inherently slow for large filter radii in three dimensions and that partial differential equation filtering methods exist in the literature \cite{lazarovFiltersTopologyOptimization2011}, which are more computationally efficient in parallel computing. Such methods were not implemented at the time of writing. Nevertheless, it may be noted that, because the mesh does not change during the optimization, the sparse matrix $\bm{\mathcal{D}}$ involved in Eqs.~\eqref{eq:density_filt} and \eqref{eq:dGammaStardGammaHat} only needs to be evaluated once. Hence, for the cases with $\varrho=\text{4E-3}$ as an example, the computation involving the projection schemes accounted for at most 3\% of the total optimization time.




\begin{figure}[h!]
\centering
\begin{subfigure}[t]{0.31\textwidth}
  \includegraphics[width=\textwidth,trim={0.cm 0.cm 0.cm 0.cm},clip]{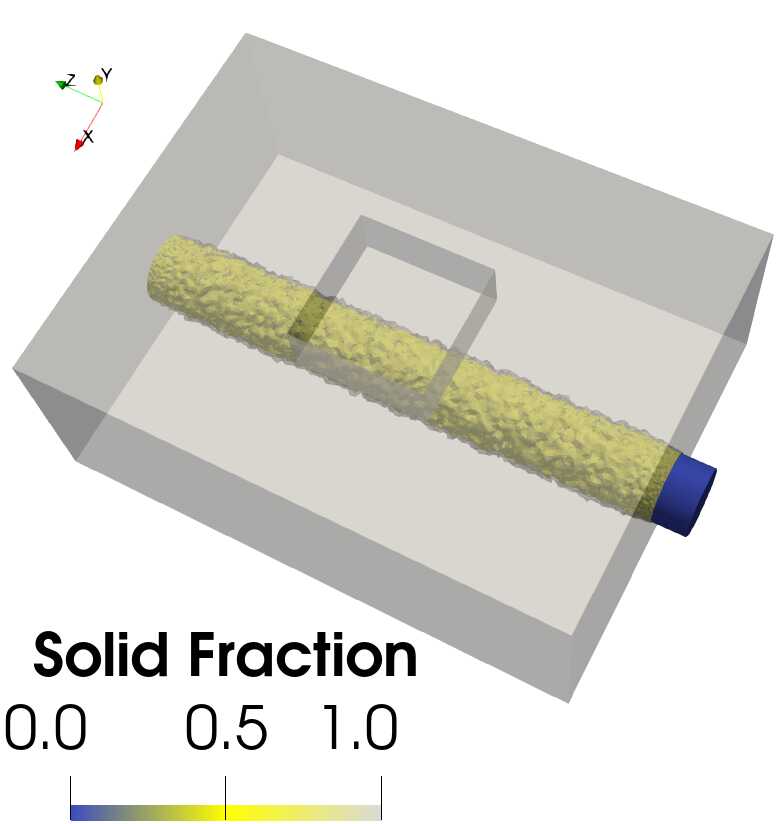}
  \caption{Solid fraction on the outer domain walls and clip on $\gamma\leq0.5$.}
  \label{fig:BaseSC_SF}
\end{subfigure}
~
\begin{subfigure}[t]{0.31\textwidth}
  \includegraphics[width=\textwidth,trim={0.cm 0.cm 0.cm 0.cm},clip]{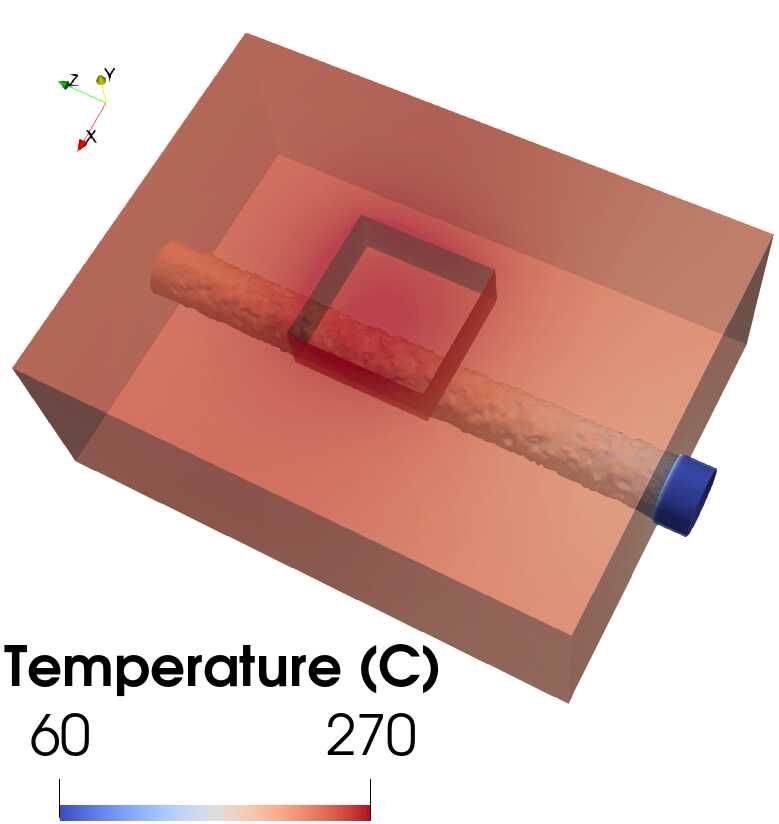}
  \caption{$\mathrm{Re}=100$; Temperature on the outer domain walls and isocontour of flow speed at $10^{-3}$\,m/s, colored by the temperature.}
  \label{fig:BaseSC_Re100}
\end{subfigure}
~
\begin{subfigure}[t]{0.31\textwidth}
  \includegraphics[width=\textwidth,trim={0.cm 0.cm 0.cm 0.cm},clip]{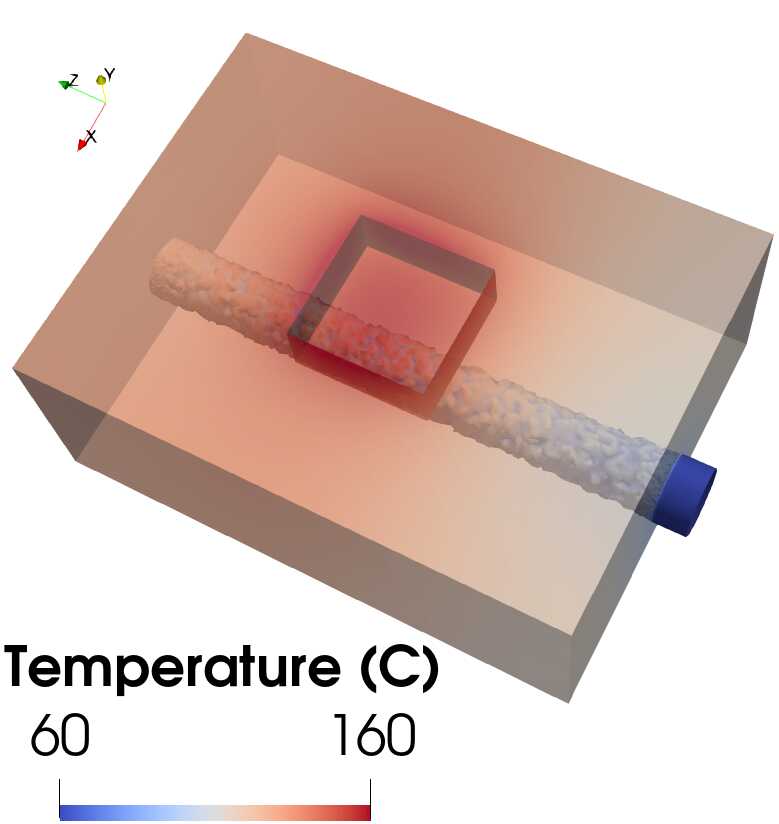}
  \caption{$\mathrm{Re}=\text{1,000}$; Temperature on the outer domain walls and isocontour of flow speed at $10^{-3}$\,m/s, colored by the temperature.}
  \label{fig:BaseSC_Re1000}
\end{subfigure}
  \caption{Baseline design.}
\label{fig:BaseSC}
\end{figure}


\begin{table}[!ht]  
  \centering
  \begin{tabular}{lcccc}
\toprule
    Baseline design					& ${\bar{T}_\Gamma}^0\,(^{\circ}\mathrm{C})$ 		& ${\bar{T}_\Omega}^0\,(^{\circ}\mathrm{C})$  & $T_\mathrm{min}^0\,(^{\circ}\mathrm{C})$	& $T_\mathrm{max}^0\,(^{\circ}\mathrm{C})$\\ 
\midrule
    $\mathrm{Re}=100$  		& 265						& 231  					&	47					& 275\\
    $\mathrm{Re}=\text{1,000}$ 		& 155						& 123 					&	53					& 164\\
\bottomrule
  \end{tabular}
  \caption{Results of the baseline designs; The average temperatures on the surfaces of the cavity as well as the average, the minimum and the maximum temperatures in the domain.}
\label{Tab:baseCosts}
\end{table}

\subsection{Baseline configurations}
In order to appreciate the improvement, in terms of the cooling efficiency of the obtained topology, we need a baseline configuration to serve as the reference for comparison. Since the current state-of-the-art cooling channel design often relies on conventional manufacturing techniques (in contrast to additive manufacturing), we adopt, as baseline, the case of a straight-drilled channel, concentric with the inlet and outlet surfaces and having a radius of 0.005\,m; the channel is represented implicitly via $\gamma$, using a Heaviside step function (there are no intermediate nodal $\gamma$ values), \ie similarly to Eq.~\eqref{eq:U_sf}. Furthermore, the baseline cases share, with the optimization cases, the same meshes as well as the same properties and inlet conditions for the respective Reynolds numbers. There is nevertheless a notable difference for the baseline cases: the filter radius is set to  $\varrho=0$, as the filter is rather useful in an optimization context. 

The variation of the Reynolds number ($\mathrm{Re}=100$ vs $\mathrm{Re}=\text{1,000}$) produces two baseline cases, for which the results are provided in Fig.~\ref{fig:BaseSC} and in Table \ref{Tab:baseCosts}. From the latter we observe that the null filter width, along with the sharp fluid to solid transition and the lack of finer elements near the pseudo-interface, cause some unphysical temperature undershoots\footnote{Our investigation have shown that the amount and magnitude of the undershoots can decrease by refining the mesh.} (below $60\, ^{\circ}\mathrm{C}$). We also note that increasing the Reynolds number produces an enhanced cooling, since the Peclet number, which indicates the ratio of advective over diffusive heat transfer rates, scales as $\mathrm{Pe} = \mathrm{Re} \, \mathrm{Pr}$.


\begin{table}[t]
	\centering
	\begin{tabular}{ccccccccc}
		\toprule
		Case & $\beta$	& $\mathcal{C}\,(^{\circ}\mathrm{C})$ & $\bar{T}_\Gamma\,(^{\circ}\mathrm{C})$ & $\frac{\bar{T}_\Gamma^0-\bar{T}_\Gamma}{\bar{T}_\Gamma^0}$		&$\bar{T}_\Omega\,(^{\circ}\mathrm{C})$ &$\frac{\bar{T}_\Omega^0-\bar{T}_\Omega}{\bar{T}_\Omega^0}$& $T_\text{min}\,(^{\circ}\mathrm{C})$ & $T_\text{max}\,(^{\circ}\mathrm{C})$ \\
		\midrule			
		A    & 8    	  	& 144           	& 139                	& 48\%	& 112	&    52\%      & 55		& 163		 \\
		B    & 8    		& 156           	& 151                	& 43\%	& 112	&    52\%      & 60		& 163		 \\
		C    & 1    		& 134           	& 131                	& 51\%	& 101	&    56\%      & 59		& 255		 \\
		D    & 8    		& 148         		& 142                	& 46\%	& 110	&    52\%      & 57		& 160		 \\
		E    & 4    		& 199           	& 174                	& 34\%	& 137	&    41\%      & 57		& 185		 \\
		F    & 4    		& 188           	& 183                	& 31\%	& 142	&    39\%      & 54		& 333		 \\
		G    & 4    		& 145           	& 138                	& 48\%	& 110	&    52\%      & 54		& 170		 \\
		H    & 4    		& 140           	& 136                	& 49\%	& 102	&    56\%      & 56		& 150		 \\
		I    & 4    		& 68            	& 193                	& 27\%   & 66	&    72\%      & 34		& 368		 \\
		\midrule		
		J    & 4    		& 98            	& 92                 	& 41\%	& 74	&    40\%      & 51		& 132		 \\
		K    & 2    		& 82            	& 79                 	& 49\%	& 64	&    48\%      & 43		& 126		 \\
		L    & 2    		& 52            	& 116                 	& 25\%  	& 51	&    59\%      & -51		& 293		 \\
		\bottomrule
	\end{tabular}
  \caption{Optimization results.} 
\label{Tab:OptResults}   
\end{table}

\begin{figure}[h!]
	\centering
	\includegraphics[width=0.8\textwidth]{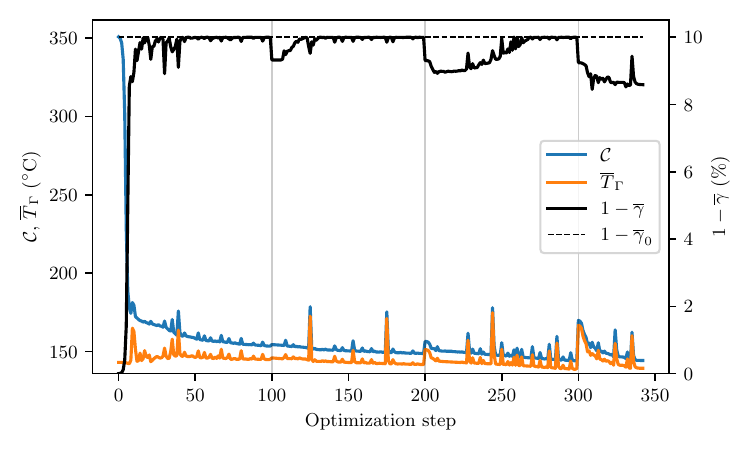}
	\caption{Evolution of the cost function $\cost$, cavity average surface temperature $\bar{T}_\Gamma$ and volume constraint $\constraint_1$, versus optimization steps, for Case A.}
	\label{fig:costEvol-A}
\end{figure}

\subsection{Optimization cases}
\begin{figure}[h!]
	\centering
	\begin{subfigure}{0.32\textwidth}
		\includegraphics[width=\textwidth]{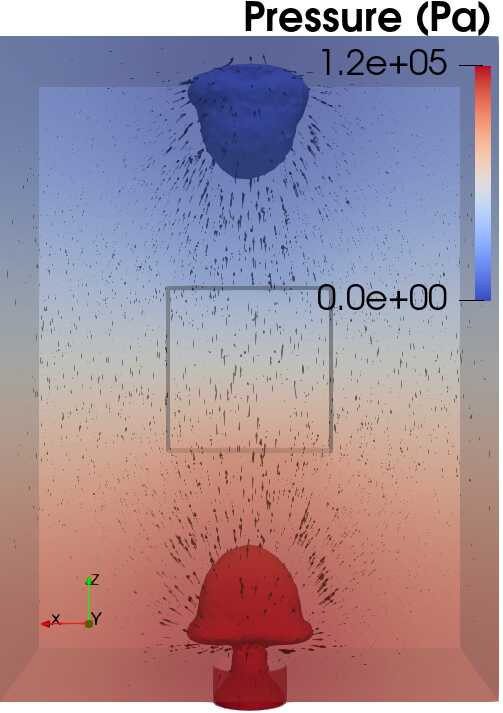}
		\caption{Step 7}
	\end{subfigure}\,
	\begin{subfigure}{0.32\textwidth}
		\includegraphics[width=\textwidth]{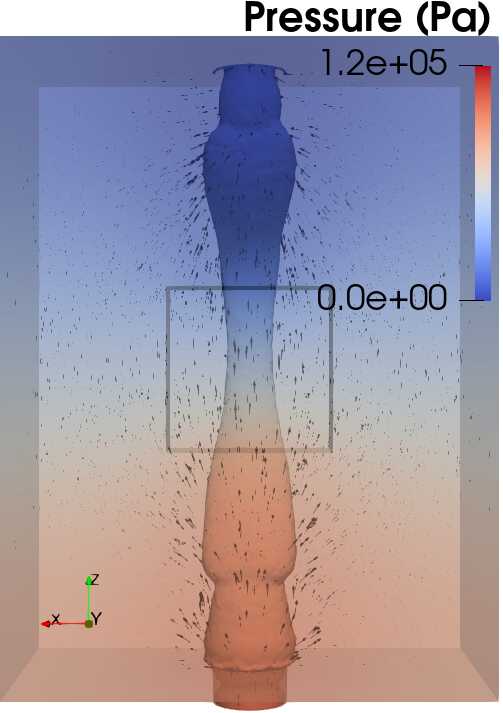}
		\caption{Step 8}
	\end{subfigure}\,
	\begin{subfigure}{0.32\textwidth}
		\includegraphics[width=\textwidth]{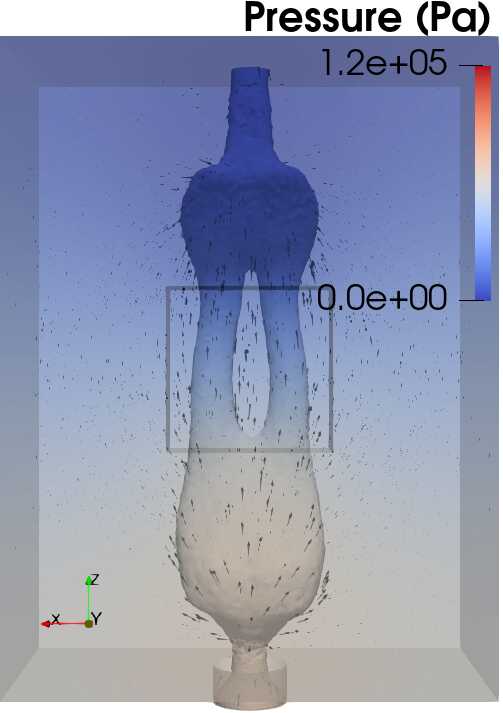}
		\caption{Step 20}
	\end{subfigure}\,
	\caption{Optimized channel designs at selected steps for Case A, as represented by the $10^{-3}$ m/s isocontour of the flow speed and colored by the pressure; Velocity vectors outside of the channels.}
	\label{fig:Viso-A-beta1}	
\end{figure}

\begin{figure}[h!]
	\centering
	\begin{subfigure}{0.49\textwidth}
		\includegraphics[width=\textwidth]{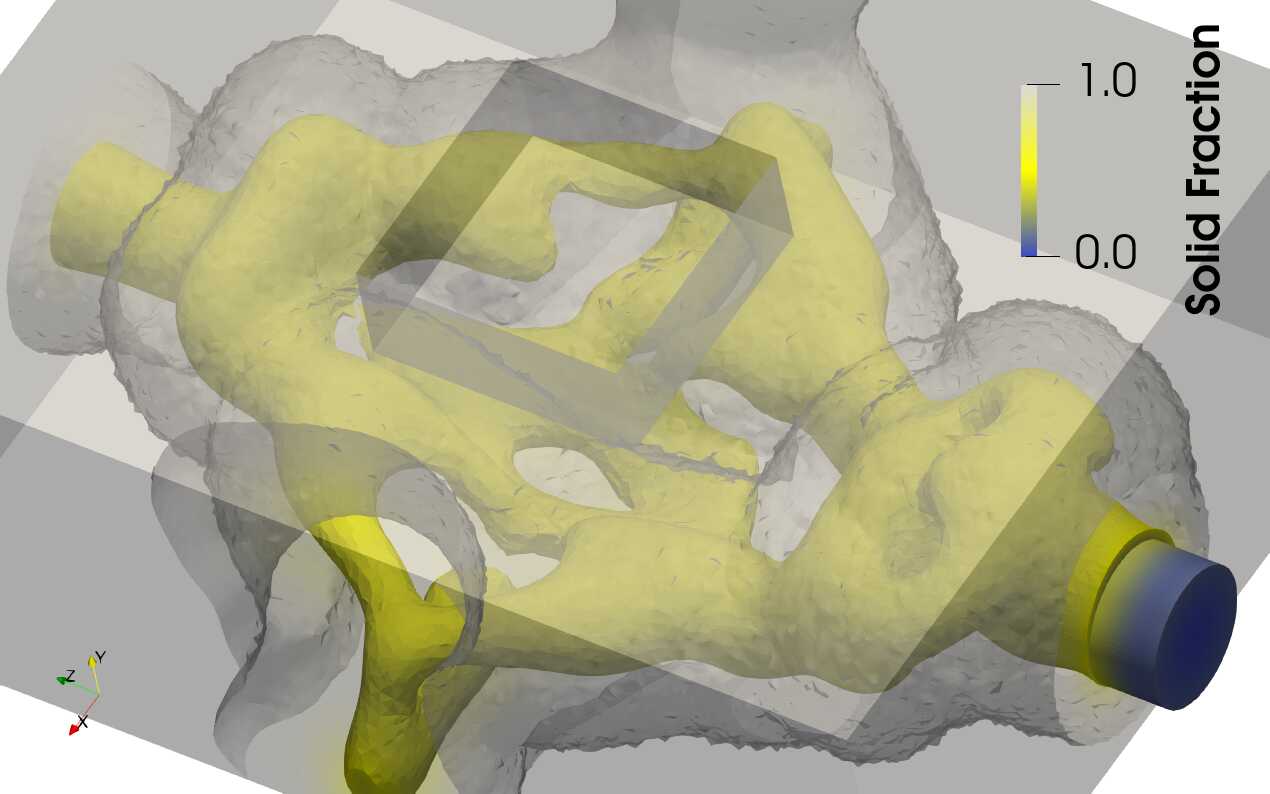}
		\caption{Step 98; $\beta=1$}
	\end{subfigure}\,
	\begin{subfigure}{0.49\textwidth}
		\includegraphics[width=\textwidth]{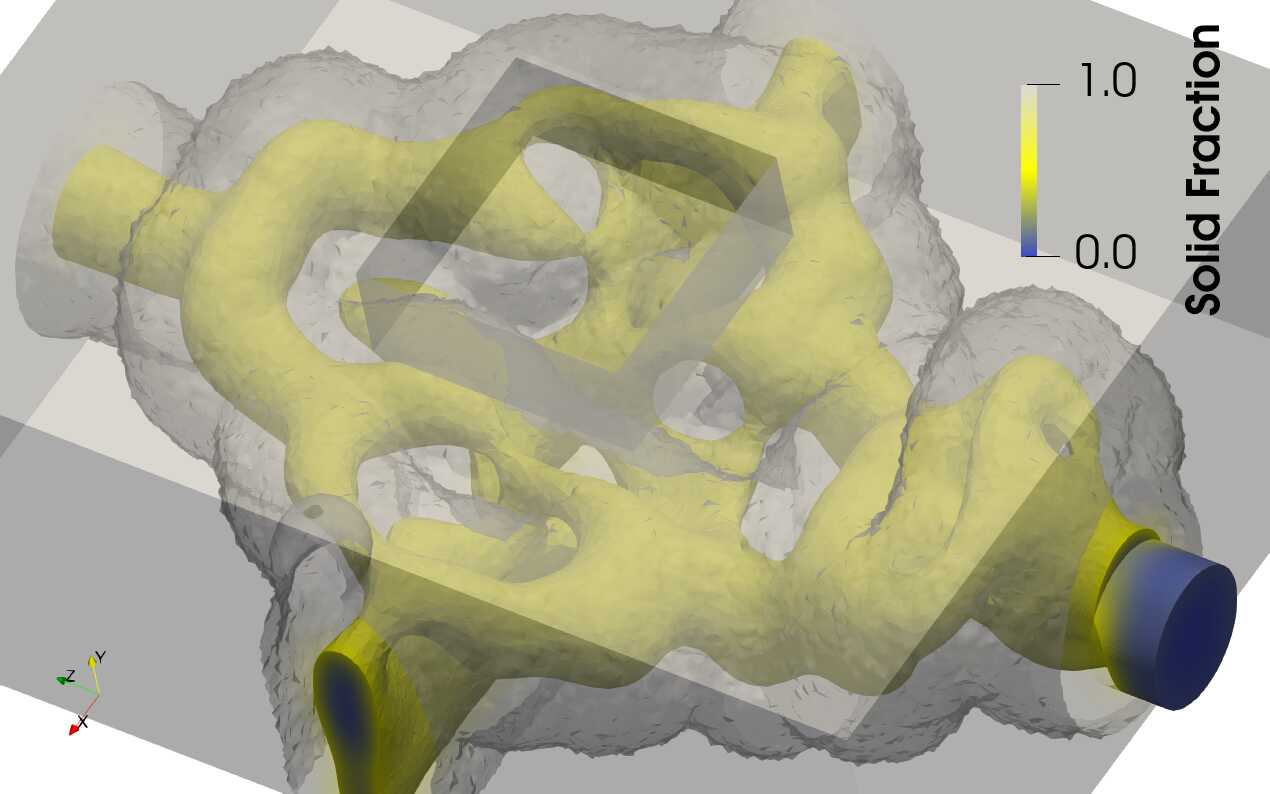}
		\caption{Step 199; $\beta=2$}
	\end{subfigure}\\
	\begin{subfigure}{0.49\textwidth}
		\includegraphics[width=\textwidth]{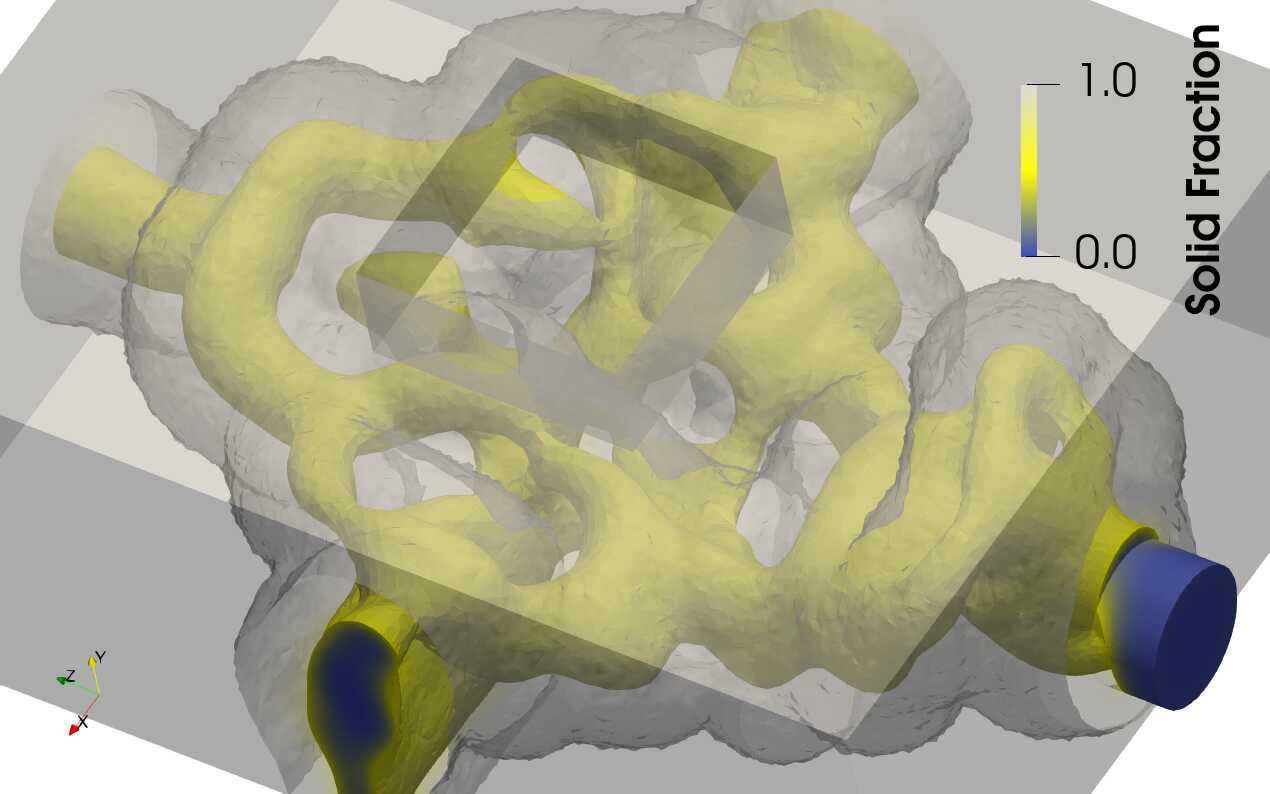}
		\caption{Step 299; $\beta=4$}
	\end{subfigure}\,
	\begin{subfigure}{0.49\textwidth}
		\includegraphics[width=\textwidth]{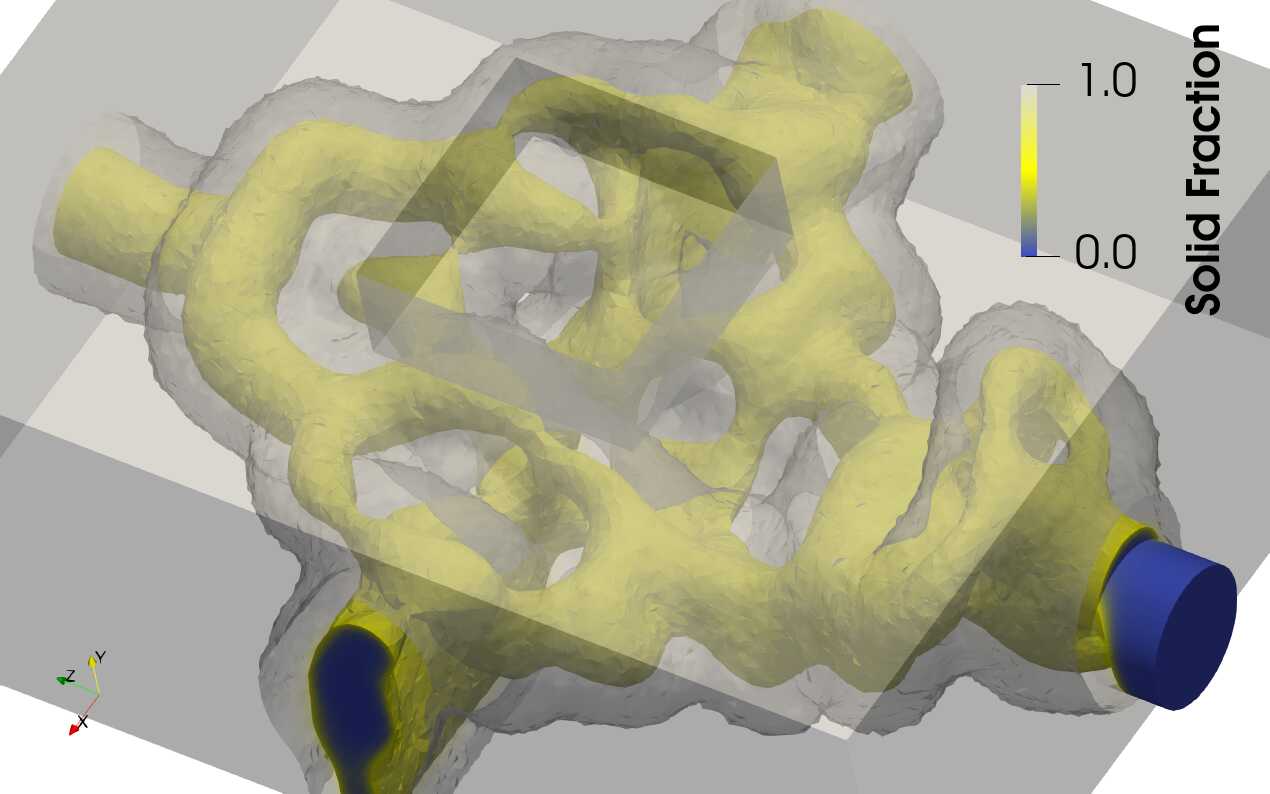}
		\caption{Step 342; $\beta=8$}
	\end{subfigure}
	\caption{Case A; Optimized designs at different $\beta$ values; Solid fraction on the outer domain walls, clip on $\gamma\leq0.5$ and isocontour of $\gamma=0.999$.}
	\label{fig:SF-A-beta}	
\end{figure}

\begin{figure}[h!]
	\centering
	\begin{subfigure}{0.49\textwidth}
		\includegraphics[width=\textwidth]{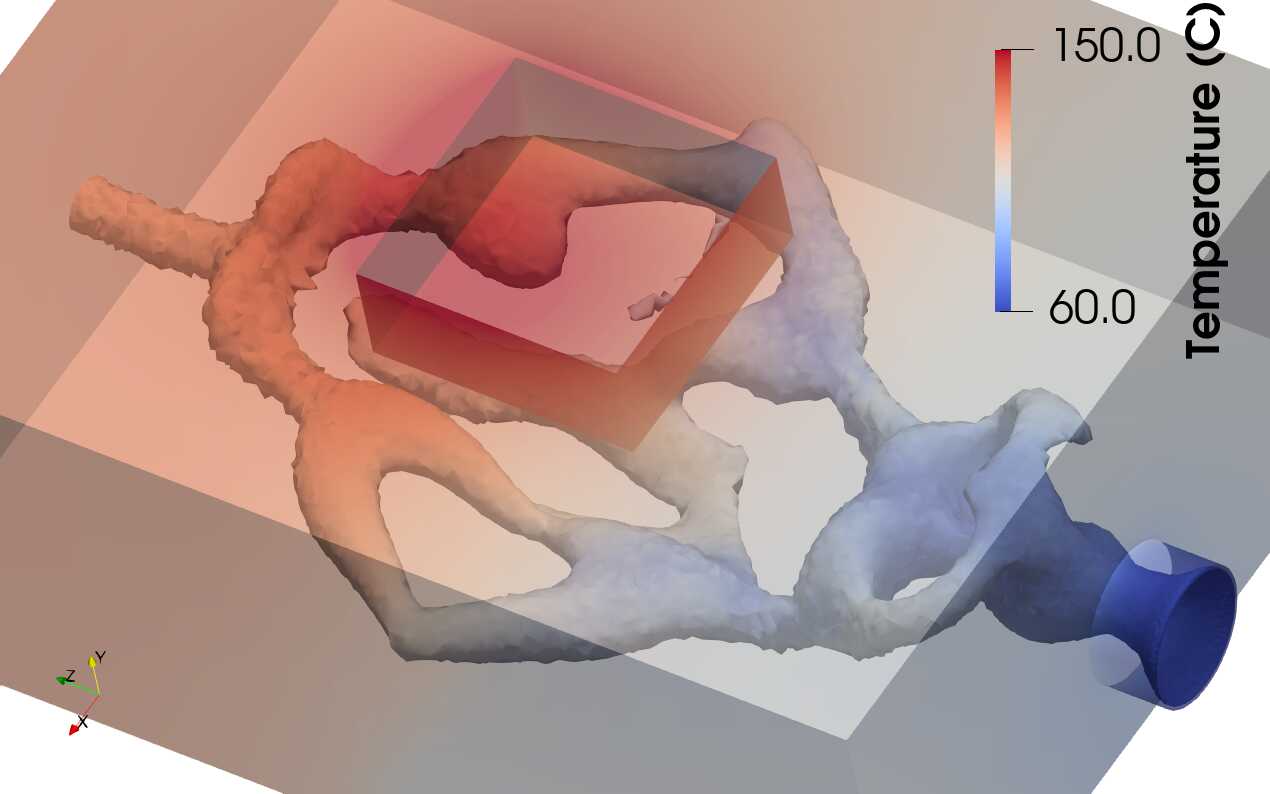}
		\caption{Step 98; $\beta=1$}
	\end{subfigure}\,
	\begin{subfigure}{0.49\textwidth}
		\includegraphics[width=\textwidth]{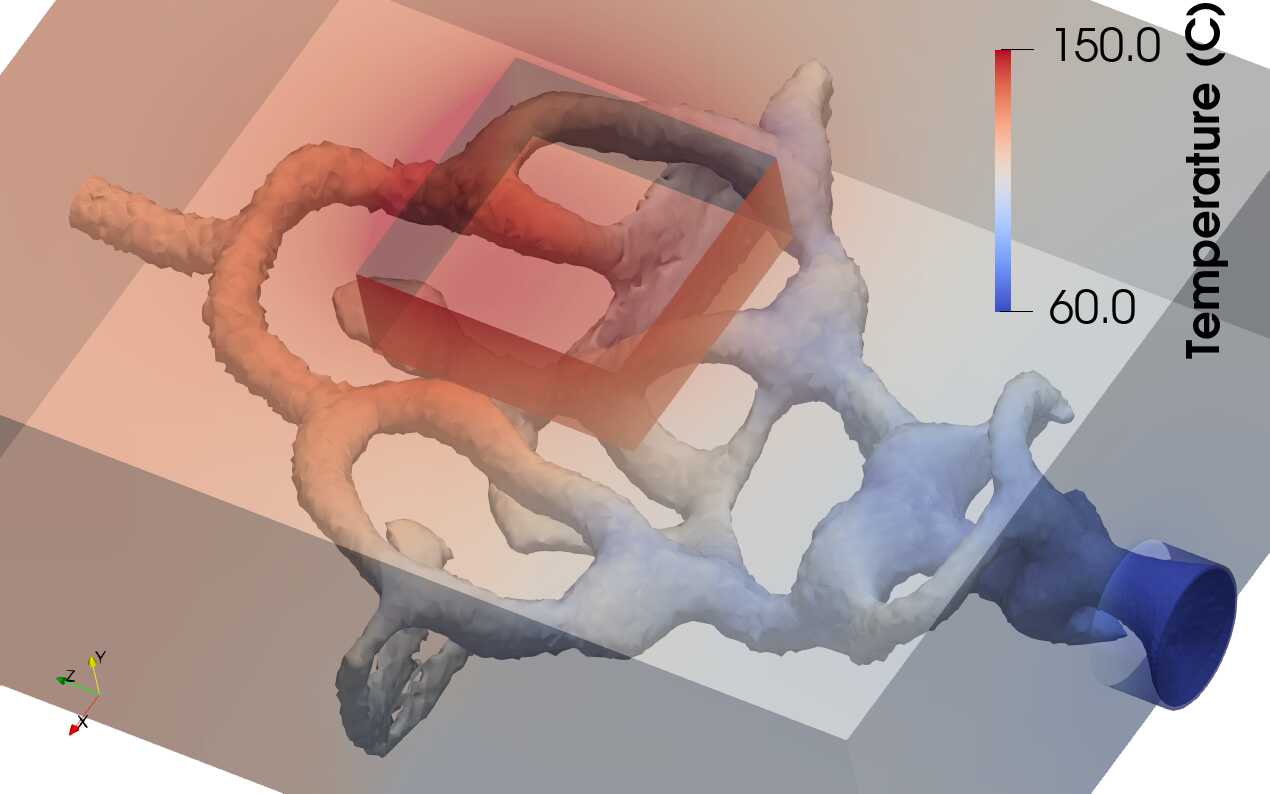}
		\caption{Step 199; $\beta=2$}
	\end{subfigure}\\
	\begin{subfigure}{0.49\textwidth}
		\includegraphics[width=\textwidth]{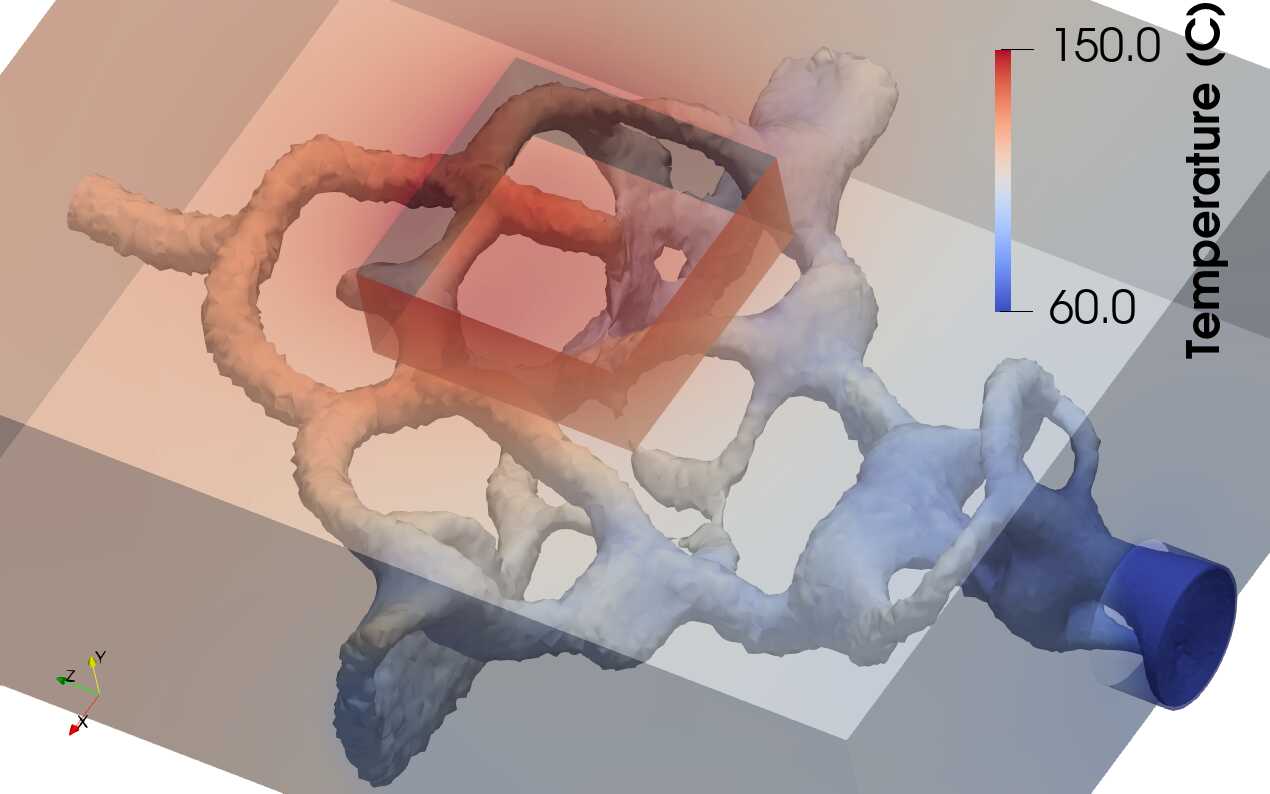}
		\caption{Step 299; $\beta=4$}
	\end{subfigure}\,
	\begin{subfigure}{0.49\textwidth}
		\includegraphics[width=\textwidth]{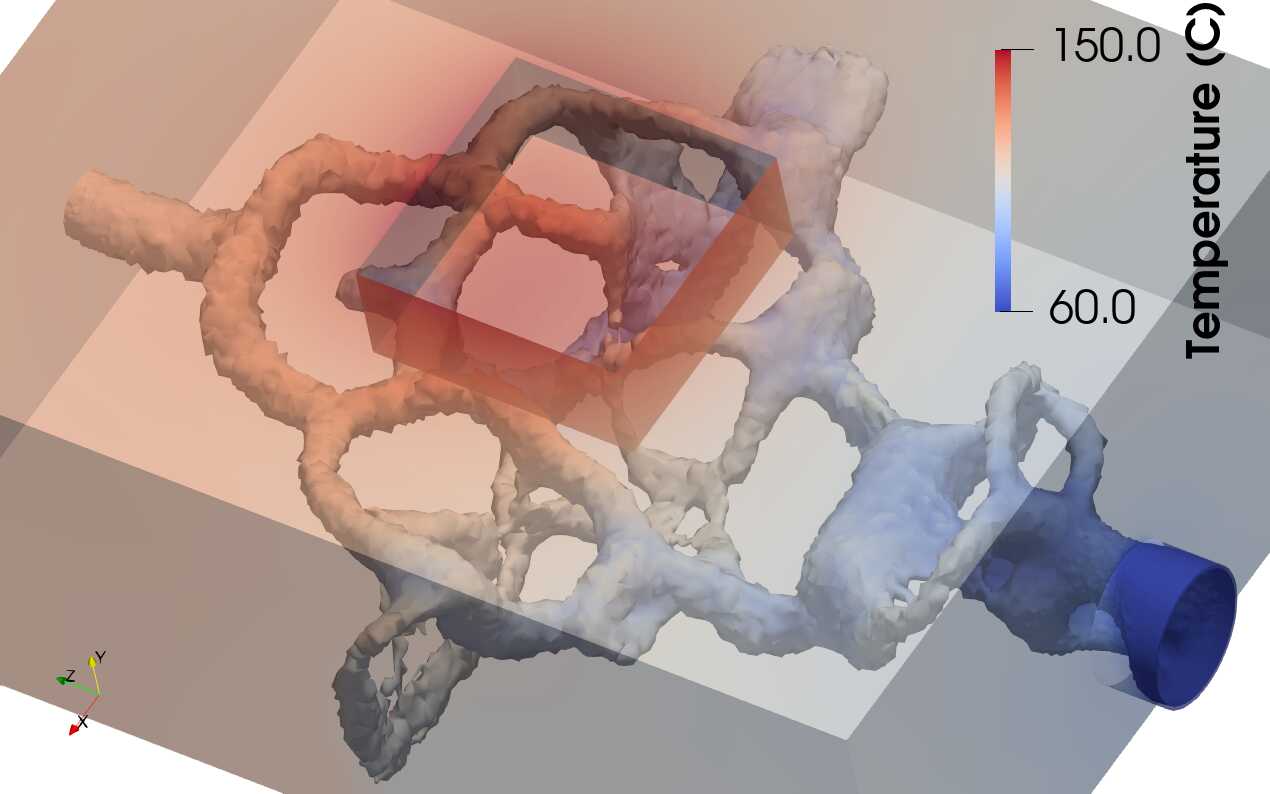}
		\caption{Step 342; $\beta=8$}
	\end{subfigure}
	\caption{Case A; Optimized designs at different $\beta$ values; Isocontours of flow speed at $10^{-3}$\,m/s, colored by the temperature.}
\label{fig:Viso-A-beta}	
\end{figure}

Now that the baseline cases are set and can serve as references, we can look into the results of the optimization cases (summarized in Table \ref{Tab:OptResults}), starting by the evolution of Case A through various optimization steps, presented in \cref{fig:costEvol-A,fig:Viso-A-beta1,fig:SF-A-beta,fig:Viso-A-beta}. From Fig.~\ref{fig:costEvol-A}, we can track the respective evolution of the cost function, $\mathcal{C}$, and its temperature component, $\bar{T}_\Gamma$, as well as the variations of the total fluid volume fraction, $1-\bar{\gamma}$ through the optimization, which is spanned by loops of  100 steps long. Each loop is characterized by a change in the value of $\beta$ via $\beta_i=2^i$ where $i$ stands for the optimization loop, starting from $i=0$ for the steps $0-99$ (first loop). For the first 10 steps or so, the sharp drop of the cost function is due to the decrease of the pressure component, as this term drops by roughly one order of magnitude. In fact, starting from a pure solid configuration, the pressure is rather large and drops as channels form to accommodate the fluid flow, as depicted in Fig.\ \ref{fig:Viso-A-beta1}. Meanwhile, the total fluid volume ratio rises to reach the maximum target value of 10\%. Once the channels are formed, the effort is rather concentrated on the minimization of the temperature component of the cost function. Indeed, as shown in Fig.\ \ref{fig:costEvol-A}, $\bar{T}_\Gamma$ rises during the first steps of the first loop as the diffuse, low-speed fluid flow through the solid gets replaced by flow through the initial channels. The average surface temperature drops slowly by the end of the loop as the channels are further optimized. 
Peaks in  $\bar{T}_\Gamma$ and drops in $1-\bar{\gamma}$ are noticeable at the beginning of the third and fourth loops. These sudden variations are caused by the exponential change in $\beta$. Indeed, as  $\beta$ follows an exponential growth through the optimization cycles, its impact becomes more tangible for the last cycles. 

Figure \ref{fig:SF-A-beta} shows the effect of the change in $\beta$, on the solid fraction field. One can note that the space between the gray isocontour of $\gamma=0.999$ and the yellow clip on $\gamma\leq0.5$ shrinks as $\beta$ rises. This effect is particularly observable in the vicinity of the domain outlet. The reduction of the extent of transitional regions with increasing $\beta$ leaves room for the generation of a more complex design. The flow speed isocontours, illustrated in Fig.~\ref{fig:Viso-A-beta}, indeed shrink and become more convoluted for higher $\beta$, although there is also an apparent roughness induced by the underlying mesh for the larger values. 

\begin{figure}
\centering
\begin{subfigure}[h]{0.31\textwidth}
  \includegraphics[width=\textwidth,trim={0.cm 0.cm 82.3cm 30.cm},clip]{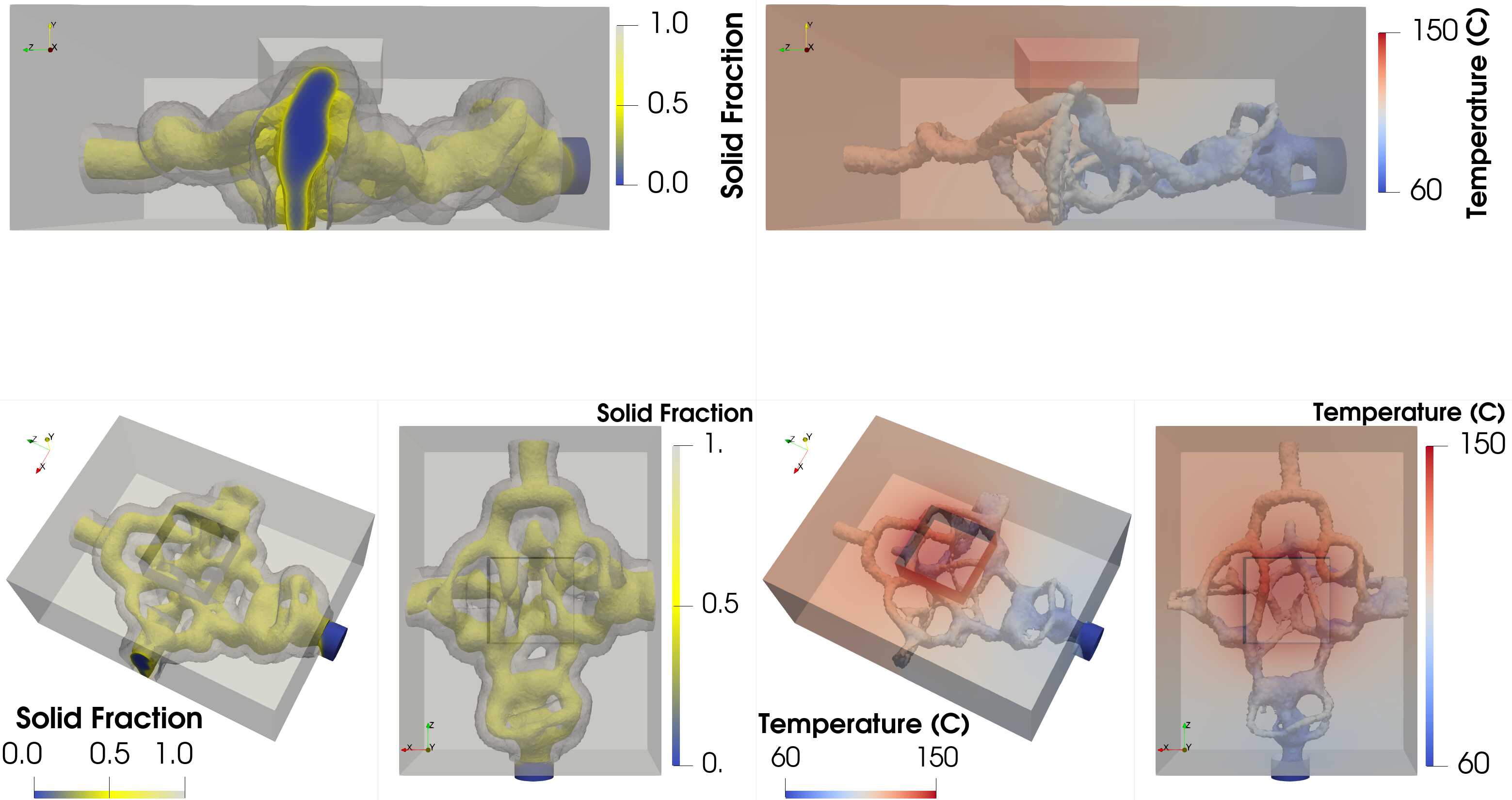}
  \caption*{Case A}
  \label{fig:SF-A}
\end{subfigure}
\\
\begin{subfigure}[h]{0.31\textwidth}
  \includegraphics[width=\textwidth,trim={0.cm 0.cm 82.3cm 30.cm},clip]{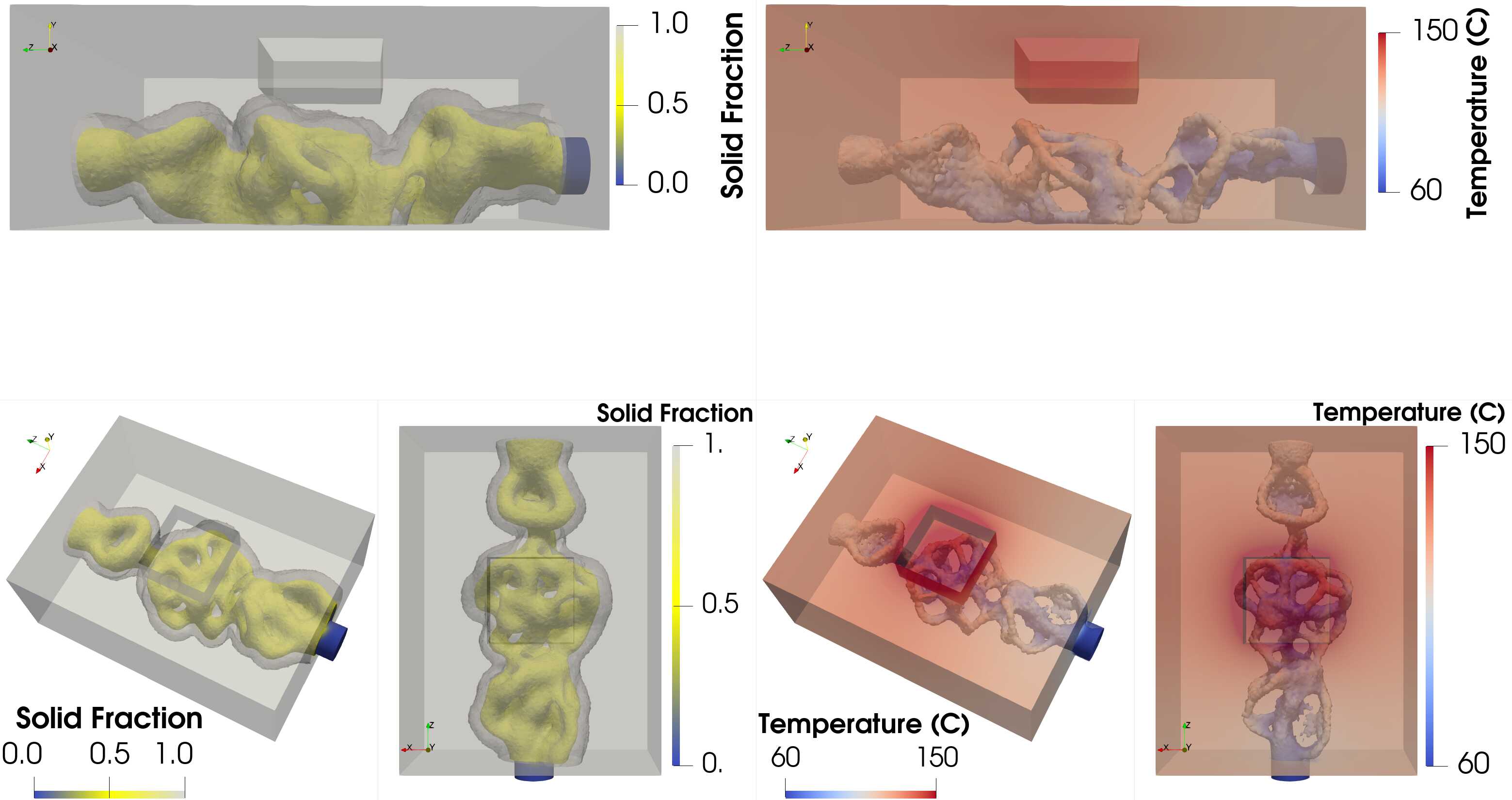}
  \caption*{Case B}
  \label{fig:SF-B}
\end{subfigure}
~
\begin{subfigure}[h]{0.31\textwidth}
  \includegraphics[width=\textwidth,trim={0.cm 0.cm 82.3cm 30.cm},clip]{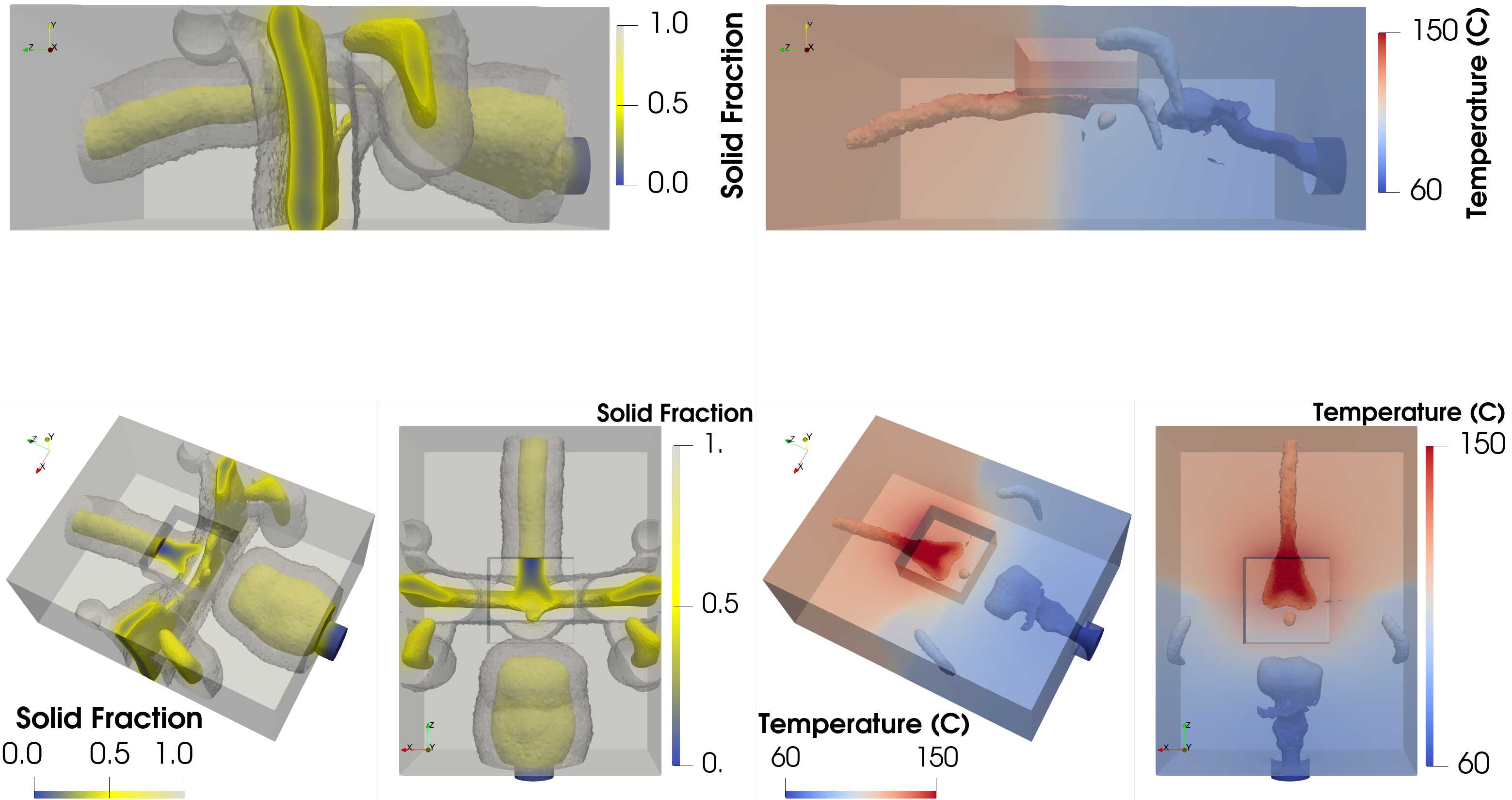}
  \caption*{Case C}  
  \label{fig:SF-C}
\end{subfigure}
~
\begin{subfigure}[h]{0.31\textwidth}
  \includegraphics[width=\textwidth,trim={0.cm 0.cm 82.3cm 30.cm},clip]{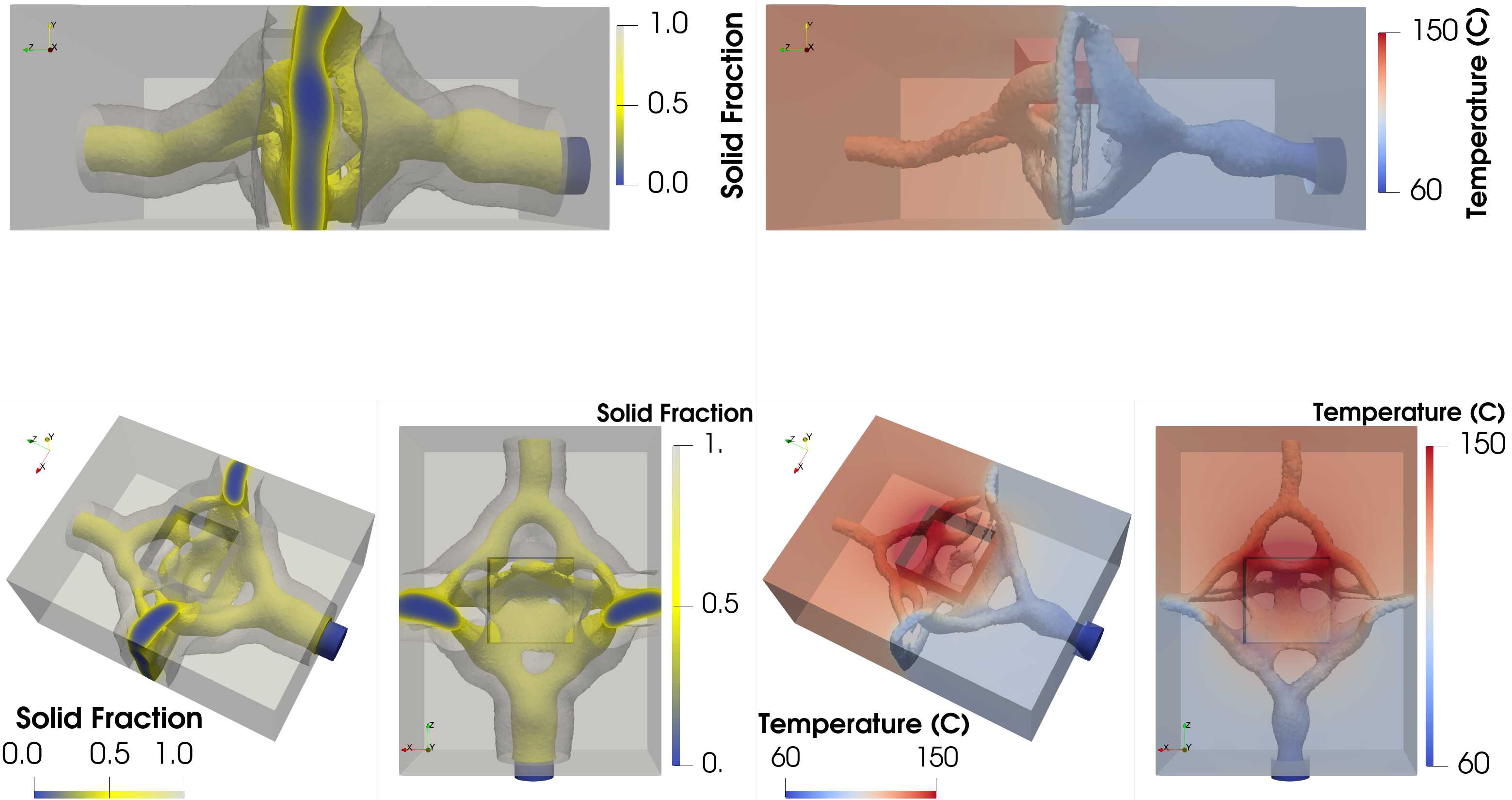}
  \caption*{Case D}
  \label{fig:SF-D}
\end{subfigure}
\\
\begin{subfigure}[h]{0.31\textwidth}
  \includegraphics[width=\textwidth,trim={0.cm 0.cm 82.3cm 30.cm},clip]{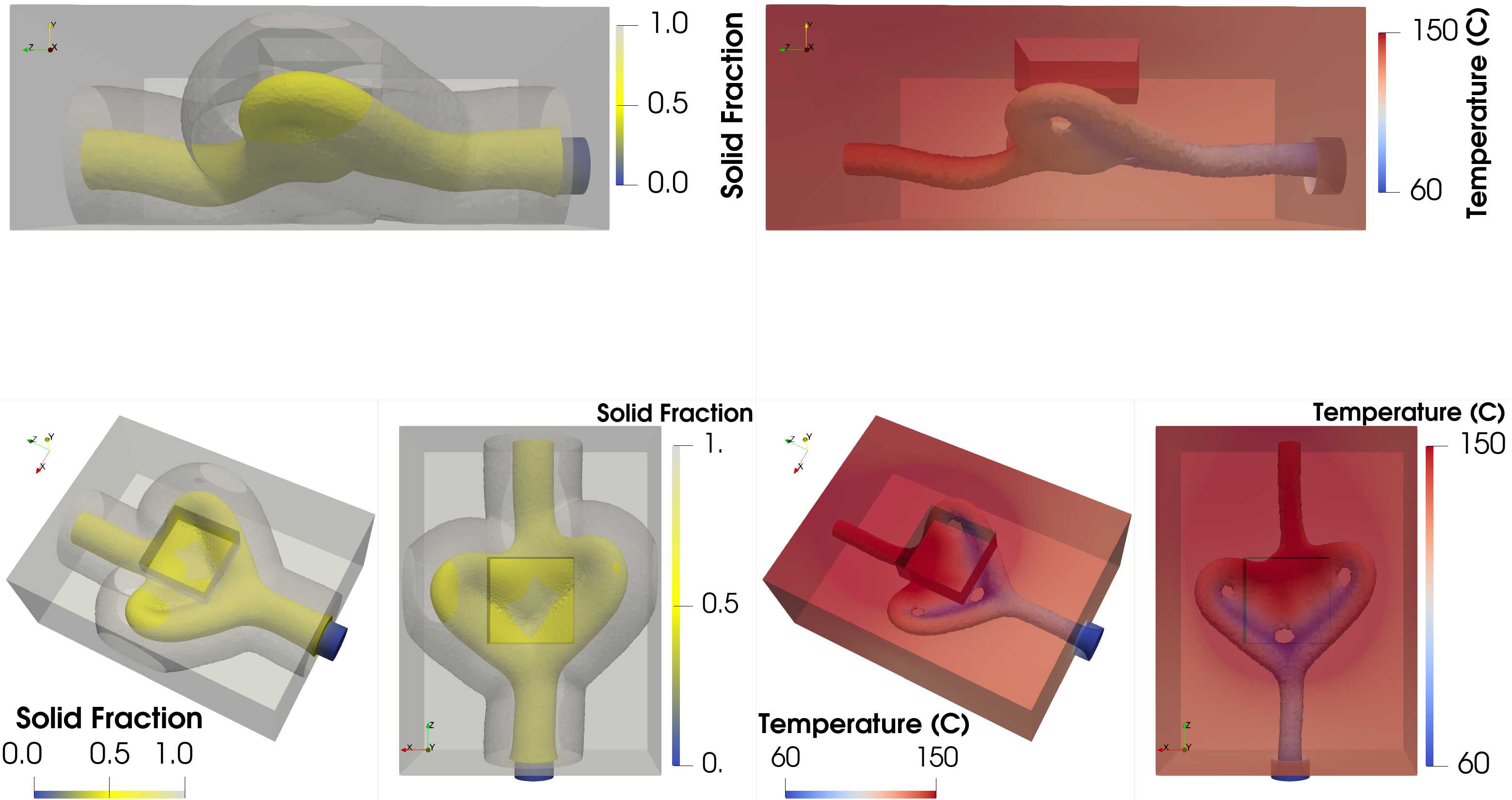}
  \caption*{Case E}  
  \label{fig:SF-E}
\end{subfigure}
~
\begin{subfigure}[h]{0.31\textwidth}
  \includegraphics[width=\textwidth,trim={0.cm 0.cm 82.3cm 30.cm},clip]{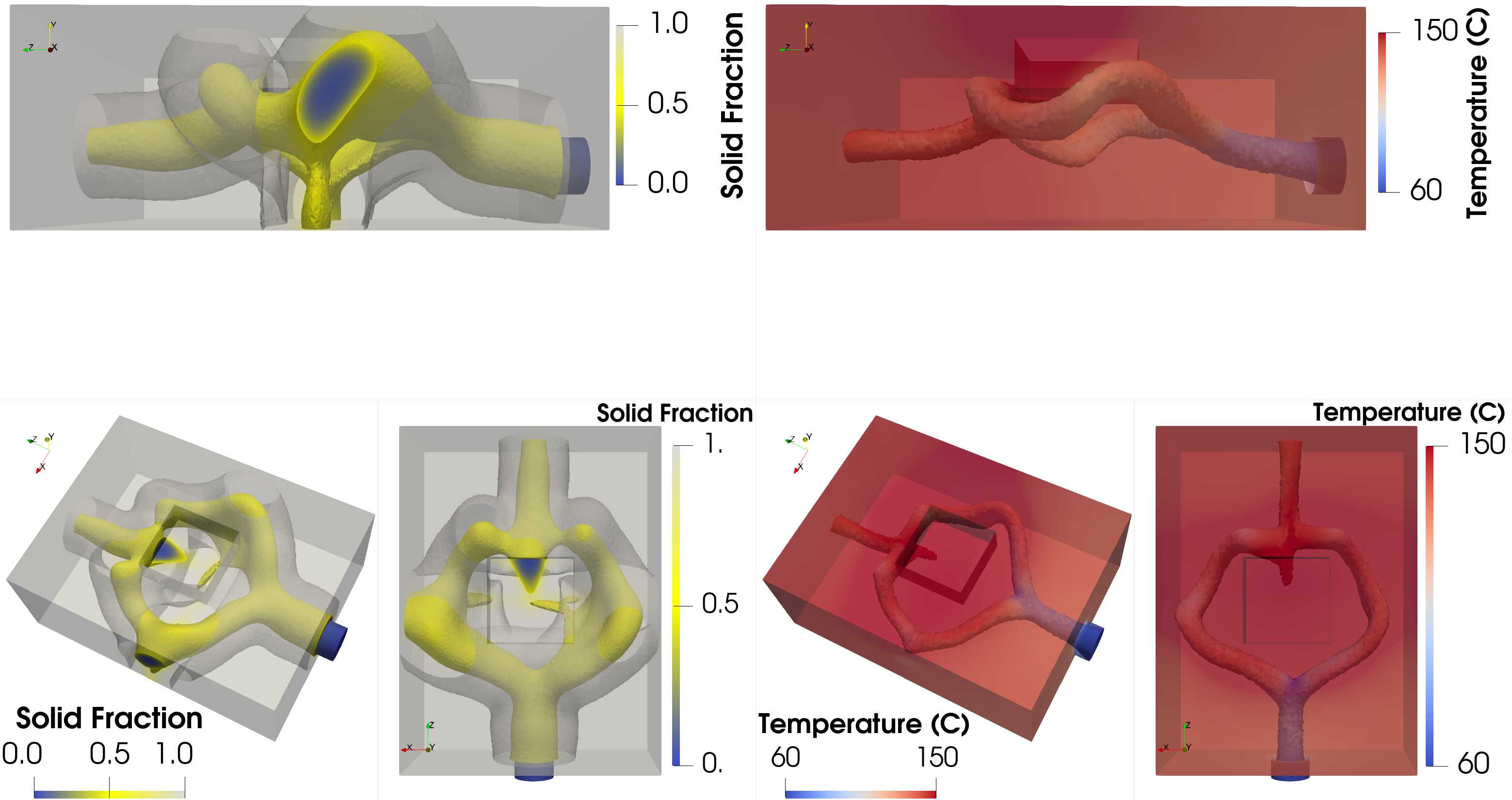}
  \caption*{Case F}
  \label{fig:SF-F}
\end{subfigure}
~
\begin{subfigure}[h]{0.31\textwidth}
  \includegraphics[width=\textwidth,trim={0.cm 0.cm 82.3cm 30.cm},clip]{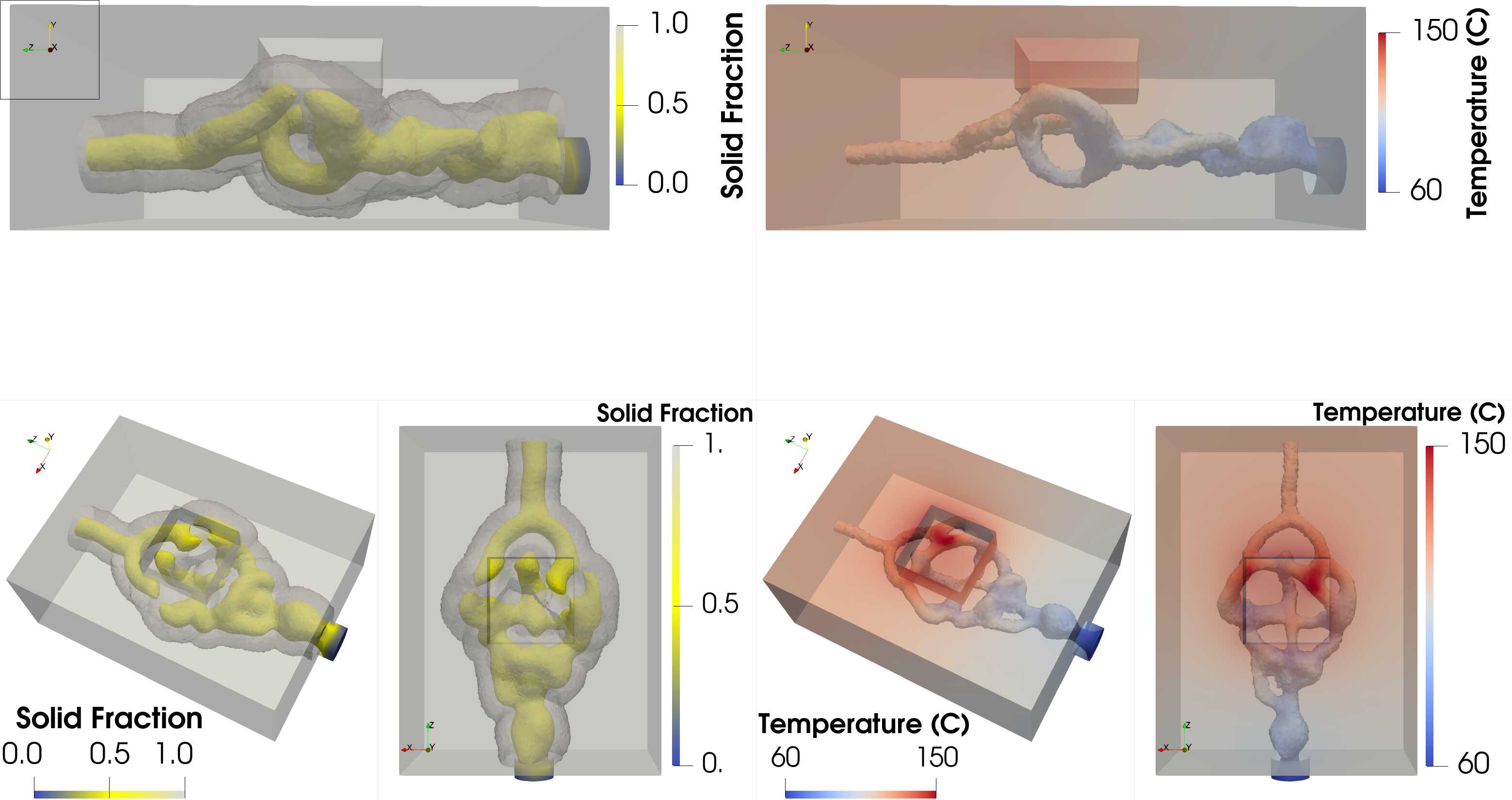}
  \caption*{Case G}  
  \label{fig:SF-G}
\end{subfigure}
\\
\begin{subfigure}[h]{0.31\textwidth}
  \includegraphics[width=\textwidth,trim={0.cm 0.cm 82.3cm 30.cm},clip]{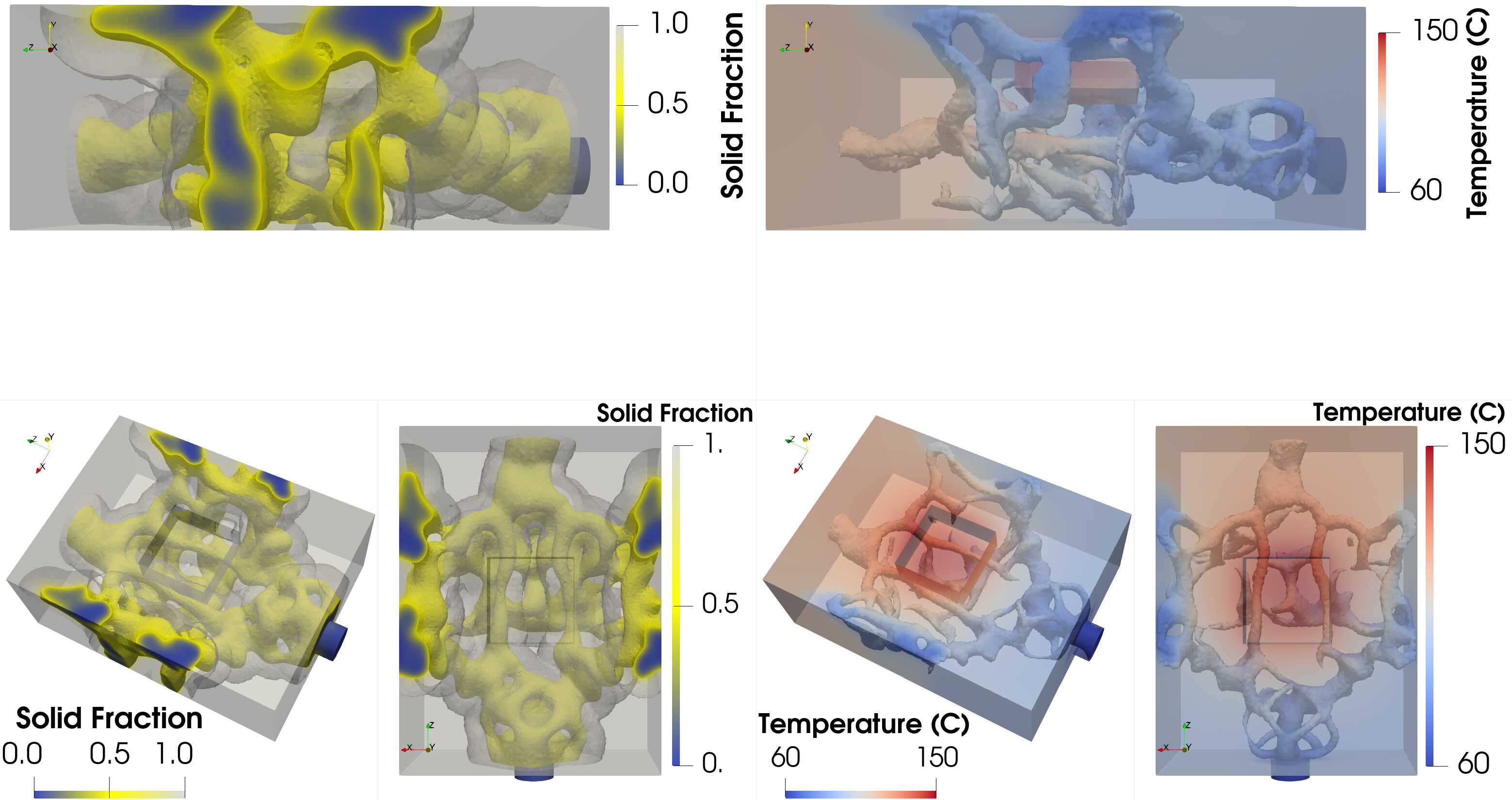}
  \caption*{Case H}
  \label{fig:SF-H}
\end{subfigure}
~
\begin{subfigure}[h]{0.31\textwidth}
  \includegraphics[width=\textwidth,trim={0.cm 0.cm 82.3cm 30.cm},clip]{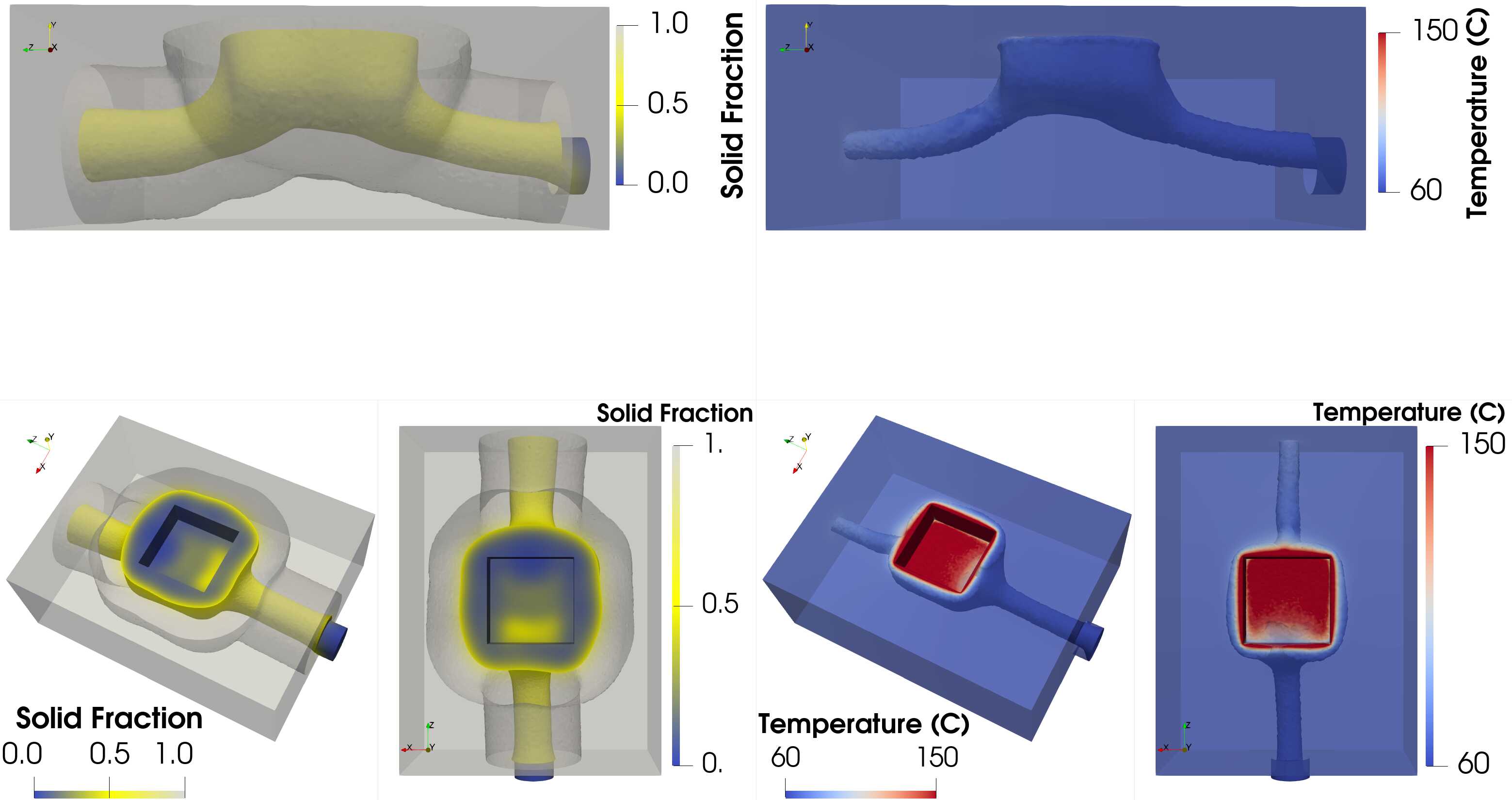}
  \caption*{Case I}
  \label{fig:SF-I}
\end{subfigure}
  \caption{Optimization Cases A--I: on the small cavity and with $\mathrm{Re}=100$; Solid fraction on the outer domain walls, clip on $\gamma\leq0.5$ and isocontour of $\gamma=0.999$.}
\label{fig:SF-A2I}
\end{figure}

\begin{figure}[]
\centering
\begin{subfigure}[h]{0.31\textwidth}
  \includegraphics[width=\textwidth,trim={84.cm 0.cm 0.cm 29.cm},clip]{{figs/mesh03_sol_costB_tgF0.1_b3_rad0.4_maxBeta8_PPen3e-4_dim_Tin60_0342-fs8}}
  \caption*{Case A}
\label{fig:Viso-A}
\end{subfigure}
\\
\begin{subfigure}[h]{0.31\textwidth}
  \includegraphics[width=\textwidth,trim={84.cm 0.cm 0.cm 29.cm},clip]{{figs/mesh03_sol_costB_tgF0.1_b3_rad0.4_maxBeta8_PPen3e-3_dim_Tin60_0399-fs8}}
  \caption*{Case B}
\label{fig:Viso-B}
\end{subfigure}
~
\begin{subfigure}[h]{0.31\textwidth}
  \includegraphics[width=\textwidth,trim={84.cm 0.cm 0.cm 29.cm},clip]{{figs/mesh03_sol_costB_tgF0.1_b3_rad0.4_maxBeta8_PPen3e-5_dim_Tin60_STOPPED_0112-fs8}}
  \caption*{Case C}
\label{fig:Viso-C}
\end{subfigure}
~
\begin{subfigure}[h]{0.31\textwidth}
  \includegraphics[width=\textwidth,trim={84.cm 0.cm 0.cm 29.cm},clip]{{figs/mesh03_sol_costB_tgF0.1_b3_rad0.6_maxBeta8_PPen3e-4_dim_Tin60_0339-fs8}}
  \caption*{Case D}
\label{fig:Viso-D}
\end{subfigure}
\\
\begin{subfigure}[h]{0.31\textwidth}
  \includegraphics[width=\textwidth,trim={84.cm 0.cm 0.cm 29.cm},clip]{{figs/mesh03_sol_costB_tgF0.1_b3_rad0.8_maxBeta16_PPen3e-3_dim_Tin60_0299-fs8}}
  \caption*{Case E}
\label{fig:Viso-E}
\end{subfigure}
~
\begin{subfigure}[h]{0.31\textwidth}
  \includegraphics[width=\textwidth,trim={84.cm 0.cm 0.cm 29.cm},clip]{{figs/mesh03_sol_costB_tgF0.1_b3_rad0.8_maxBeta16_PPen3e-4_dim_Tin60_0299-fs8}}
  \caption*{Case F}
\label{fig:Viso-F}
\end{subfigure}
~
\begin{subfigure}[h]{0.31\textwidth}
  \includegraphics[width=\textwidth,trim={84.cm 0.cm 0.cm 29.cm},clip]{{figs/mesh03_sol_costB_tgF0.05_b3_rad0.4_maxBeta8_PPen3e-4_dim_Tin60_0299-fs8}}
  \caption*{Case G}
\label{fig:Viso-G}
\end{subfigure}
\\
\begin{subfigure}[h]{0.31\textwidth}
  \includegraphics[width=\textwidth,trim={84.cm 0.cm 0.cm 29.cm},clip]{{figs/mesh03_sol_costB_tgF0.2_b3_rad0.4_maxBeta8_PPen3e-4_dim_Tin60_0299-fs8}}
  \caption*{Case H}
\label{fig:Viso-H}
\end{subfigure}
~
\begin{subfigure}[h]{0.31\textwidth}
  \includegraphics[width=\textwidth,trim={84.cm 0.cm 0.cm 29.cm},clip]{{figs/mesh03_sol_tgF0.1_b3_rad0.8_maxBeta16_PPen3e-4_dim_Tin60_0299-fs8}}
  \caption*{Case I}
\label{fig:Viso-I}
\end{subfigure}
  \caption{Optimization Cases A--I: on the small cavity and with $\mathrm{Re}=100$; Temperature on the outer domain walls and isocontour of flow speed at $10^{-3}$\,m/s, colored by the temperature. }
\label{fig:Viso-A2I}
\end{figure}

\begin{figure}[]
\centering
\begin{subfigure}[h]{0.48\textwidth}
  \includegraphics[width=\textwidth,trim={56.cm 40.cm 0.cm 0.cm},clip]{{figs/mesh03_sol_costB_tgF0.1_b3_rad0.4_maxBeta8_PPen3e-4_dim_Tin60_0342-fs8}}
  \caption*{Case A}
\label{fig:Viso2-A}
\end{subfigure}
\\
\begin{subfigure}[h]{0.48\textwidth}
  \includegraphics[width=\textwidth,trim={56.cm 40.cm 0cm 0.cm},clip]{{figs/mesh03_sol_costB_tgF0.1_b3_rad0.4_maxBeta8_PPen3e-3_dim_Tin60_0399-fs8}}
  \caption*{Case B}
\label{fig:Viso2-B}
\end{subfigure}
~
\begin{subfigure}[h]{0.48\textwidth}
  \includegraphics[width=\textwidth,trim={56.cm 40.cm 0.cm 0.cm},clip]{{figs/mesh03_sol_costB_tgF0.1_b3_rad0.4_maxBeta8_PPen3e-5_dim_Tin60_STOPPED_0112-fs8}}
  \caption*{Case C}
\label{fig:Viso2-C}
\end{subfigure}
\\
\begin{subfigure}[h]{0.48\textwidth}
  \includegraphics[width=\textwidth,trim={56.cm 40.cm 0.cm 0.cm},clip]{{figs/mesh03_sol_costB_tgF0.1_b3_rad0.6_maxBeta8_PPen3e-4_dim_Tin60_0339-fs8}}
  \caption*{Case D}
\label{fig:Viso2-D}
\end{subfigure}
~
\begin{subfigure}[h]{0.48\textwidth}
  \includegraphics[width=\textwidth,trim={56.cm 40.cm 0.cm 0.cm},clip]{{figs/mesh03_sol_costB_tgF0.1_b3_rad0.8_maxBeta16_PPen3e-3_dim_Tin60_0299-fs8}}
  \caption*{Case E}
\label{fig:Viso2-E}
\end{subfigure}
\\
\begin{subfigure}[h]{0.48\textwidth}
  \includegraphics[width=\textwidth,trim={56.cm 40.cm 0.cm 0.cm},clip]{{figs/mesh03_sol_costB_tgF0.1_b3_rad0.8_maxBeta16_PPen3e-4_dim_Tin60_0299-fs8}}
  \caption*{Case F}
\label{fig:Viso2-F}
\end{subfigure}
~
\begin{subfigure}[h]{0.48\textwidth}
  \includegraphics[width=\textwidth,trim={56.cm 40.cm 0.cm 0.cm},clip]{{figs/mesh03_sol_costB_tgF0.05_b3_rad0.4_maxBeta8_PPen3e-4_dim_Tin60_0299-fs8}}
  \caption*{Case G}
\label{fig:Viso2-G}
\end{subfigure}
\\
\begin{subfigure}[h]{0.48\textwidth}
  \includegraphics[width=\textwidth,trim={56.cm 40.cm 0.cm 0.cm},clip]{{figs/mesh03_sol_costB_tgF0.2_b3_rad0.4_maxBeta8_PPen3e-4_dim_Tin60_0299-fs8}}
  \caption*{Case H}
\label{fig:Viso2-H}
\end{subfigure}
~
\begin{subfigure}[h]{0.48\textwidth}
  \includegraphics[width=\textwidth,trim={56.cm 40.cm 0.cm 0.cm},clip]{{figs/mesh03_sol_tgF0.1_b3_rad0.8_maxBeta16_PPen3e-4_dim_Tin60_0299-fs8}}
  \caption*{Case I}
\label{fig:Viso2-I}
\end{subfigure}
  \caption{Optimization Cases A--I: on the small cavity and with $\mathrm{Re}=100$; Temperature on the outer domain walls and isocontour of flow speed at $10^{-3}$\,m/s, colored by the temperature. }
\label{fig:Viso2-A2I}
\end{figure}

Next, we look into the effect of the variations of the parameters of Table \ref{Tab:OptCases}, starting by $\zeta$, which is changed over several orders of magnitude in Cases A to C or in Cases E and F. \Cref{fig:SF-A2I,fig:Viso-A2I,fig:Viso2-A2I} show comparative views of the optimal topologies obtained for these variations, where it can be seen that overly large $\zeta$ values tend to reduce the number and the lateral spread of the ramifications and increase the channel diameter. These effects are consistent with a reduction of pressure losses in the system. On the other hand,  an insufficiently large $\zeta$ produces unconnected channels and leads the optimizer to use the low conductivity of the fluid to generate a thermal barrier, which separates the domain into two regions with notable temperature differences, as observed especially for Case C, but also for Case D, in \cref{fig:SF-A2I,fig:Viso-A2I,fig:Viso2-A2I}.

The effects of the filter radius, $\varrho$, are exhibited via Cases A, D and F or Cases B and E in \cref{fig:SF-A2I,fig:Viso-A2I,fig:Viso2-A2I}. These results show that increasing $\varrho$, similarly to increasing $\zeta$, tends to simplify the optimal topologies by reducing the number of branches and thickening the channels. Nevertheless, in contrast to increasing $\zeta$, larger $\varrho$ values do not necessarily reduce the lateral span of the channels. From Fig.~\ref{fig:SF-A2I}, we can also note that for higher $\varrho$, the volume of the spatial gap between $\gamma=0.5$ and $\gamma=0.999$ isocontours is enlarged. We further notice  from Table \ref{Tab:OptResults} that an increase of the filter radius generally results in a reduction of the cooling efficiency. For instance, the factor of improvement, over the baseline design (\ie $\frac{\bar{T}_\Gamma^0-\bar{T}_\Gamma}{\bar{T}_\Gamma^0}$), drops from $48\%$ in Case A to $31\%$ in Case F.

Changing the target fluid volume ratio, $1-\bar{\gamma}_0$, for example from Case A to Cases G or H, modifies the available volume of the fluid which can be exploited by the optimizer to achieve optimal channel topologies. From \cref{fig:SF-A2I,fig:Viso-A2I,fig:Viso2-A2I}, it can be observed that indeed, as the available fluid volume is increased, more convoluted shapes are made possible. 

So far, we can conclude that achieving the right combination of $1-\bar{\gamma}_0$, $\zeta$ and $\varrho$ is crucial in order to obtain a well-connected network of channels with a given complexity, ramification and average diameter, which altogether define the efficiency, as well as the feasibility of the design in terms of manufacturability constraints. 

Another impactful variable of the problem is the cost function in terms of the measure of temperature to be minimized. In this regard, we have accounted for two different objectives: minimizing the average temperature of the insert cavity, designated by $\bar{T}_\Gamma$, or the average temperature of the domain, noted by $\bar{T}_\Omega$. These two objectives result in dramatically different design paradigms as illustrated by Cases F and I in \cref{fig:SF-A2I,fig:Viso-A2I,fig:Viso2-A2I}. For $\bar{T}_\Gamma$ (e.g. Case F), the channels tend to span the body of the domain by keeping a certain distance to the cavity such that the heat can penetrate into the domain first and get absorbed by the channels, thus minimizing the temperature of the part ($31\%$ improved over the baseline design), be it at the expense of a higher domain-wise average temperature ($\bar{T}_\Omega = 142 \,^{\circ}\mathrm{C}$),  compared to that for Case I ($\bar{T}_\Omega = 66\,^{\circ}\mathrm{C}$). On the other hand, the use of $\bar{T}_\Omega$ (e.g. Case I) often results in a topology that envelops the cavity in order to constitute a thermal barrier, preventing the penetration of the heat into the domain, although causing a higher temperature on the surfaces of the cavity ($\bar{T}_\Gamma=193\,^{\circ}\mathrm{C}$) compared to Case F ($\bar{T}_\Gamma=183\,^{\circ}\mathrm{C}$).

\begin{figure}[h!]
	\centering
	\begin{subfigure}[h]{0.32\textwidth}
		\centering
		\includegraphics[width=\textwidth] {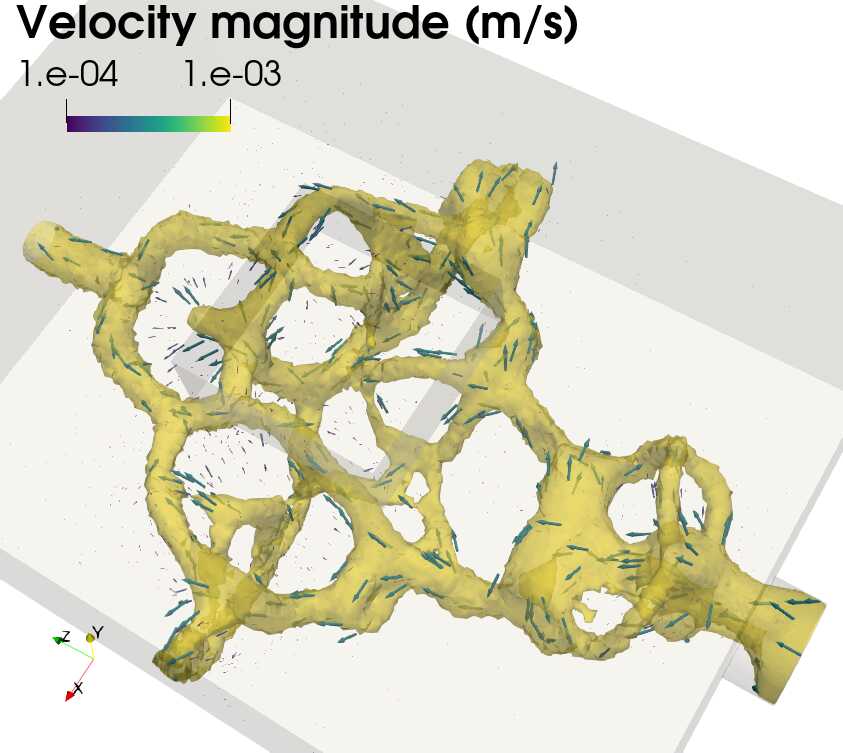}
		\caption*{Case A}
		\label{fig:glyphs01}
	\end{subfigure}\,
	\begin{subfigure}[h]{0.32\textwidth}
		\centering
		\includegraphics[width=\textwidth] {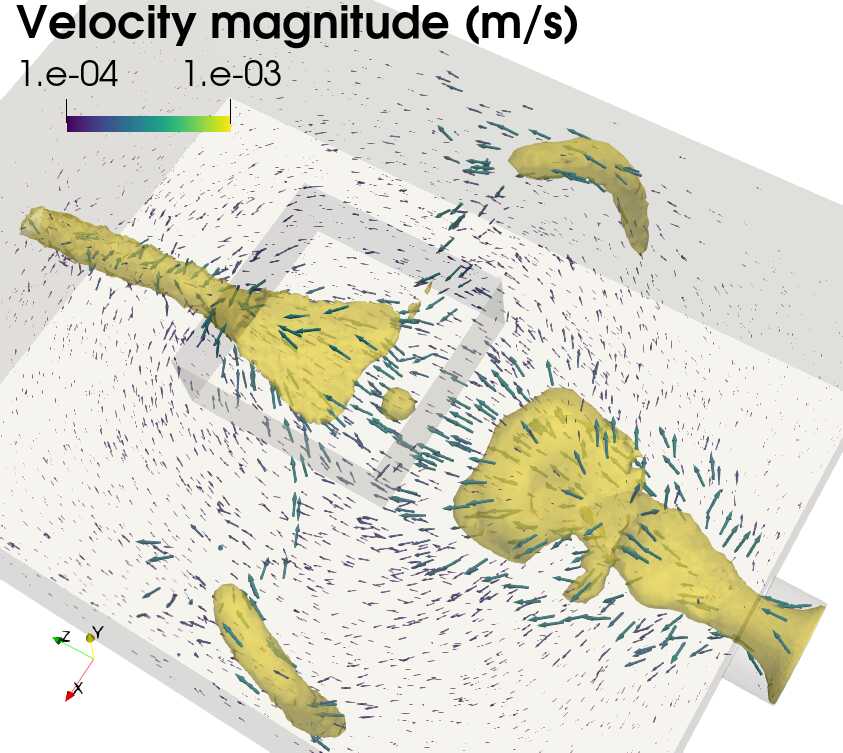}
		\caption*{Case C}
		\label{fig:glyphs02}
	\end{subfigure}\,
	\begin{subfigure}[h]{0.32\textwidth}
		\centering
		\includegraphics[width=\textwidth] {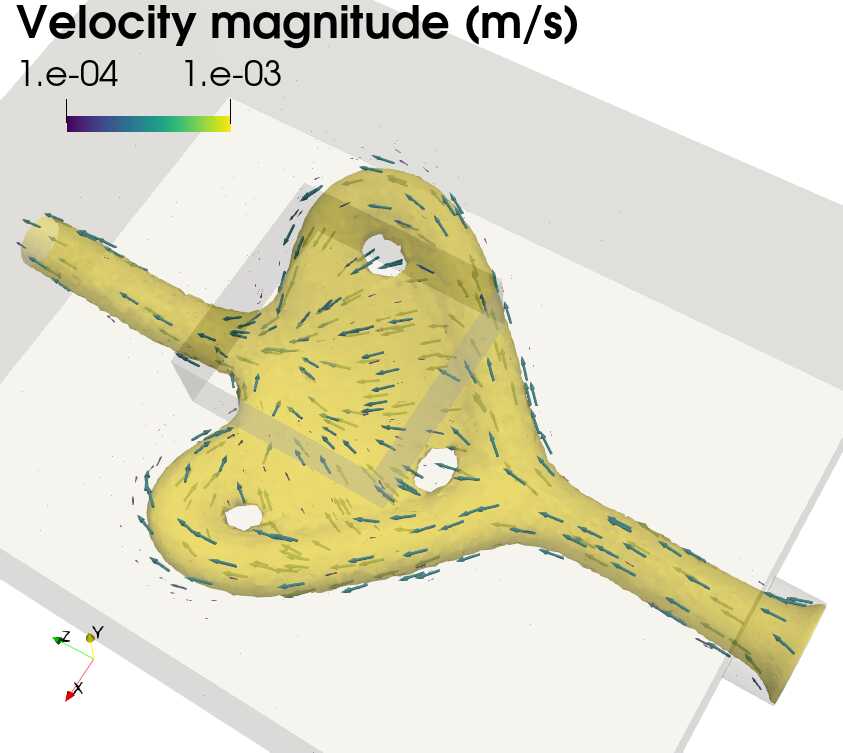}
		\caption*{Case E}
		\label{fig:glyphs03}
	\end{subfigure}
	\caption{Isocontours of flow speed at $10^{-3}$\,m/s and velocity vectors outside of the channels, colored by the velocity magnitude.}
	\label{fig:glyphs}
\end{figure}

Comparing the optimal designs obtained so far and reviewing their quantitative performances in Table \ref{Tab:OptResults}, we can identify some outstanding candidates. For example, according to the criterion on the improvement over the baseline design, of the cooling efficiency for the cavity surfaces (i.e. $\frac{(\bar{T}_\Gamma^0-\bar{T}_\Gamma)}{\bar{T}_\Gamma^0}$), we can select Cases A, C, G and H as best performers. The best performance is in fact achieved by Case C ($51\%$). However, a closer look at the topology obtained for this case (Fig.\ \ref{fig:SF-A2I}) shows unconnected channels due to an insufficient pressure penalization. These unconnected channels result in a diffuse flow through the solid regions, as shown in Fig.~\ref{fig:glyphs}, which generates a spurious cooling effect. Hence, Case C is disqualified. Figure \ref{fig:glyphs} also shows that whenever the channels are entirely, or mostly connected, the flow leakage outside of the channels is minimal. In fact, integrating the norm of the velocity magnitude outside the channels and dividing it by its counterpart in the entire domain indicates that, apart from Case C where this ratio amounts to $55\%$, all cases feature a ratio of at most $20\%$. Hence by discarding Case C, we also note that Case A produces a more uniform cooling compared to Case G, since we have $T_\mathrm{max}=163\,^{\circ}\mathrm{C}$ for the latter versus $T_\mathrm{max}=170\,^{\circ}\mathrm{C}$ for the former, for an almost equal average surface temperature on the insert cavity. As for Case H, its superior cooling is jeopardized by its much higher design complexity, compared to Case A for example.

\begin{figure}[h!]
\centering
\begin{subfigure}[h]{0.31\textwidth}
  \includegraphics[width=\textwidth,trim={0.cm 0.cm 82.3cm 30.cm},clip]{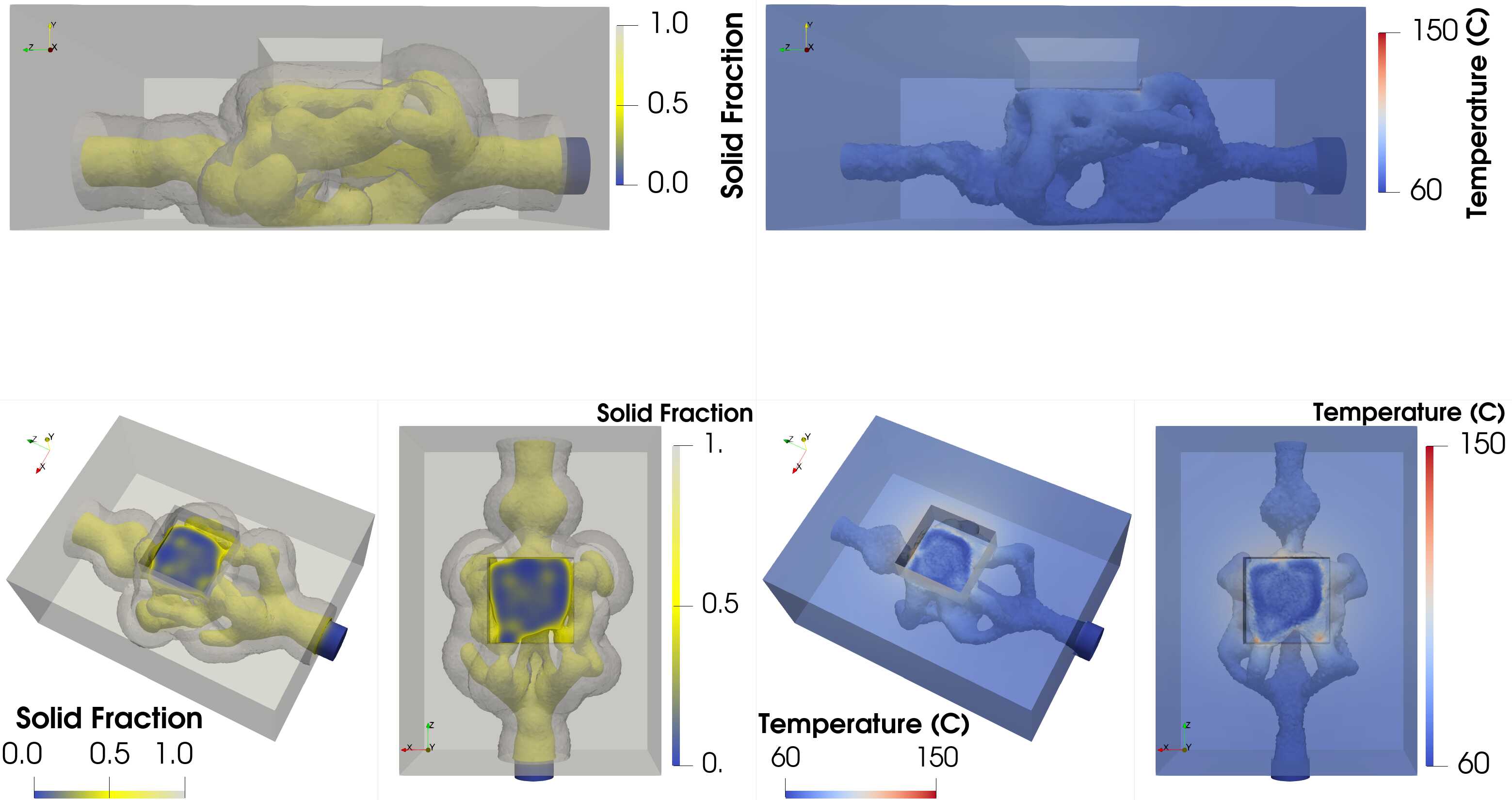}
  \caption*{Case J}
  \label{fig:SF-J}
\end{subfigure}
~
\begin{subfigure}[h]{0.31\textwidth}
  \includegraphics[width=\textwidth,trim={0.cm 0.cm 82.3cm 30.cm},clip]{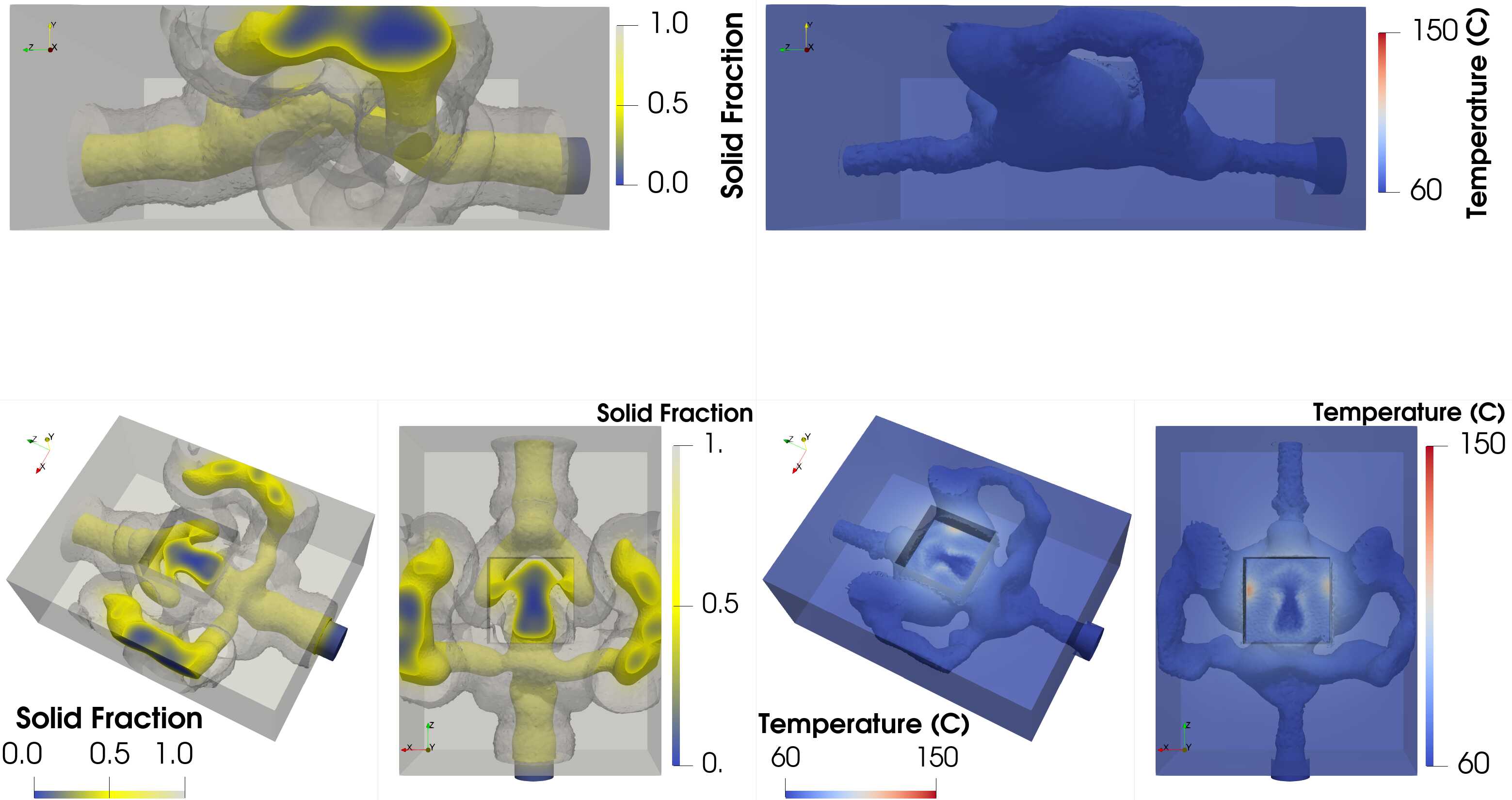}
  \caption*{Case K}
  \label{fig:SF-K}
\end{subfigure}
\\
\begin{subfigure}[h]{0.31\textwidth}
  \includegraphics[width=\textwidth,trim={0.cm 0.cm 82.3cm 30.cm},clip]{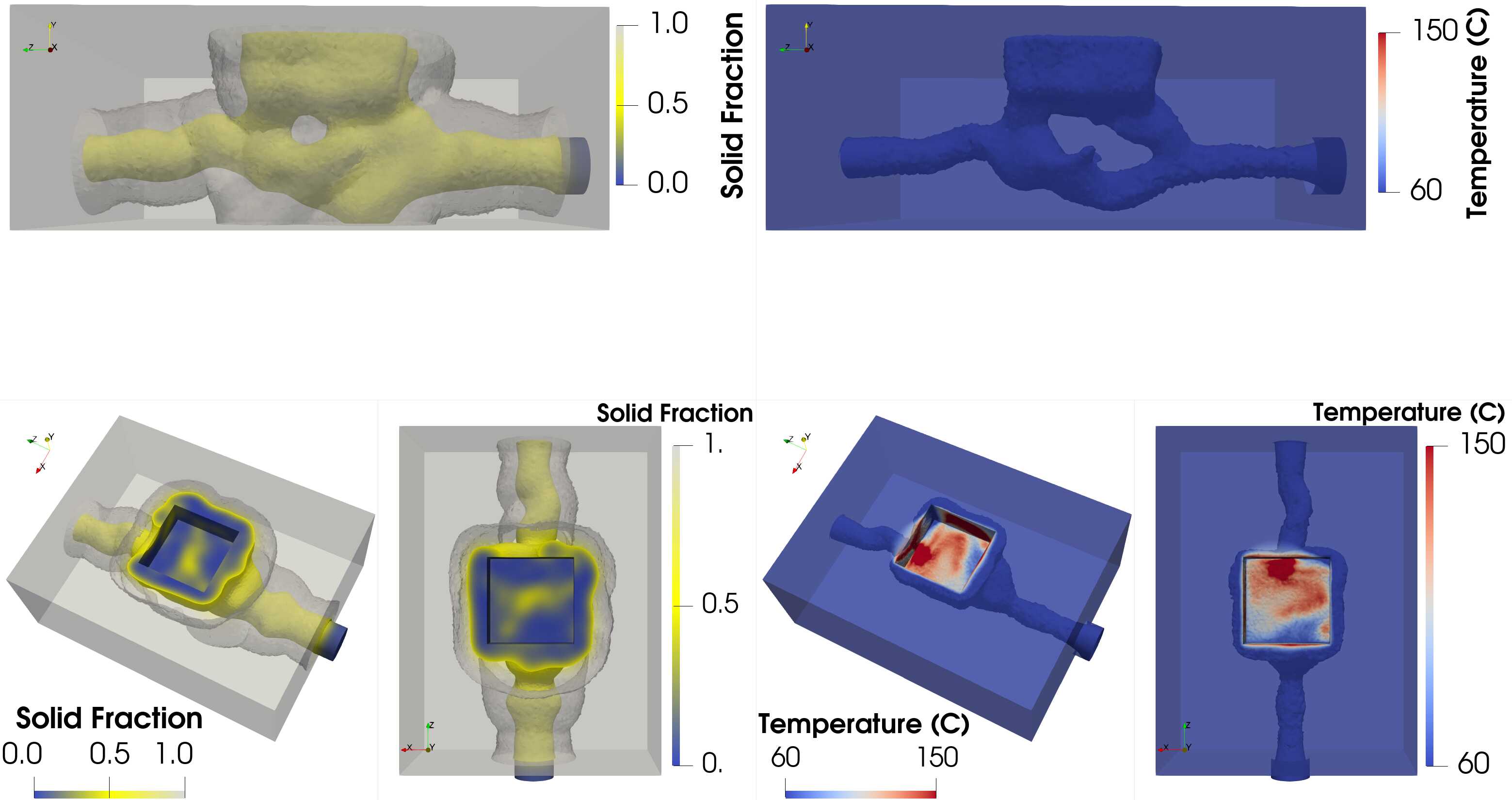}
  \caption*{Case L}
  \label{fig:SF-L}
\end{subfigure}
  \caption{Optimization Cases J--L: on the small cavity and with $\mathrm{Re}=\text{1,000}$; Solid fraction on the outer domain walls, clip on $\gamma\leq0.5$ and isocontour of $\gamma=0.999$.}
\label{fig:SF-J2L}
\end{figure}

\begin{figure}[h!]
\centering
\begin{subfigure}[h]{0.31\textwidth}
  \includegraphics[width=\textwidth,trim={84.cm 0.cm 0.cm 29.cm},clip]{{figs/mesh03_sol_costB_tgF0.1_b3_rad0.4_maxBeta16_PPen3e-4_dim_Tin60_re1000_0278-fs8}}
  \caption*{Case J}
\label{fig:Viso-J}
\end{subfigure}
~
\begin{subfigure}[h]{0.31\textwidth}
  \includegraphics[width=\textwidth,trim={84.cm 0.cm 0.cm 29.cm},clip]{{figs/mesh03_sol_costB_tgF0.1_b3_rad0.4_maxBeta16_PPen3e-5_dim_Tin60_re1000_0178-fs8}}
  \caption*{Case K}
\label{fig:Viso-K}
\end{subfigure}
\\
\begin{subfigure}[h]{0.31\textwidth}
  \includegraphics[width=\textwidth,trim={84.cm 0.cm 0.cm 29.cm},clip]{{figs/mesh03_sol_tgF0.1_b3_rad0.4_maxBeta16_PPen6e-5_dim_Tin60_re1000_0110-fs8}}
  \caption*{Case L}
\label{fig:Viso-L}
\end{subfigure}
  \caption{Optimization Cases J--L: on the small cavity and with $\mathrm{Re}=\text{1,000}$; Temperature on the outer domain walls and isocontour of flow speed at $10^{-3}$\,m/s, colored by the temperature. }
\label{fig:Viso-J2L}
\end{figure}

\begin{figure}[h!]
\centering
\begin{subfigure}[h]{0.48\textwidth}
  \includegraphics[width=\textwidth,trim={56.cm 40.cm 0cm 0.cm},clip]{{figs/mesh03_sol_costB_tgF0.1_b3_rad0.4_maxBeta16_PPen3e-4_dim_Tin60_re1000_0278-fs8}}
\caption*{Case J}
\label{fig:Viso2-J}
\end{subfigure}
~
\begin{subfigure}[h]{0.48\textwidth}
  \includegraphics[width=\textwidth,trim={56.cm 40.cm 0.cm 0.cm},clip]{{figs/mesh03_sol_costB_tgF0.1_b3_rad0.4_maxBeta16_PPen3e-5_dim_Tin60_re1000_0178-fs8}}
\caption*{Case K}
\label{fig:Viso2-K}
\end{subfigure}
\\
\begin{subfigure}[h]{0.58\textwidth}
  \includegraphics[width=\textwidth,trim={56.cm 40.cm 0.cm 0.cm},clip]{{figs/mesh03_sol_tgF0.1_b3_rad0.4_maxBeta16_PPen6e-5_dim_Tin60_re1000_0110-fs8}}
\caption*{Case L}
\label{fig:Viso2-L}
\end{subfigure}
  \caption{Optimization Cases J--L: on the small cavity and with $\mathrm{Re}=\text{1,000}$; Temperature on the outer domain walls and isocontour of flow speed at $10^{-3}$\,m/s, colored by the temperature.  }
\label{fig:Viso2-J2L}
\end{figure}

As the Reynolds number is increased to $\mathrm{Re}=\text{1,000}$, the cooling capacity of the fluid is enhanced. Consequently, when the objective is to cool down the surfaces of the insert cavity, the channels are pushed towards the cavity and touch its surfaces, in comparison with lower Reynolds topologies, which are at a distance to the cavity for comparable problem hyper-parameters. This effect is for example visible when comparing Case J (\cref{fig:SF-J2L,fig:Viso-J2L,fig:Viso2-J2L}) to its counterpart for $\mathrm{Re}=100$, i.e. Case A (\cref{fig:SF-A2I,fig:Viso-A2I,fig:Viso2-A2I}). 

Another interesting effect concerns the $\zeta$ value in adequation with the Reynolds number: as the higher cooling capacity of $\mathrm{Re}=\text{1,000}$ reduces the average and maximum temperatures in the domain, and since at higher speeds, the pressure losses are amplified, the relative weight of the pressure penalty term decreases with regards to the term on average temperature. Consequently, for a given $\zeta$ used for $\mathrm{Re}=100$, a lower $\zeta$ should be used for $\mathrm{Re}=\text{1,000}$  to produce a similar balance between the cost function components. To illustrate this idea, we can compare the Cases C and K which have the same hyper-parameters, but at two different Reynolds numbers. While the channels are not fully connected in Case C, signifying an insufficient pressure penalization, the same $\zeta$ produces well-connected channels in case K.

For $\mathrm{Re}=\text{1,000}$, changing the cost function (Cases J or K vs L) produces similar effects as for $\mathrm{Re}=100$ (Cases F vs I): when the objective shifts to minimizing the domain temperature in an average sense, the channels envelop the cavity such that the domain is isolated from the hot cavity.

Finally, we can observe from Table \ref{Tab:OptResults}, that except for Case L, which features the domain-wise average temperature in its cost function, the temperature undershoots are minimal, confirming that the mesh is generally sufficiently fine for the considered optimization cases.

\subsection{Design validation using the body-fitted solver}
\begin{figure}[t]
	\centering
	\begin{subfigure}{0.49\textwidth}
		\centering
		\includegraphics[width=\linewidth]{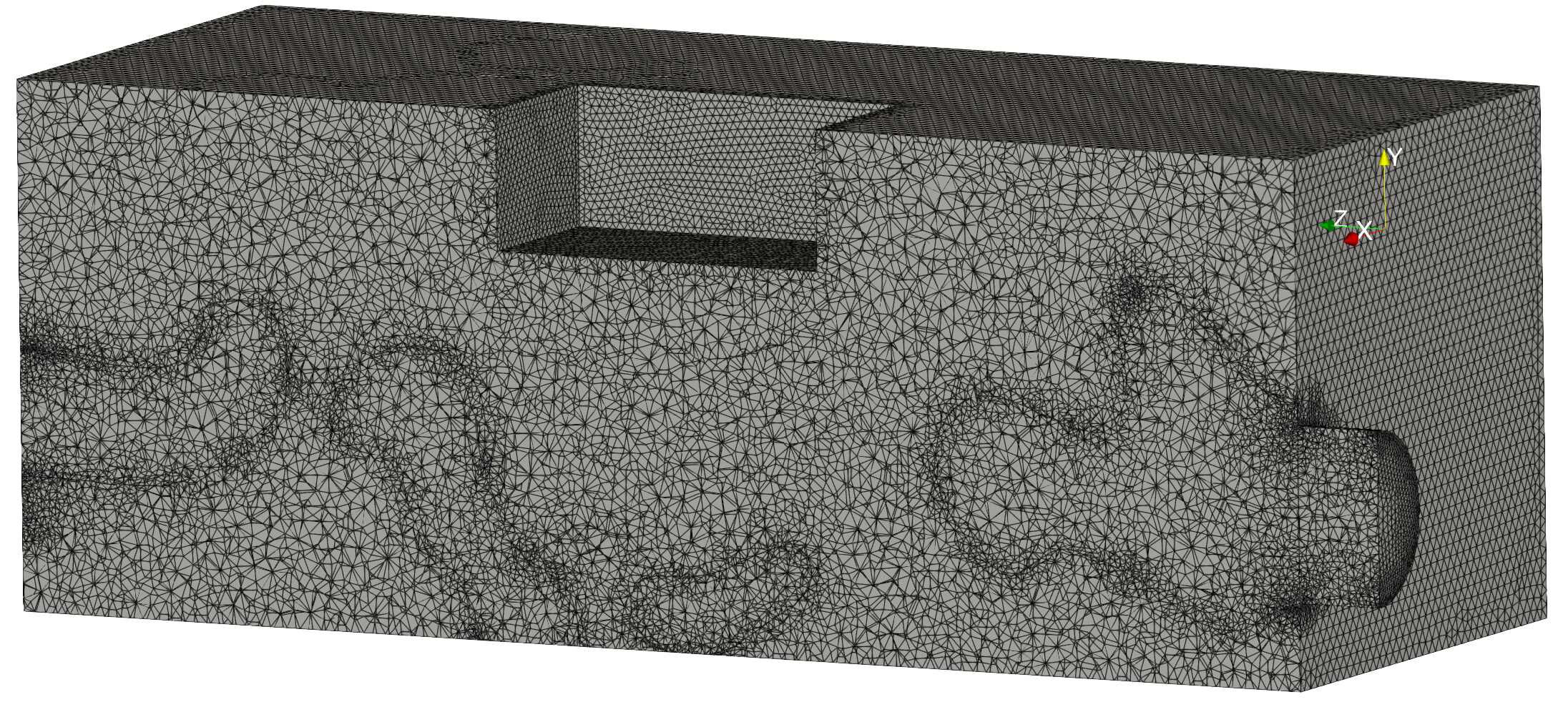}
		\caption{Cut view along the $y$--$z$ symmetry plane}
		\label{fig:caseA_conformalmeshA}
	\end{subfigure}
	\begin{subfigure}{0.49\textwidth}
		\centering
		\includegraphics[width=\linewidth]{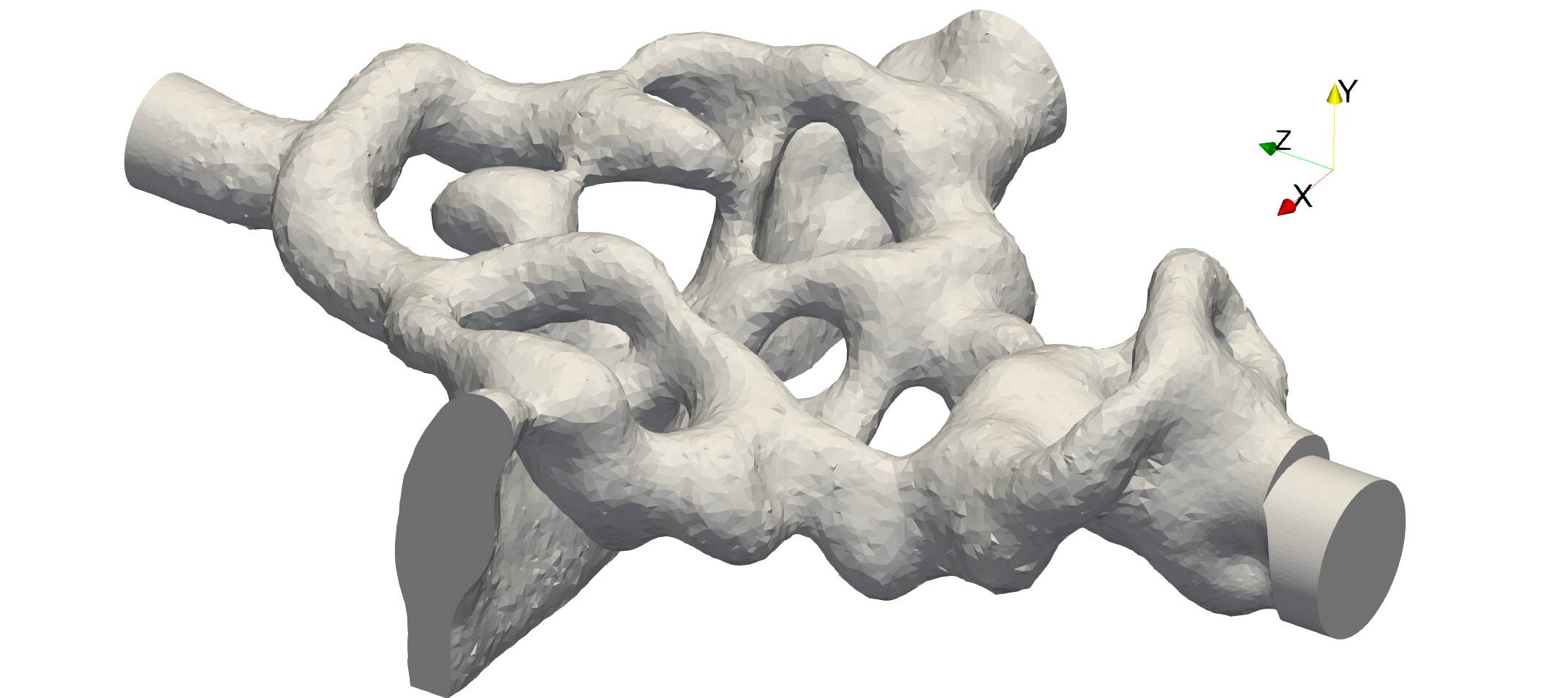}
		\caption{Fluid domain}
		\label{fig:caseA_conformalmeshB}
	\end{subfigure}
	\caption{Body-fitted mesh generated using the final design for Case A. The refined regions correspond to the position of the interface.}
	\label{fig:caseA_conformalmesh}
\end{figure}

\begin{figure}
	\centering
	\includegraphics[width=\linewidth]{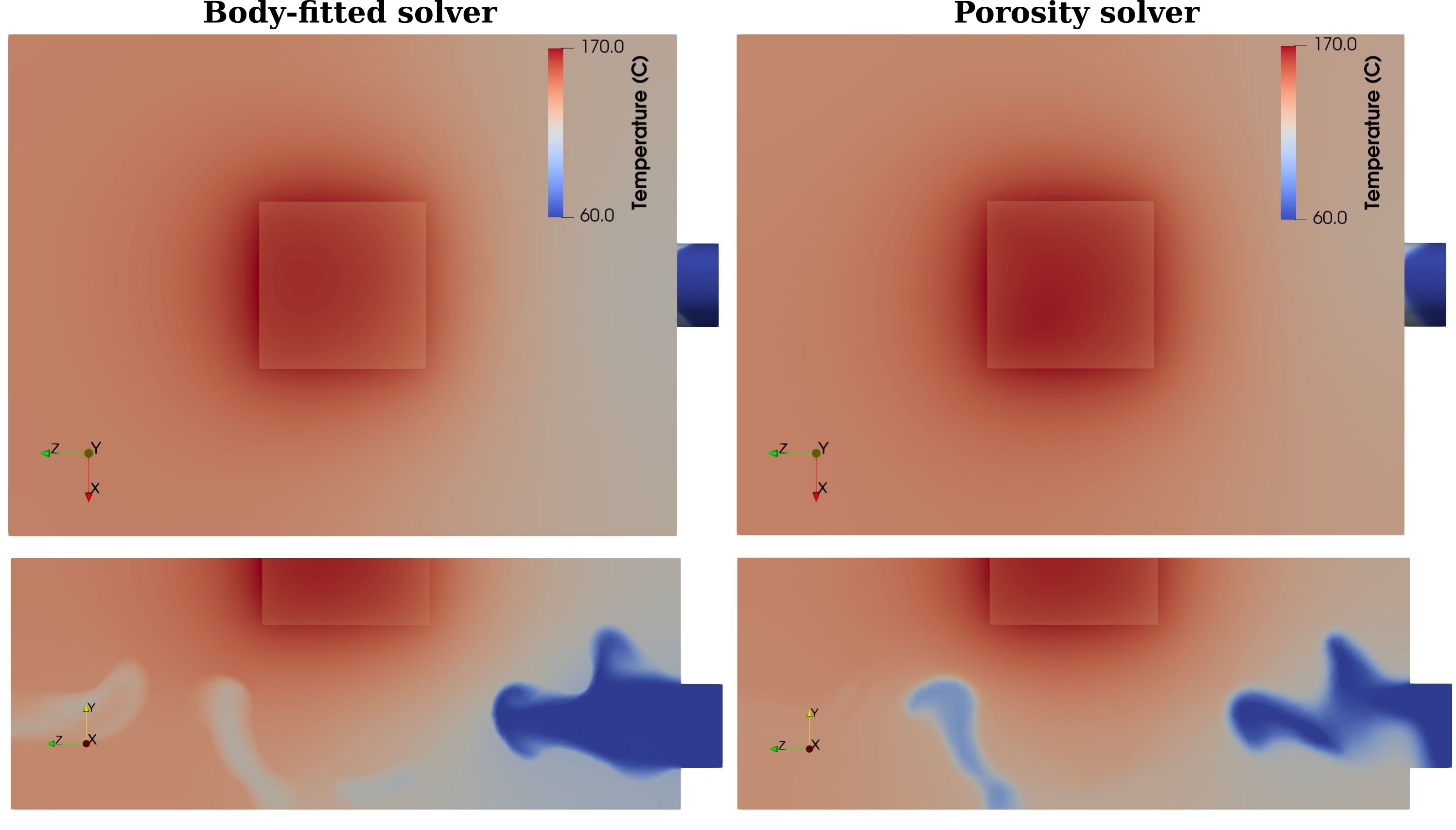}
	\caption{Top and cut (along the $y$--$z$ symmetry plane) views comparing the body-fitted (left) and porosity (right) temperature distributions for Case A. Both solutions were obtained using the body-fitted mesh (cf. Fig~\ref{fig:caseA_conformalmesh}.)}
	\label{fig:caseA_conformal-vs-diffuse}
\end{figure}

\newcommand{\degC}{\,^{\circ}\mathrm{C}}
A body-fitted design for Case A is generated using the Mmg \cite{dapognyThreedimensionalAdaptiveDomain2014} remeshing library, by splitting the mesh along the $\gamma=0.55$ isosurface. The latter value is chosen such that the total solid fraction of the resulting design is approximately 0.1 (\ie equal to $1-\bar{\gamma}_0$). The result can be visualized in \cref{fig:caseA_conformalmesh}. This mesh is now utilized to solve the Navier-Stokes and energy conservation equations for the conditions of Case A using both the body-fitted and porosity solvers. The resulting temperature distributions are compared in \cref{fig:caseA_conformal-vs-diffuse}. Discrepancies in the temperature can be observed close to the inlet and outlet. However, the top temperature distributions compare remarkably well. Furthermore, the domain-averaged temperature, $\bar{T}_\Omega$, is $127.9\degC$ for the body-fitted solver and $130.0\degC$ for the porosity one, \ie a discrepancy of less than 2\,\%. This further validates the porosity solver and confirms that the porous design flow leakage, which was identified in \Cref{fig:glyphs}, has a minimal effect on the domain-wise heat transfer.

\section{Conclusion}
In this work, a topology optimization methodology tailored to tackle the design of cooling channels in die casting molds is proposed. This purpose is motivated by the new design avenues made possible by additive manufacturing techniques. Our approach relies on a porous modeling of the conjugate  heat transfer, which produces a flexible and robust framework. The optimization regularization and control mechanisms adopted in this work are described. The scientific computing software arising from the discretization of the equations is numerically validated against a body-fitted (non-porous) solver on an industrially-relevant cooling channel case. A thorough investigation is performed to calibrate the Darcy friction factor based on its effect upon the pressure, temperature and velocity fields. Furthermore, the details of the adjoint-based gradient computation and the verification of the CHT solver via the method of manufactured solutions are presented in the appendices for the sake of completeness. Finally, the optimization results are presented for two Reynolds numbers, i.e. $\mathrm{Re} = 100$ and $\mathrm{Re} = 1,000$, and for two objective functions, \ie the domain-averaged and the cavity surface-averaged temperature. It is shown that the latter lead to designs with completely different paradigms, especially when the Reynolds number is low. Indeed, optimizations using the former objective function tend to generate designs which enclose the heated cavity in order to prevent heat from diffusing within the domain, at the cost of a much higher cavity temperature, which is undesirable for cooling channels in molds. The effects of the most impactful hyper-parameters of the problem are also discussed. A body-fitted mesh is generated from a selected porous solution and used to validate the optimized design. These results show the robustness and proven capabilities of the proposed framework for automated design of optimal cooling channels. To pursue this project further, we intent to undertake endeavors in the following directions: 1) inclusion of constraints considering additive manufacturing capabilities (e.g. prevent overhanging), 2) integration of structural optimization in a thermal-fluid-structural multi-objective framework with the goal of incorporating criteria on structural integrity and longevity, 3) development of mesh adaptation strategies to tackle the computational burden, and 4) full- or reduced-order modeling to account for turbulent effects.

\section{Acknowledgment}
This project is made possible through funding by the Office of Energy Research and Development of Canada, the Centre Qu\'ebecois de Recherche et de D\'eveloppement de l'Aluminium, the National Research Council of Canada (NRC) and the METALTec--NRC industrial consortium. The authors would like to thank NRC colleagues, Dr.\@ Marjan Molavi Zarandi, Mr.\@ Martin Audet and Mr.\@ Vincent Raymond, for their support of the project.

\appendix
\section{Sensitivity computation via the adjoint method}
\label{sec:adjoint}
In order to solve the optimization Problem \eqref{eq:optimProblem}, a gradient-based approach is adopted in this work which relies on the adjoint method to yield the sensitivity of the cost function to the nodal solid fraction values. In this section, we discuss the details of this methodology, applied to the loosely coupled system of Eqs. \eqref{eq:weak_cont_frac}, \eqref{eq:weak_mom_frac} and \eqref{eq:weak_energy_frac}, which has the following unified solution vector:
\begin{equation}
	\bm{U}(\bm{\gamma}):=[\bm{U}^\mathrm{NS}(\bm{\gamma}),\bm{U}^\mathrm{E}(\bm{\gamma})],
	\label{eq:unknows_vec}
\end{equation}
where $^\mathrm{NS}$ and $^\mathrm{E}$ exponents respectively refer to global unknowns of the Navier--Stokes (conservation of mass and momentum) and the (conservation of) energy equations, which more specifically read
\begin{align}
	\bm{U}^\mathrm{NS}(\bm{\gamma})&:=\left[\,\bm{p}(\bm{\gamma}),\bm{u_1}(\bm{\gamma}),\bm{u_2}(\bm{\gamma}),\bm{u_3}(\bm{\gamma})\,\right],
	\label{eq:unknows_vecNS}
	\\
	\bm{U}^\mathrm{E}(\bm{\gamma})&:=[\,\bm{T}(\bm{\gamma})\,].
	\label{eq:unknows_vecE}
\end{align}
Similarly, we can regroup the residual equations into a unified global vector as
\begin{equation}
	\begin{split}
		\bm{R}\,(\bm{\gamma},\,\bm{U}(\bm{\gamma})):=
		\left[\bm{R}^\mathrm{NS}\left(\bm{\gamma},\,\bm{U}^\mathrm{NS}(\bm{\gamma}),\,{\bm{U}^\mathrm{E}(\bm{\gamma})}\right),\, 
		\bm{R}^\mathrm{E}\left(\bm{\gamma},\,\bm{U}^\mathrm{NS}(\bm{\gamma}),\,\bm{U}^\mathrm{E}(\bm{\gamma})\right)\,\right].
		\label{eq:Res_sys}
	\end{split}
\end{equation}

In order to solve the system of Eq.~\eqref{eq:Res_sys},  i.e. $\bm{R}\,(\bm{\gamma},\bm{U}(\bm{\gamma}))$, we proceed in two steps: we first solve for  $\bm{R}^\mathrm{NS}\left(\bm{\gamma},\,\bm{U}^\mathrm{NS}(\bm{\gamma}),\,{ \bm{U}^\mathrm{E}(\bm{\gamma})}\right)$, and then for $ \bm{R}^\mathrm{E}\left(\bm{\gamma},\,\bm{U}^\mathrm{NS}(\bm{\gamma}),\,\bm{U}^\mathrm{E}(\bm{\gamma})\right)$. This process arises from the loose, one-way coupling of the $^\mathrm{NS}$ and $^\mathrm{E}$ subsystems and is justified by the fact that in the problems considered in this work, the fluid properties do not depend on the temperature.

Let us recall that the objective is to minimize the cost function of \eqref{eq:optimProblem}, concisely noted as $ \mathcal{C} \left(\bm{\gamma}, \bm{U}(\bm{\gamma}) \right)$, subject to the constraint that the sought minimum cancels the residuals. To enforce this constraint, we adopt a \textit{reduced-space approach} and leverage  the method of Lagrange multipliers \cite{gilesIntroductionAdjointApproach2000} to constitute the following Lagrangian function:
\begin{equation}
	\begin{split}
		&\mathcal{L}\left(\bm{\gamma}, \bm{U}(\bm{\gamma}) , \bm{\lambda}^\mathrm{NS},\bm{\lambda}^\mathrm{E} \right):=
		\mathcal{C} \left(\bm{\gamma}, \bm{U}^\mathrm{NS}(\bm{\gamma}),\,\bm{U}^\mathrm{E}(\bm{\gamma}) \right) 
		\\&+ {\bm{\lambda}^\mathrm{NS}}^\top \bm{R}^\mathrm{NS}\left(\bm{\gamma},\,\bm{U}^\mathrm{NS}(\bm{\gamma}),\,\bm{U}^\mathrm{E}(\bm{\gamma})\right)
		+ {\bm{\lambda}^\mathrm{E}}^\top \bm{R}^\mathrm{E}\left(\bm{\gamma},\,\bm{U}^\mathrm{NS}(\bm{\gamma}),\,\bm{U}^\mathrm{E}(\bm{\gamma})\right),
		\label{eq:lagrangien_decpld}
	\end{split}
\end{equation}
where  $\bm{\lambda}^\mathrm{NS}$ and $\bm{\lambda}^\mathrm{E}$ are the Lagrange multipliers of their respective residuals and are also referred to as \textit{adjoint solution}, or \textit{dual solution} in the literature. By deriving the Lagrangian function with regards to the design variables, we obtain
\begin{equation}
	\begin{split}
		\frac{\mathrm{d} \mathcal{L}}{\mathrm{d}\bm{\gamma}} &= \frac{\partial \mathcal{C}}{\partial \bm{\gamma}} + \frac{\partial \mathcal{C}}{\partial \bm{U}^\mathrm{NS}} \frac{\mathrm{d} \bm{U}^\mathrm{NS}}{\mathrm{d} \bm{\gamma}} +
		\frac{\partial \mathcal{C}}{\partial \bm{U}^\mathrm{E}} \frac{\mathrm{d} \bm{U}^\mathrm{E}}{\mathrm{d} \bm{\gamma}} \\ &+{\bm{\lambda}^\mathrm{NS}}^\top \left(
		\frac{\partial \bm{R}^\mathrm{NS}}{\partial \bm{\gamma}} + \frac{\partial \bm{R}^\mathrm{NS}}{\partial \bm{U}^\mathrm{NS}}
		\frac{\mathrm{d} \bm{U}^\mathrm{NS}}{\mathrm{d} \bm{\gamma}}
		+ \frac{\partial \bm{R}^\mathrm{NS}}{\partial \bm{U}^\mathrm{E}}
		\frac{\mathrm{d} \bm{U}^\mathrm{E}}{\mathrm{d} \bm{\gamma}}
		\right)\\
		&+{\bm{\lambda}^\mathrm{E}}^\top \left(
		\frac{\partial \bm{R}^\mathrm{E}}{\partial \bm{\gamma}} + \frac{\partial \bm{R}^\mathrm{E}}{\partial \bm{U}^\mathrm{NS}}
		\frac{\mathrm{d} \bm{U}^\mathrm{NS}}{\mathrm{d} \bm{\gamma}}
		+ \frac{\partial \bm{R}^\mathrm{E}}{\partial \bm{U}^\mathrm{E}}
		\frac{\mathrm{d} \bm{U}^\mathrm{E}}{\mathrm{d} \bm{\gamma}}
		\right).
	\end{split}
\end{equation}
These terms can be reorganized such that the coefficients of the solution sensitivities ($\frac{\mathrm{d} \bm{U}^\mathrm{NS}}{\mathrm{d}\bm{\gamma}}$ and $\frac{\mathrm{d} \bm{U}^\mathrm{E}}{\mathrm{d} \bm{\gamma}})$ are gathered together, viz.
\begin{equation}
	\begin{split}
		\frac{\mathrm{d} \mathcal{L}}{\mathrm{d}\bm{\gamma}} &= 
		\frac{\partial\mathcal{C}}{\partial \bm{\gamma}} 
		+ {\bm{\lambda}^\mathrm{NS}}^\top \, \frac{\partial \bm{R}^\mathrm{NS}}{\partial \bm{\gamma}} 
		+ {\bm{\lambda}^\mathrm{E}}^\top \, \frac{\partial \bm{R}^\mathrm{E}}{\partial \bm{\gamma}} \\
		&+  \left( \underbrace{\frac{\partial \mathcal{C}}{\partial \bm{U}^\mathrm{NS}}  
			+ {\bm{\lambda}^\mathrm{NS}}^\top\, \frac{\partial \bm{R}^\mathrm{NS}}{\partial \bm{U}^\mathrm{NS}}
			+ {\bm{\lambda}^\mathrm{E}}^\top\, \frac{\partial \bm{R}^\mathrm{E}}{\partial \bm{U}^\mathrm{NS}}}_\text{=0} \right)
		\frac{\mathrm{d} \bm{U}^\mathrm{NS}}{\mathrm{d} \bm{\gamma}}\\
		&+  \left( \underbrace{\frac{\partial \mathcal{C}}{\partial \bm{U}^\mathrm{E}}  
			+ {\bm{\lambda}^\mathrm{NS}}^\top\, \frac{\partial \bm{R}^\mathrm{NS}}{\partial \bm{U}^\mathrm{E}}
			+ {\bm{\lambda}^\mathrm{E}}^\top\, \frac{\partial \bm{R}^\mathrm{E}}{\partial \bm{U}^\mathrm{E}}}_\text{=0} \right)
		\frac{\mathrm{d} \bm{U}^\mathrm{E}}{\mathrm{d} \bm{\gamma}}.
	\end{split}
\end{equation}
The purpose is to cancel these sensitivity coefficients, such that the need for computing the sensitivities is avoided as, often, there is a considerable cost associated with their computation. Requiring the cancellation of the terms between parentheses, provides us with a system of two coupled algebraic equations, which reads
\begin{align}
	\left(\frac{\partial \bm{R}^\mathrm{NS}}{\partial \bm{U}^\mathrm{NS}} \right)^\top {\bm{\lambda}^\mathrm{NS}}
	+ \left(\frac{\partial \bm{R}^\mathrm{E}}{\partial \bm{U}^\mathrm{NS}} \right)^\top {\bm{\lambda}^\mathrm{E}}
	&= - \left(\frac{\partial \mathcal{C}}{\partial \bm{U}^\mathrm{NS}} \right)^\top,
	\label{eq:adjoint_decpld_NS}
	\\
	\left(\frac{\partial \bm{R}^\mathrm{NS}}{\partial \bm{U}^\mathrm{E}} \right)^\top {\bm{\lambda}^\mathrm{NS}}
	+ \left(\frac{\partial \bm{R}^\mathrm{E}}{\partial \bm{U}^\mathrm{E}} \right)^\top {\bm{\lambda}^\mathrm{E}}
	&= -\left( \frac{\partial \mathcal{C}}{\partial \bm{U}^\mathrm{E}} \right)^\top.
	\label{eq:adjoint_decpld_E}
\end{align}
The Eqs.~\eqref{eq:adjoint_decpld_NS} and \eqref{eq:adjoint_decpld_E} are called the \textit{adjoint equations}. The solution of this system necessitates an iterative approach or a strong coupling into a monolithic form. However, as in the present work $\frac{\partial \bm{R}^\mathrm{NS}}{\partial \bm{U}^\mathrm{E}}=\bm{0}$, the coupling becomes unidirectional, enabling us to proceed in two consecutive steps, without further iterations, i.e.
\begin{align}
	\left(\frac{\partial \bm{R}^\mathrm{E}}{\partial \bm{U}^\mathrm{E}} \right)^\top {\bm{\lambda}^\mathrm{E}}
	&= -\left( \frac{\partial \mathcal{C}}{\partial \bm{U}^\mathrm{E}} \right)^\top,
	\label{eq:adjoint_decpld_E2}
	\\
	\left(\frac{\partial \bm{R}^\mathrm{NS}}{\partial \bm{U}^\mathrm{NS}} \right)^\top {\bm{\lambda}^\mathrm{NS}}
	&= -\left(\frac{\partial \bm{R}^\mathrm{E}}{\partial \bm{U}^\mathrm{NS}} \right)^\top {\bm{\lambda}^\mathrm{E}} - \left(\frac{\partial \mathcal{C}}{\partial \bm{U}^\mathrm{NS}} \right)^\top,
	\label{eq:adjoint_decpld_NS2}
\end{align}
The solution of Eq. \eqref{eq:adjoint_decpld_E2}, followed by that of Eq. \eqref{eq:adjoint_decpld_NS2}, allow us to compute the Lagrangian function gradient as
\begin{equation}
	\frac{\mathrm{d} \mathcal{L}}{\mathrm{d}\bm{\gamma}} = 
	\frac{\partial\mathcal{C}}{\partial \bm{\gamma}} 
	+ {\bm{\lambda}^\mathrm{NS}}^\top \, \frac{\partial \bm{R}^\mathrm{NS}}{\partial \bm{\gamma}} 
	+ {\bm{\lambda}^\mathrm{E}}^\top \, \frac{\partial \bm{R}^\mathrm{E}}{\partial \bm{\gamma}},
	\label{eq:DCDgamma_decpld}
\end{equation}
which is further complemented in Section \ref{sec:smthdheavy} to yield the Lagrangian function gradient with regards to the design parameters and serve to drive the optimization.

The adjoint solver implementation was verified by sensitivity analysis using finite differencing, as explained in \ref{sec:verif_adjoint}.

Finally, note that we opted to impose an inequality constraint on a functional of $\gamma$, via the optimizer (see Section \ref{sec:FSconst}). Alternatively, one could also opt for equality constraints on the same functional, but imposed via a penalization of the cost function.  In the former case, the term $\frac{\partial\mathcal{C}}{\partial \bm{\gamma}}$ in Eq.\@ \eqref{eq:DCDgamma_decpld} cancels, i.e. the cost function does not explicitly depend on the solid fraction.

\section{Verification of the adjoint solver}
\label{sec:verif_adjoint}
Prior to proceeding to optimization, one needs to ensure that the implementation of the adjoint method for the computation of the Lagrangian function gradient (LFG) (see Section \ref{sec:adjoint}) is sound. To this end, and taking into account that the method of manufactured solutions is not available for the discrete adjoint approach, we compare the LFG computed by the adjoint method, i.e. Eq. \eqref{eq:DCDgamma_decpld}, to the total cost function gradient, $\d{\mathcal{C}}{\bm{\gamma}}$, approximated by sensitivity analysis via finite differencing, as
\begin{equation}
	\pd{\mathcal{C}}{{\gamma_i}} \approx  \frac{\mathcal{C} \left(\bm{\acute{\gamma}}^{i}, \,\bm{U}(\bm{\acute{\gamma}}^{i})  \right)  - \mathcal{C} \left(\bm{\gamma}, \,\bm{U}(\bm{\gamma})  \right) } {\epsilon},
	\label{eq:fd}
\end{equation}
where ${\acute{\gamma}_j^{i}} := \gamma_j + \epsilon \, \delta_{ij}$ with $\epsilon$ and $\delta_{ij}$ respectively designating a small perturbation and the Kronecker delta. We adopt a dynamic approach for determining the value of the former, which for a given $\gamma_j$, is set to $\epsilon:=c_\epsilon\max(\gamma_j\,\epsilon_0,\epsilon_0\,\epsilon_s)$, where $c_\epsilon=+1$ if $\gamma_j+\epsilon\leq1$ and $c_\epsilon=-1$ otherwise,
$\epsilon_0=10^{-6}$ is a default perturbation and $\epsilon_s=0.1$.

Note that the computation of the cost function necessitates the resolution of the system represented by Eq.~\eqref{eq:Res_sys}. The sensitivity-based approach is hence particularly costly as it requires $N_\mathrm{DOFs}+1$ system solutions. Consequently, we consider a CHT problem on a coarse mesh which consists of a cuboid geometry heated on the top and traversed by an implicitly represented circular cylinder, where the cooling fluid flows at $\mathrm{Re}=100$.

\begin{figure}[h!]
	\centering
	\includegraphics[width=0.5\textwidth,trim={0.cm 0.cm 0.cm 0.cm},clip]{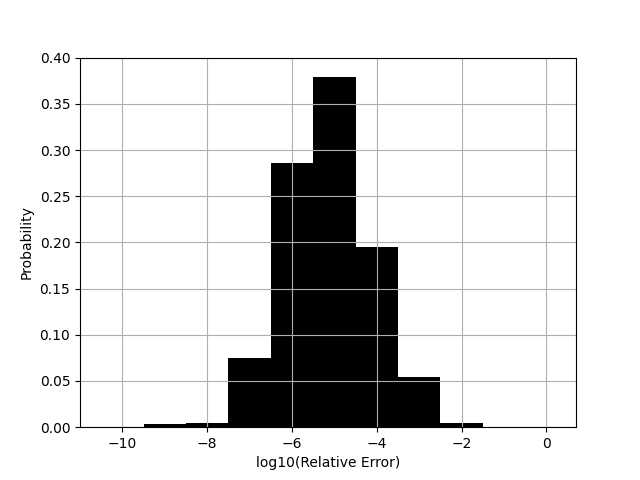}
	\caption{Histogram of matching significant digits between the cost function computation via the adjoint solver versus the sensitivity analysis. }
	\label{fig:adjverif_histo}
\end{figure}
\begin{figure}[h!]
	\centering
	\begin{subfigure}[h]{0.31\textwidth}
		\includegraphics[width=\textwidth,trim={0.cm 0.cm 0.cm 0.cm},clip]{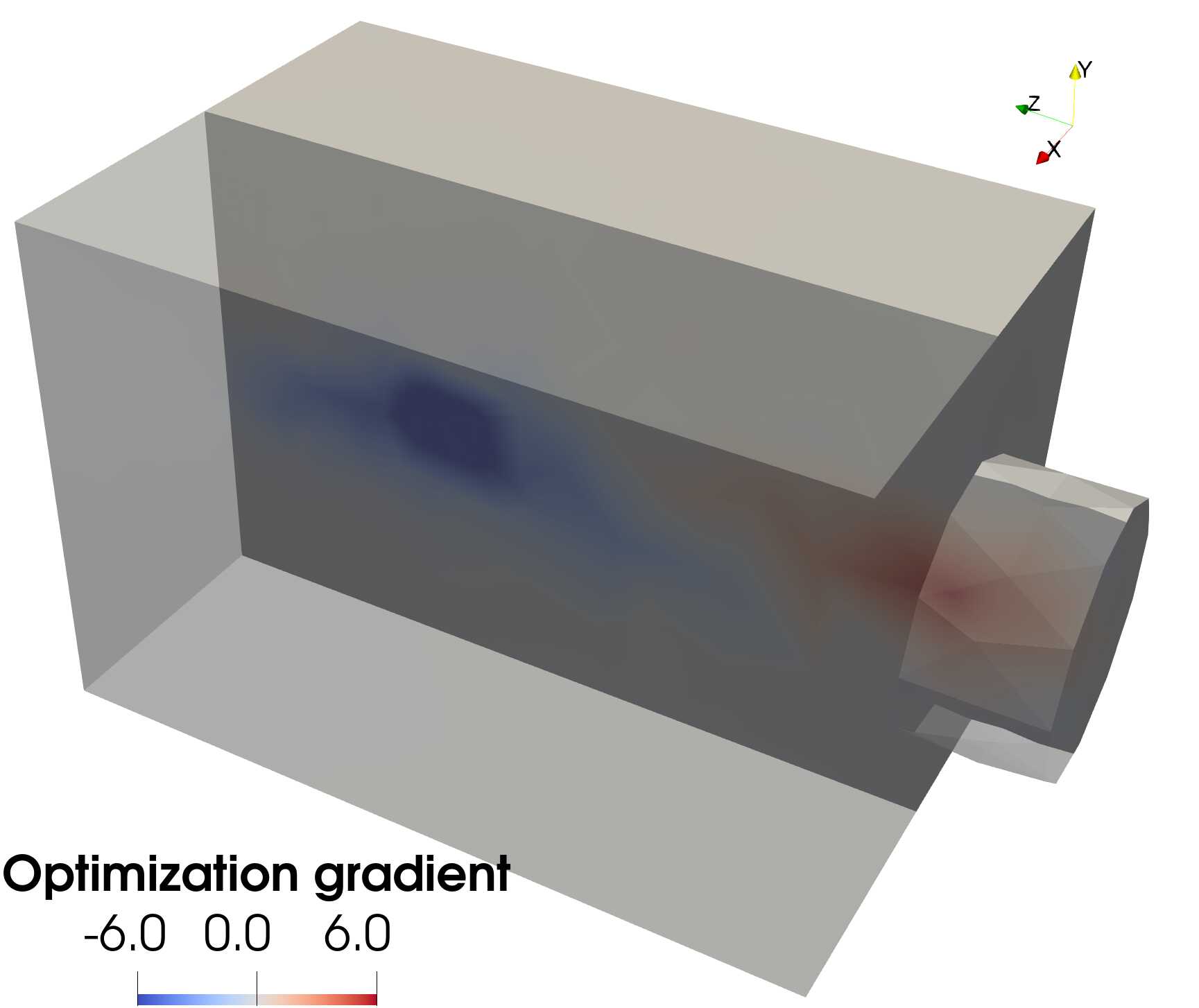}
		\caption{}
		\label{fig:LFG}
	\end{subfigure}
	~
	\begin{subfigure}[h]{0.31\textwidth}
		\includegraphics[width=\textwidth,trim={0.cm 0.cm 0.cm 0.cm},clip]{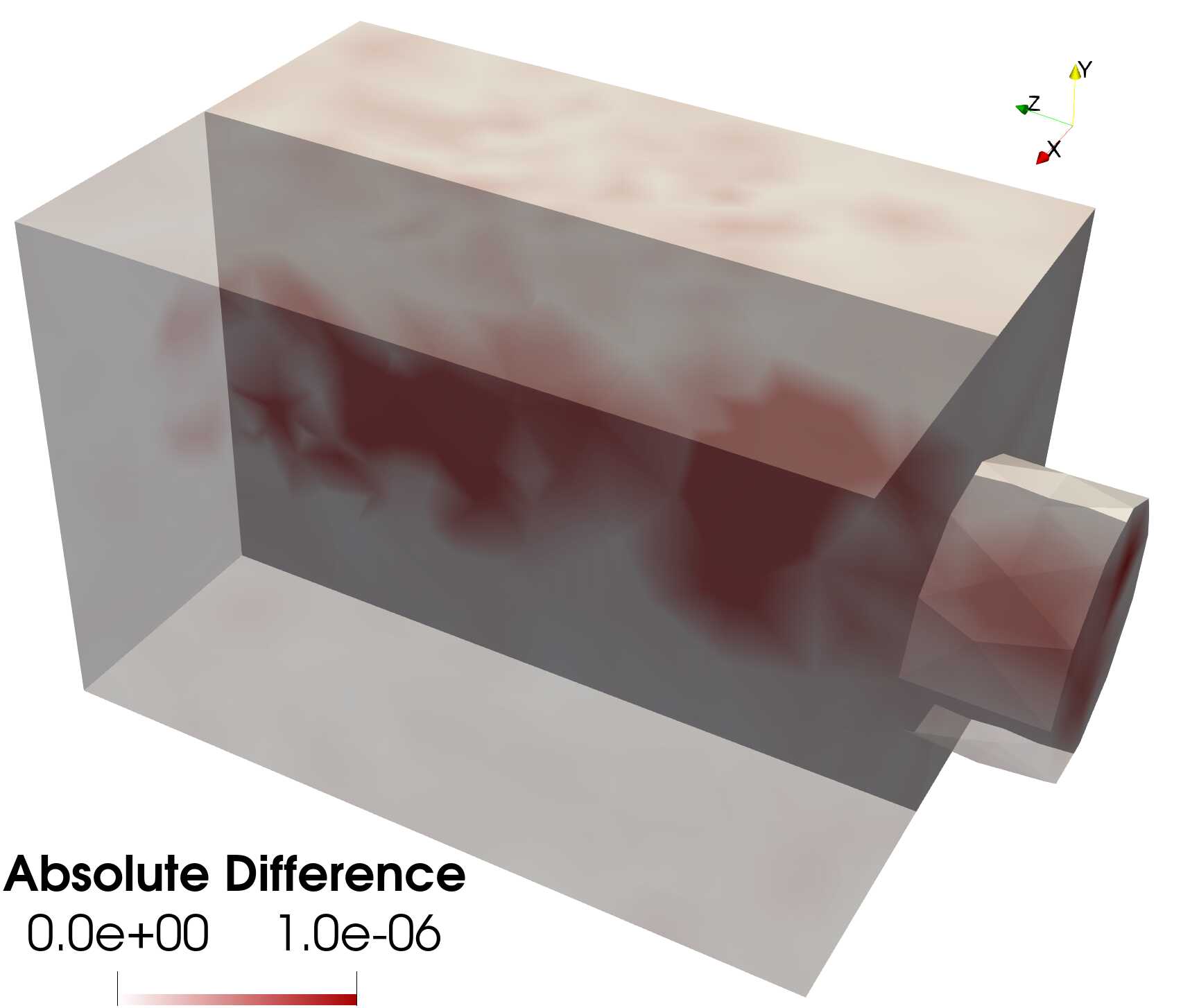}
		\caption{}
		\label{fig:LFGabs}
	\end{subfigure}
	~
	\begin{subfigure}[h]{0.31\textwidth}
		\includegraphics[width=\textwidth,trim={0.cm 0.cm 0.cm 0.cm},clip]{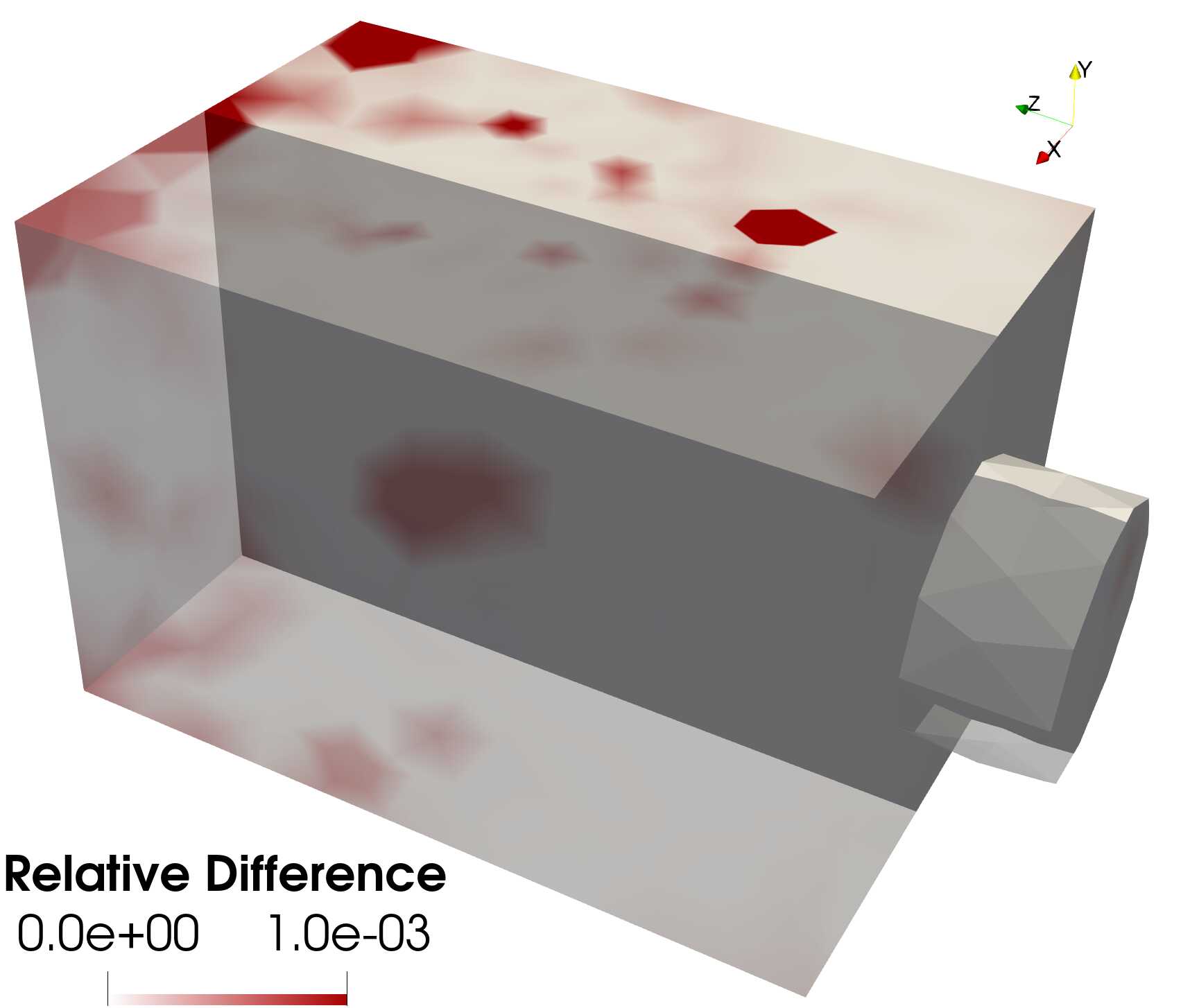}
		\caption{}
		\label{fig:LFGrel}
	\end{subfigure}
	\caption{Lagrangian function gradient as computed by the adjoint method and its absolute and relative differences with regards to the cost function computed by sensitivity analysis.}
	\label{fig:adjverif}
\end{figure}

Figure \ref{fig:adjverif_histo} shows the probability distribution of the number of matching significant digits between the adjoint-based LFG and the sensitivity-based gradient, based on 961 nodal values. The distribution has a maximum of nine, an average of five and a minimum of two matching significant digits. Figure \ref{fig:adjverif} provides a closer look at the spatial distribution of the LFG and its absolute and relative differences with regards to its sensitivity-based counterpart. As expected, the absolute differences are the most important in regions of significant gradient value, whereas the relative differences seem to follow a random pattern and are most notable in regions of low gradient magnitude. We have noted that varying the values of $\epsilon_0$ and $\epsilon_s$ affects the distribution of these differences and thus points at the inaccuracy of the finite differencing approach as the explanation of the observed differences.

In the light of these observations, we consider the LFG computed by adjoint solver to be sufficiently reliable in driving the optimization.

\section{Verification of the CHT solver via the method of manufactured solutions}
\label{sec:mms}
In order to verify the implementation of the conjugate heat transfer model described in Section \ref{sec:cht_mod}, we leverage the method of manufactured solutions (MMS) via the auto-generative scripts provided in \cite{navahComprehensiveHighorderSolver2018}. The first step of the MMS consists of setting a manufactured solution, which in our case is described by the following fields:
\begin{equation}
	\begin{split}
		u^\mathrm{MS} (\bm{x}) &:= \sin(x)\,\cos(y)\,\cos(z), \\
		v^\mathrm{MS} (\bm{x}) &:= \cos(x)\,\sin(y)\,\cos(z), \\
		w^\mathrm{MS} (\bm{x}) &:= -2\,\cos(x)\,\cos(y)\,\sin(z),\\
		p^\mathrm{MS} (\bm{x}) &:= \sin(x)\,\cos(y)\,\cos(z),\\
		T^\mathrm{MS} (\bm{x}) &:= -\sin(2\,\pi\,x)\,\sin(2\,\pi\,y)\,\sin(2\,\pi\,z),\\
		\gamma^\mathrm{MS} (\bm{x}) &: = 0.5\,(-0.5\,x - 0.5\,y + z)+0.5.
	\end{split}
\end{equation}
Furthermore, we set the fluid and solid properties, the Darcy coefficient, as well as the parameters of the interpolation laws, to the following values:
$\mu = 1$, $\rho = 1$, $c_p = 1$, $C_\alpha = 10$,
$\alpha_f = 1$, $\alpha_s = 0$, $P_\alpha = 2$,
$\kappa_f = 1$, $\kappa_s = 100$ and  $P_\kappa = 3$.
The next step of the MMS is to inject these manufactured fields and values into the governing equations and thus find the expressions of the forcing functions ($s_i$ in \eqref{eq:mom_base} and $s_q$ in \eqref{eq:energ}) which balance the equations out. The forcing functions for the chosen manufactured solution are expressed by:

\begin{multline}
	s_1^\mathrm{MS}(\bm{x})  := 2.5\,(0.5\,x + 0.5\,y - z - 1)^2\,\sin(x)\,\cos(y)\,\cos(z) -\sin(x)\,\sin(y)^2\,\cos(x)\,\cos(z)^2 \\
	+ 2\,\sin(x)\,\sin(z)^2\,\cos(x)\,\cos(y)^2 + \sin(x)\,\cos(x)\,\cos(y)^2\,\cos(z)^2 \\
	+ 3\,\sin(x)\,\cos(y)\,\cos(z) + \cos(x)\,\cos(y)\,\cos(z),
\end{multline}

\begin{multline}
	s_2^\mathrm{MS}(\bm{x})  :=(2.5\,(0.5\,x + 0.5\,y - z - 1)^2\,\cos(x)\,\cos(z) - \sin(x)^2\,\cos(y)\,\cos(z)^2 - \sin(x)\,\cos(z) \\
	+ 2\,\sin(z)^2\,\cos(x)^2\,\cos(y) + \cos(x)^2\,\cos(y)\,\cos(z)^2 + 3\,\cos(x)\,\cos(z))\,\sin(y),
\end{multline}

\begin{multline}
	s_3^\mathrm{MS}(\bm{x})  :=(-5\,(0.5\,x + 0.5\,y - z - 1)^2\,\cos(x)\,\cos(y) + 2\,\sin(x)^2\,\cos(y)^2\,\cos(z) \\
	- \sin(x)\,\cos(y) + 2\,\sin(y)^2\,\cos(x)^2\,\cos(z) + 4\,\cos(x)^2\,\cos(y)^2\,\cos(z) - 6\,\cos(x)\,\cos(y))\,\sin(z),
\end{multline}

\begin{multline}
	s_q^\mathrm{MS}(\bm{x})  :=\pi\,(-12\,\pi\,(12.375\,(-0.5\,x - 0.5\,y + z + 1)^3 + 1.0)\,\sin(2\,\pi\,x)\,\sin(2\,\pi\,y)\,\sin(2\,\pi\,z) \\
	+ 74.25\,(-0.5\,x - 0.5\,y + z + 1)^2\,\sin(2\,\pi\,x)\,\sin(2\,\pi\,y)\,\cos(2\,\pi\,z) \\
	- 37.125\,(-0.5\,x - 0.5\,y + z + 1)^2\,\sin(2\,\pi\,x)\,\sin(2\,\pi\,z)\,\cos(2\,\pi\,y) \\
	- 37.125\,(-0.5\,x - 0.5\,y + z + 1)^2\,\sin(2\,\pi\,y)\,\sin(2\,\pi\,z)\,\cos(2\,\pi\,x) \\
	- 2\,\sin(x)\,\sin(2\,\pi\,y)\,\sin(2\,\pi\,z)\,\cos(y)\,\cos(z)\,\cos(2\,\pi\,x) \\
	- 2\,\sin(y)\,\sin(2\,\pi\,x)\,\sin(2\,\pi\,z)\,\cos(x)\,\cos(z)\,\cos(2\,\pi\,y) \\
	+ 4\,\sin(z)\,\sin(2\,\pi\,x)\,\sin(2\,\pi\,y)\,\cos(x)\,\cos(y)\,\cos(2\,\pi\,z)).
\end{multline}

Note that the manufactured velocity field is divergence-free such that the equation of conservation of mass is inherently balanced without the need of a forcing function.

\begin{figure}[h]
	\centering
	\begin{subfigure}[h]{0.48\textwidth}
		\includegraphics[width=\textwidth,trim={0.4cm 0.5cm 0.35cm 0.35cm},clip]{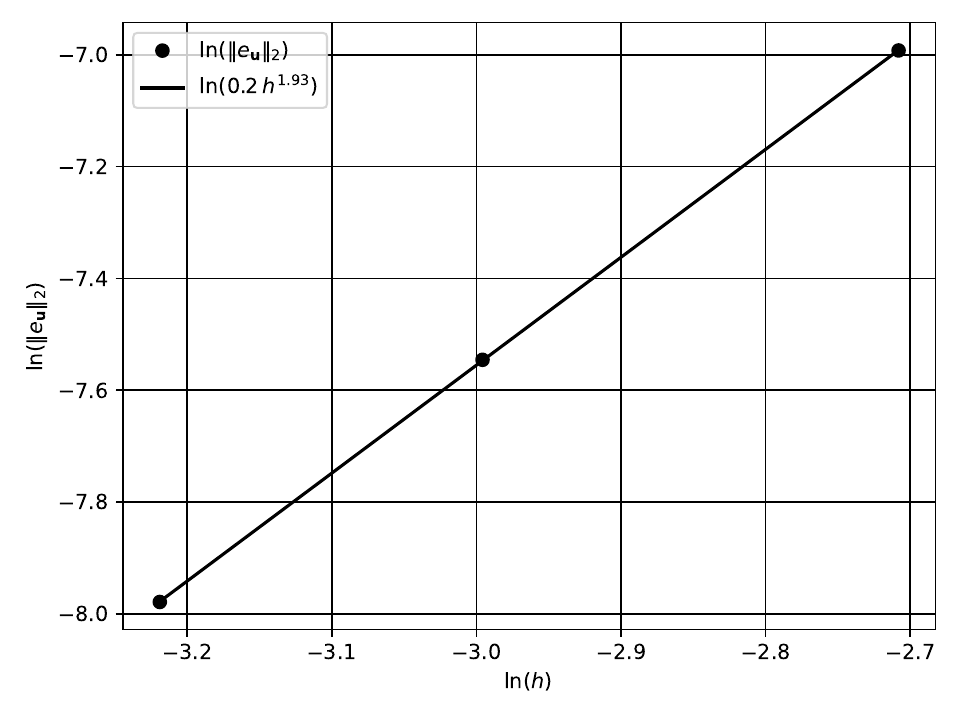}
		\caption{Velocity}
	\end{subfigure}
	~
	\begin{subfigure}[h]{0.48\textwidth}
		\includegraphics[width=\textwidth,trim={0.4cm 0.5cm 0.35cm 0.35cm},clip]{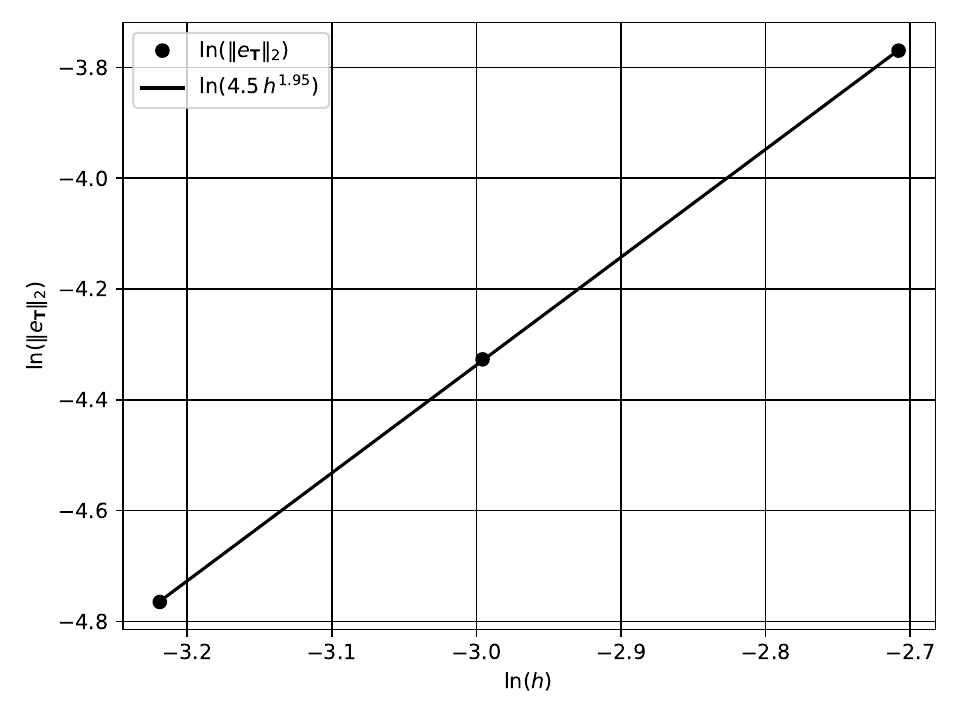}
		\caption{Temperature}
	\end{subfigure}
	\caption{Evolution of the errors  of $\bm{u}$ and $\bm{T}$  in $L^2$ norm and with mesh refinement, as a function of $h$, the element size.}
	\label{fig:mms_errors}
\end{figure}

The domain of the problem is $\bm{\Omega}:=(x,y,z)\in[0,1]^3$ and the manufactured velocity and temperature values are applied on all boundaries via Dirichlet conditions. We consider a sequence of three uniform and structured meshes, each containing in the order of increasing refinement, $5\times15^3$, $5\times20^3$ and $5\times25^3$ tetrahedra. The results of this verification exercise are shown in Fig.\ \ref{fig:mms_errors}, where the $L^2$ norms of the errors in velocity and temperature, versus element size, are shown to diminish with mesh refinement as expected, and more importantly, with a sustained quadratic rate. In the light of these outcomes, we can establish a good level of confidence that the implementation of the conjugate heat transfer model of Section \ref{sec:cht_mod} is sound.

\end{document}